\documentclass[twocolumn]{aastex61}
\pdfoutput=1 
\usepackage{amsmath,amstext}
\usepackage[T1]{fontenc}
\usepackage{apjfonts} 
\usepackage[figure,figure*]{hypcap}

\newcommand{\mh}[0]{[M/H]}
\newcommand{\afe}[0]{[$\alpha$/Fe] }
\newcommand{\afens}[0]{[$\alpha$/Fe]} 

\newcommand{\amns}[0]{[$\alpha$/M]} 
\newcommand{\feh}[0]{[Fe/H] }
\newcommand{\fehns}[0]{[Fe/H]} 
\newcommand{\teff}[0]{$T_{\rm eff}$ }
\newcommand{\teffns}[0]{$T_{\rm eff}$} 
\newcommand{\logg}[0]{$\log{g}$ }
\newcommand{\loggns}[0]{$\log{g}$} 
\newcommand{\dnu}[0]{\ensuremath{\Delta \nu}}
\newcommand{\fdnu}[0]{\ensuremath{f_{\Delta \nu}}}
\newcommand{\nmax}[0]{\ensuremath{\nu_{\rm \max}}}
\newcommand{\fnmax}[0]{\ensuremath{f_{\nu \max}}}
\newcommand{\rsun}[0]{\ensuremath{R_{\odot}}}
\newcommand{\msun}[0]{\ensuremath{M_{\odot}}}

\newcommand{\kep}[0]{\textsl{Kepler}}
\newcommand{\gaia}[0]{\textsl{Gaia}}

\shorttitle{APOKASC-3}
\shortauthors{Pinsonneault et al.}

\begin{document}

\title{APOKASC-3: The Third Joint Spectroscopic and Asteroseismic Catalog for Evolved Stars in the \kep\ Fields}

\author[0000-0002-7549-7766]{Marc H. Pinsonneault}
\affiliation{Department of Astronomy, The Ohio State University, Columbus, OH 43210, USA}

\author[0000-0002-7550-7151]{Joel C. Zinn}
\affiliation{Department of Physics and Astronomy, California State University, Long Beach, Long Beach, CA 90840, USA}
\author[0000-0002-4818-7885]{Jamie Tayar}
\affiliation{Department of Astronomy, University of Florida, Gainesville, FL 32611, USA }
\author{Aldo Serenelli}
\affiliation{Institute of Space Sciences (ICE, CSIC), Carrer de Can Magrans S/N, Campus UAB, E-08193 Bellaterra, Spain}
\affiliation{Institut d'Estudis Espacials de Catalunya, Carrer Gran Capitá 2, E-08034, Barcelona, Spain}
\author[0000-0002-8854-3776]{Rafael A. Garc\'ia}
\affiliation{Universit\'e Paris-Saclay, Universit\'e Paris Cit\'e, CEA, CNRS, AIM, 91191, Gif-sur-Yvette, France}
\author[0000-0002-0129-0316]{Savita Mathur}
\affiliation{Instituto de Astrofsica de Canarias (IAC), C/Va Lactea, s/n, E-38200 La Laguna, Tenerife, Spain}
\affiliation{Universidad de La Laguna (ULL), Departamento de Astrof\'isica, E-38206 La Laguna, Tenerife, Spain}
\affiliation{Space Science Institute, 4750 Walnut street, Suite 205, Boulder, CO 80301, USA}
\author{Mathieu Vrard}
\affiliation{Department of Astronomy, The Ohio State University, Columbus, OH 43210, USA}
\affiliation{Observatoire de la Côte d’Azur, CNRS, Laboratoire Lagrange, Bd de l’Observatoire, CS 34229, 06304 Nice Cedex 4, France}
\author{Yvonne P. Elsworth}
\affiliation{University of Birmingham, School of Physics and Astronomy, Edgbaston, Birmingham B15 2TT, UK}
\affiliation{Stellar Astrophysics Centre, Department of Physics and Astronomy, Aarhus University, Ny Munkegade 120, DK-8000 Aarhus C, Denmark}
\author[0000-0002-7547-1208]{Benoit Mosser}
\affiliation{LESIA, Observatoire de Paris, Universit\'e PSL, CNRS, Sorbonne Universit\'e, Universit\'e de Paris, 92195 Meudon, France}
\author[0000-0002-4879-3519]{Dennis Stello}
\affiliation{School of Physics, University of New South Wales, NSW 2052, Australia}
\affiliation{Sydney Institute for Astronomy (SIfA), School of Physics, University of Sydney, NSW 2006 Australia}
\affiliation{ARC Centre of Excellence for All Sky Astrophysics in 3 Dimensions (ASTRO 3D), Australia}
\author[0000-0002-0656-032X]{Keaton J. Bell}
\affiliation{Department of Physics, Queens College, City University of New York, Flushing, NY-11367, USA}
\author[0000-0003-0142-4000]{Lisa Bugnet}
\affiliation{Institute of Science and Technology Austria (ISTA), Am campus 1, Klosterneuburg, Austria}
\author[0000-0001-8835-2075]{Enrico Corsaro}
\affiliation{INAF - Osservatorio Astrofisico di Catania, via S. Sofia 78, 95123 Catania, Italy}
\author[0000-0001-8330-5464]{Patrick Gaulme}
\affiliation{Th\"{u}ringer Landessternwarte, Sternwarte 5, 07778 Tautenburg, Germany}
\affiliation{Department of Astronomy, New Mexico State University, P.O. Box 30001, MSC 4500, Las Cruces, NM 88003-8001, USA}
\author[0000-0002-1463-726X]{Saskia Hekker}
\affiliation{Heidelberg University (ZAH/LSW), K\"onigstuhl 12, 69118 Heidelberg, Germany}
\affiliation{Heidelberg Institute for Theoretical Studies, Schloss-Wolfsbrunnenweg 35, 69118 Heidelberg, Germany}
\affiliation{Stellar Astrophysics Centre, Department of Physics and Astronomy, Aarhus University, Ny Munkegade 120, DK-8000 Aarhus C, Denmark}
\author[0000-0003-2400-6960]{Marc Hon}
\affiliation{Kavli Institute for Astrophysics and Space Research, Massachusetts Institute of Technology,
77 Massachusetts Avenue, Cambridge, MA 02139, USA}
\affiliation{Institute for Astronomy, University of Hawai`i, 2680 Woodlawn Drive, Honolulu, HI 96822, USA}
\author[0000-0001-8832-4488]{Daniel Huber}
\affiliation{Institute for Astronomy, University of Hawai`i, 2680 Woodlawn Drive, Honolulu, HI 96822, USA}
\affiliation{Sydney Institute for Astronomy (SIfA), School of Physics, University of Sydney, NSW 2006, Australia}
\author{Thomas Kallinger}
\affiliation{Institute for Astronomy, University of Vienna, T\"urkenschanzstrasse 17, A-1180 Vienna, Austria}
\author[0000-0002-1699-6944]{Kaili Cao}
\affiliation{Center for Cosmology and AstroParticle Physics (CCAPP), The Ohio State University, 191 West Woodruff Ave, Columbus, OH 43210, USA}
\affiliation{Department of Physics, The Ohio State University, 191 West Woodruff Ave, Columbus, OH 43210, USA}
\author{Jennifer A. Johnson}
\affiliation{Department of Astronomy, The Ohio State University, Columbus, OH 43210, USA}
\affiliation{Center for Cosmology and AstroParticle Physics (CCAPP), The Ohio State University, 191 West Woodruff Ave, Columbus, OH 43210, USA}
\author[0009-0008-7869-7430]{Bastien Liagre}
\affiliation{ENS Paris-Saclay, Université Paris-Saclay, 91190, Gif-sur-Yvette, France}
\affiliation{Universit\'e Paris-Saclay, Universit\'e Paris Cit\'e, CEA, CNRS, AIM, 91191, Gif-sur-Yvette, France}
\affiliation{Instituto de Astrofsica de Canarias (IAC), C/Va Lactea, s/n, E-38200 La Laguna, Tenerife, Spain}
\author{Rachel A. Patton}
\affiliation{Department of Astronomy, The Ohio State University, Columbus, OH 43210, USA}
\affiliation{Center for Cosmology and AstroParticle Physics (CCAPP), The Ohio State University, 191 West Woodruff Ave, Columbus, OH 43210, USA}
\author[0000-0001-7195-6542]{\^Angela R. G. Santos}
\affiliation{Instituto de Astrof\'isica e Ci\^encias do Espa\c{c}o, Universidade do Porto, CAUP, Rua das Estrelas, PT4150-762 Porto, Portugal}
\author[0000-0002-6163-3472]{Sarbani Basu}
\affiliation{Department of Astronomy, Yale University, PO Box 208101, New Haven CT, 06520-8101}
\author[0000-0003-4745-2242]{Paul G. Beck}
\affiliation{Instituto de Astrofsica de Canarias (IAC), C/Va Lactea, s/n, E-38200 La Laguna, Tenerife, Spain}
\affiliation{Universidad de La Laguna (ULL), Departamento de Astrof\'isica, E-38206 La Laguna, Tenerife, Spain}
\author[0000-0003-4573-6233]{Timothy C. Beers}
\affiliation{Dept. of Physics and Astronomy, and JINA Center for the Evolution of the Elements
University of Notre Dame, Notre Dame, IN 46556  USA}
\author{William J. Chaplin}
\affiliation{University of Birmingham, School of Physics and Astronomy, Edgbaston, Birmingham B15 2TT, UK}
\affiliation{Stellar Astrophysics Centre, Department of Physics and Astronomy, Aarhus University, Ny Munkegade 120, DK-8000 Aarhus C, Denmark}
\author[0000-0001-6476-0576]{Katia Cunha}
\affiliation{University of Arizona, Steward Observatory, Tucson, AZ 85719, USA}
\affiliation{Observat\'orio Nacional, S\~ao Crist\'ov\~ao, Rio de Janeiro, Brazil}
\author[0000-0002-0740-8346]{Peter M. Frinchaboy}
\affiliation{Department of Physics and Astronomy, Texas Christian University, TCU Box 298840 Fort Worth, TX 76129, USA}
\affiliation{Maunakea Spectroscopic Explorer, Canada-France-Hawaii-Telescope, Kamuela, HI 96743, USA}
\author[0000-0002-6301-3269]{L\'eo Girardi}
\affiliation{Osservatorio Astronomico di Padova, INAF, Vicolo dell Osservatorio 5, I35122 Padova, Italy}
\author[0000-0003-4556-1277]{Diego Godoy-Rivera}
\affiliation{Instituto de Astrofsica de Canarias (IAC), C/Va Lactea, s/n, E-38200 La Laguna, Tenerife, Spain}
\affiliation{Universidad de La Laguna (ULL), Departamento de Astrof\'isica, E-38206 La Laguna, Tenerife, Spain}
\author[0000-0002-9771-9622]{Jon A. Holtzman}
\affiliation{Department of Astronomy, MSC 4500, New Mexico State University, P.O. Box 30001, Las Cruces, NM 88003, USA}
\author[0000-0002-4912-8609]{Henrik J\"onsson}
\affiliation{Materials Science and Applied Mathematics, Malm\"o University, SE-205 06 Malm\"o, Sweden}

\author[0000-0001-8237-5209]{Szabolcs M{\'e}sz{\'a}ros}
\affiliation{ELTE E\"otv\"os Lor\'and University, Gothard Astrophysical Observatory, Szombathely, Hungary}
\affiliation{MTA-ELTE Lend{\"u}let "Momentum" Milky Way Research Group, Hungary}
\author[0000-0001-9632-2706]{Claudia Reyes}
\affiliation{School of Physics, University of New South Wales, NSW 2052, Australia}
\author[0000-0003-4996-9069]{Hans-Walter Rix}
\affiliation{Max Planck Institute for Astronomy, K\"onigstuhl 17, 69117 Heidelberg, Germany}
\author[0000-0003-0509-2656]{Matthew Shetrone}
\affiliation{University of California Observatories, University of California Santa Cruz, 1156 High St, Santa Cruz, CA 95064, USA}
\author[0000-0002-0134-2024]{Verne V. Smith}
\affiliation{NSF's NOIRLab, Tucson, AZ 85719 USA}
\author[0000-0003-4019-5167]{Taylor Spoo}
\affiliation{Department of Physics and Astronomy, Texas Christian University, TCU Box 298840 Fort Worth, TX 76129, USA}
\author[0000-0002-3481-9052]{Keivan G.\ Stassun}
\affiliation{Department of Physics and Astronomy, Vanderbilt University, Nashville, TN 37235, USA}
\author{Ji Wang}
\affiliation{Department of Astronomy, The Ohio State University, Columbus, OH 43210, USA}

\begin{abstract}

In the third APOKASC catalog, we present data for the complete sample of 15,808 evolved stars with APOGEE spectroscopic parameters and \kep\ asteroseismology. We used ten independent asteroseismic analysis techniques and anchor our system on fundamental radii derived from \gaia\ $L$ and spectroscopic $T_{\rm eff}$.  We provide evolutionary state, asteroseismic surface gravity, mass, radius, age, and the spectroscopic and asteroseismic measurements used to derive them for 12,418 stars. This includes 10,036 exceptionally precise measurements, with median fractional uncertainties in \nmax, \dnu, mass, radius and age of 0.6\%, 0.6\%, 3.8\%, 1.8\%, and 11.1\% respectively. We provide more limited data  for 1,624 additional stars which either have lower quality data or are outside of our primary calibration domain. Using lower red giant branch (RGB) stars, we find a median age for the chemical thick disk of $9.14 \pm 0.05 ({\rm ran}) \pm 0.9 ({\rm sys})$ Gyr with an age dispersion of 1.1 Gyr, consistent with our error model. We calibrate our red clump (RC) mass loss to derive an age consistent with the lower RGB and provide asymptotic GB and RGB ages for luminous stars. We also find a sharp upper age boundary in the chemical thin disk. We find that scaling relations are precise and accurate on the lower RGB and RC, but they become more model dependent for more luminous giants and break down at the tip of the RGB. We recommend the usage of multiple methods, calibration to a fundamental scale, and the usage of stellar models to interpret frequency spacings.  

\end{abstract}


\section{Introduction}


Stellar oscillations reveal the inner workings of stars, and time domain surveys from space have made it possible to measure them in exquisite detail for large numbers of stars. Asteroseismology, the study of stellar pulsations, has therefore emerged as a transformational tool for understanding stellar astrophysics and stellar populations. 

The \kep\ satellite had an especially dramatic impact on asteroseismology due to a combination of unprecedented precision, sample size, and long duration observations. Over the past decade the community has steadily refined the interpretation of this landmark data set.  Time-resolved space data does not stand alone; its power is amplified by complementary ground-based campaigns. High-resolution spectroscopy, in particular, is essential for stellar characterization, population studies, and chemical evolution. The combined power of time domain and spectroscopic data has resulted in catalogs of thousands of masses and radii for evolved red giant stars, opening up a completely new age dimension that has changed our understanding of the formation and evolution of the Milky Way galaxy. A detailed analysis of oscillation frequency patterns has also yielded novel insights into stellar physics. Two examples show the depth of the potential insights. Core He-burning stars with non-degenerate cores, hereafter red clump or RC stars, can be distinguished from shell H-burning stars, hereafter red giant branch or RGB stars, with degenerate cores \citep{Bedding11}, 
and internal rotation rates have been 
measured for large numbers of giants \citep{Gehan2018,Mosser2024,LiG2024}.

In this paper we perform a comprehensive analysis of the full joint sample of cool evolved stars with \kep\ time domain data and APOGEE high-resolution spectroscopy, hereafter APOKASC-3. This is the third paper in a series.  \citet[APOKASC-1]{APOKASC1} released the first set of data combining a large (1916 star) high-resolution spectroscopic survey with asteroseismic data. We followed up with \citet[APOKASC-2]{APOKASC2}, which enlarged the sample considerably to 6676 stars. APOKASC-2 also introduced an explicit calibration of the results to a fundamental system, in that case giants in open clusters. Although powerful, the latter data set was incomplete, both in terms of spectroscopic and asteroseismic analysis. We can also now take advantage of data from the \gaia\ DR3 release \citep{GaiaDR3}, which permits a much more precise calibration of stellar properties.

\subsection{Population Asteroseismology in the \kep\ Fields}

To place APOKASC-3 in context, a brief summary of the properties of solar-like oscillations is in order. Cool stars have deep convection zones in their outer layers. The turbulence within them generates waves, which are in turn refracted by the enormous density gradients in stars. Sufficiently low frequency waves are reflected at the surface, producing an acoustic cavity. This combination naturally produces standing wave patterns at discrete frequencies, which can be detected from brightness or radial velocity variations. These oscillations were first detected in the Sun \citep{1962ApJ...135..474L,1962ApJ...135..812E}, and helioseismology yielded crucial data on the internal structure and global properties of the Sun; for a comprehensive review, see \citet{2016LRSP...13....2B}. 
These solar-like oscillations are a nearly universal feature of cool stars, making their study of general interest.  We focus here on inferring global stellar properties – in particular, mass, radius, and age.

Solar-like oscillations are well-described by spherical harmonics. The number of nodes in the radial, latitudinal, and longitudinal directions are given, respectively, by quantum numbers $n$, $l$, and $m$. Global inferences rely on $l$ and $n$. Spatially resolved modes can 
be detected in the Sun, but for unresolved stars, only the lowest order modes can be detected ($l = 0, 1, 2, 3$) because of cancellation effects. 

Early ground-based campaigns had some success in detecting oscillations in other stars \citep{2014aste.book...60B}, creating the field of asteroseismology. However, Sun-like stars have low amplitudes, and it was difficult to detect oscillations in more evolved stars with ground-based telescopes because their slow oscillations required long uninterrupted time series data. It was only with the advent of time-domain space missions, primarily designed for exoplanet transit studies, that the real potential of asteroseismology could be realized.
Data from the \textit{CoRoT} satellite \citep{hekker2009,Deridder2009} established conclusively that evolved giants were solar-like oscillators. The \kep\ mission dramatically expanded the sample size, with unprecedented data quality and light curve duration as well \citep{Bedding10,Stello2013,Yu18}. 

Sound waves spend most of their time in the outermost layers of stars ($ c_s^2 \sim \frac{T}{\mu}$), where $c_s$ is the sound speed, $T$ is the temperature, and $\mu$ is the mean molecular weight.  The absolute frequencies are therefore difficult to predict accurately from theoretical models, 
because they depend on difficult outer layer physics, such as the details of how modes reflect close to the surface. However, in Sun-like stars even and odd $l$ modes naturally separate into close pairs with a nearly uniform spacing between modes of the same degree $l$ but different degree $n$. In asymptotic theory, this large frequency spacing \dnu\ is related to the mean density \citep{tassoul1980}. The observed pattern for giants is more complex because of mixed modes arising from interactions between p-modes, analogs of sound waves, and g-modes, analogs of water waves \citep{dziembowski2001,Dupret2009,Bedding11,Grosjean14}. However, the radial ($l=0$) modes retain a regular structure, allowing a precise measurement for \dnu. The trapping of sound waves in an acoustic cavity is related to the density scale height near the surface of the star, which is tied to the surface gravity through hydrostatic balance. The frequency of maximum power \nmax\ therefore scales with the surface gravity \citep{Kjeldsen1995,Belkacem2011}, so \dnu\ and \nmax\ can be combined to infer masses and radii.

After the initial burst of discovery, it became apparent that complementary data on stellar properties was essential for further progress. The scalings between \dnu, \nmax, and stellar variables depend on effective temperature \teff and chemical abundances, as do stellar ages.

The next step to reliable masses and radii was establishing a robust \teff scale. The groundbreaking \kep\ Input Catalog \citep{Brown2011}, hereafter KIC, developed to identify stars that might host transiting planets, was tied to an absolute temperature scale by \citet{Pinsonneault2012}. Further updates dramatically improved stellar characterization of targets \citep{Mathur2017}. 

Initial stellar population studies used photometric metallicities, such as those in the original KIC. 
Although powerful in principle, such estimates are sensitive to extinction and the choice of filters \citep{Casagrande2014}. In particular, the $griz$ filters adopted by the KIC yield only coarse metallicity data because they were primarily designed to characterize galaxies, not stars \citep{Pinsonneault2012}.

The solution for the determination of reliable abundances was the development of massive high resolution spectroscopic surveys, which came of age in the last 10-15 years. Large data sets, and automated pipeline analysis, yielded abundance mixtures, \teffns, and other spectroscopic parameters of an unprecedented quality and sample size. 
We focus here on the  Apache Point Observatory Galactic Evolution Experiment \citep{Majewski2017}, hereafter APOGEE, which is our reference spectroscopic data set.

APOGEE was part of the Sloan Digital Sky Survey (SDSS; \citealt{York2000}), and in both SDSS-III \citep{sdss3} and SDSS-IV \citep{sdss4}. It was envisioned as an engine for Galactic 
Archaeology, the study of the formation and evolution of the Milky Way galaxy. It is a multi-fiber spectroscopic survey that uses a moderately high resolution ($R\sim 22,000$) infrared spectrograph \citep{Wilson2019} on the SDSS telescope \citep{Gunn2006}, with stellar parameters inferred using the ASPCAP pipeline \citep{aspcap}. 
APOGEE targeted red giants in the \kep\ field to take advantage of the overlap with seismology \citep{zasowski_2013,zasowski_2017, Beaton2021}.
Asteroseismic surface gravities were invaluable calibrators for spectroscopic ones \citep{Holtzman2018}, while spectroscopic data allowed asteroseismic data to be used to infer mass, radius, and age.

The evolutionary state -- whether we are observing stars before or after He burning has ignited -- is also an important stellar property. Fortunately, evolutionary state can be inferred from a detailed study of the oscillation frequency pattern \citep{Mosser2014, Elsworth19}. Stars with spectra and asteroseismic states can also be used as a training set to infer spectroscopic evolutionary states. When asteroseismic states are not available, these spectroscopic evolutionary states can be used to distinguish RGB and RC stars, with typical recovery rates of order 93\% \citep{Holtzman2018}. 

The final major observational advance is the immense power of the \gaia\ mission \citep{GaiaDR3}, which permits precise and accurate distance measurements for the large majority of \kep\ targets. In turn, these distances can be combined with \gaia\ photometry and 3D extinction maps, producing exquisitely precise and accurate luminosities. Luminosities, $L$, combined with absolute \teff from APOGEE  then can be used to compute fundamental radii, $R$, which are an invaluable cross-check on asteroseismic radius inferences \citep{Zinn2019Rtest}. We therefore believe that this is an ideal time for a comprehensive analysis of \kep\ asteroseismology.

\subsection{APOKASC-3 Goals}

Our effort has several important goals. First, we want to provide a legacy data set of the highest quality \kep\ asteroseismic measurements. To do this, we use a number of distinct analysis pipelines; targets with a large number of consistent measurements are the core sample for this purpose. This represents a significant advantage over single-method surveys, as it allows us to discover outliers and unusual light curves that can be difficult to analyze with automated methods.

Second, we want to anchor our mass, radius, and age measurements firmly to an absolute scale, which we can derive from a combination of \gaia\ DR3 and APOGEE data. This absolute reference system also allows us to quantify the domains in which asteroseismic inferences are the most reliable, and the domains where the assumptions in asteroseismic scaling relations break down. As an example, one important lesson from APOKASC-1 and APOKASC-2 was the need to use stellar models to map the observed frequency pattern onto the mean density. Here we explore the impact of different choices for this mapping on the results. Even for a given mass, different ages are inferred from different stellar interiors codes and different choices for input physics, an additional subject that we explore here.

A third major goal is to be comprehensive. To that end, we present results for all \kep\ stars with APOGEE spectra here. In a companion paper, we provide stellar properties for asteroseismic detections in stars without APOGEE spectra; the combination of the two represents the full census of asteroseismic detections of evolved cool stars in the \kep\ fields.

Our final goal is transparency; we therefore provide both recommended mean values, alternative averages, and the individual results used to derive them. We also summarize a set of recommended best practices for population asteroseismology and outline areas for future research.

\subsection{Changes From APOKASC-2}

Our overall method is similar to that employed in APOKASC-2. We use multiple pipelines to interpret time series data from \kep. The global oscillation properties \dnu\ and \nmax\ are related to mass and radii through scaling relations, of the form: 
\begin{equation}
\frac{M}{\msun} = \left( \frac{f_{\nu \max} \nu_{\rm max}}{\nu_{\rm max, \odot}} \right)^3 \left( \frac{f_{\Delta \nu} \Delta \nu}{\Delta \nu_{\odot}} \right)^{-4}
\left( \frac{T_{\rm eff}}{T_{\rm eff,\odot}} \right)^{1.5} 
\label{eqt-scalingM}
\end{equation}
and
\begin{equation}
\frac{R}{\rsun} = \left( \frac{f_{\nu \max} \nu_{\rm max}}{\nu_{\rm max,\odot}} \right) \left( \frac{f_{\Delta \nu} \Delta \nu}{\Delta \nu_{\odot}} \right)^{-2}
\left( \frac{T_{\rm eff}}{T_{\rm eff,\odot}} \right)^{0.5}. 
\label{eqt-scalingR}
\end{equation}

The \nmax\ and \dnu\ measurements from the pipelines are placed on a common zero-point, and the measurement scatter between methods is taken as an error estimate. Asteroseismic diagnostics are used to infer evolutionary states where possible. Where they are not available, we use spectroscopic evolutionary-state diagnostics calibrated with asteroseismic data. Effective temperatures are taken from APOGEE spectroscopic data. The \fdnu\ term relates the mean density to the observed frequency spacing, and is computed from stellar interiors models (see Section \ref{sec:dnu}). The \fnmax\  term is an empirical calibration function inferred from comparisons to fundamental data, and is used to place our results on an absolute scale (see Section \ref{sec:nmax}). 
However, there are significant differences with APOKASC-2 as well:

\begin{itemize}

\item {This paper uses the full sample with spectroscopic data from the APOGEE survey and time domain data from the \kep\ survey.} 

\item {We used different techniques for preparing the light curves, and a larger number of pipelines for light curve analysis.}

\item {We adopted outlier rejection tools for asteroseismic measurements inconsistent either with a broad spectroscopic prior or the median results from other methods. This allowed us to identify background stars and deal with lower quality time series data in some stars.}

\item {We used radii from a combination of \gaia\ $L$ and APOGEE \teff to anchor our fundamental scale, rather than masses in open cluster stars. In APOKASC-2, $f_{\nu \max} =1$ with an adopted $\nu_{\rm max,\odot} = 3076 \;\rm{\mu Hz}$. 
Here \fnmax\ is a function of \nmax, with the same solar benchmark. These radii are also provided in the catalog.}

\item {We considered three distinct methods for inferring the mean density from the measured frequency spacings, and two methods for inferring ages.}

\item{We separate our detections into three categories: Gold, Silver, and Detections (Section \ref{sec:seis_sample}). Both \nmax\ and \dnu\ are detected for Gold and Silver sample stars. They differ in that Gold sample stars have the most precise and accurate data, and Silver sample stars are ones with larger uncertainties. Detections are cases where we can only measure \nmax.}

\item{We provide flags for non-detections as well as sources that we attribute to background stars, and define domains where the scaling relations are valid.}

\end{itemize}

\subsection{Roadmap of this Paper and Related Publications}

We discuss the 10 pipelines used to analyze the \kep\ light curves in Appendix \ref{app:pipelines}, and our method for combining them into the global asteroseismic properties \dnu\ and \nmax\ in Appendix \ref{app:merging}. In Section \ref{sec:sample} we describe the APOKASC-3 sample selection criterion, spectroscopic properties, and our method for assigning evolutionary states. We infer masses, radii, and ages in Section \ref{sec:massradiusage}. The catalog itself is presented in Section \ref{sec:catalog}. We also illustrate stellar physics and stellar population applications of the data there. Our key results are summarized in Section \ref{sec:discuss}, including recommended practices for population asteroseismology. 

We also note some related papers here. The evolutionary states of the APOKASC-3 sample are the focus of a separate paper \citep{vrard2024}. \citet{Roberts2024} describes the empirical first dredge-up pattern using APOKASC-3 data. A theoretical treatment of both the first dredge-up and the red giant branch bump is provided by a separate paper Cao et al. 2024). Stars with oscillation frequencies close to the long cadence \kep\ Nyquist sampling frequency ($\sim ~283.2~ \mu$Hz) require special analysis techniques, and they are the focus of Liagre et al.(in prep).
The binary population in this sample was investigated by \cite{Beck2024}, using \gaia\ and APOKASC data. The kinematic classification of this sample is presented in \citet{godoy2024}.

We present asteroseismic data for stars without APOGEE spectra in Garc\'\i a et al. (in prep), with more than 9,100 additional targets. The \kep-APOGEE sample was chosen from targets in the KIC that 
have time domain data. Sustained efforts by the APOGEE team have resulted in substantial, but not complete, overlap between the two data sets. The large majority of the missing stars were too faint for APOGEE. Of the Garc\'\i a et al. (in prep) stars, 6,663 were detected by 5 or more pipelines in the initial analysis. 
The medium and low resolution LAMOST survey \citep{Cui2012} is a valuable resource for these fainter stars \citep{Fu2020LAMOSTKepler}. It provides optical coverage that complements the IR data from APOGEE for a large data set. LAMOST results are also tied to the APOGEE scale; see for example \citet{Xiang2019}, which provided abundance mixtures for 6 million stars.

\section{The APOKASC-3 Sample} \label{sec:sample}
The \kep\ field has been the subject of intensive study, and it was naturally a high priority for the APOGEE spectroscopic survey. Asteroseismology was a major focus but not the only one, so the APOGEE selection function needs to be considered. As discussed in APOKASC-1, the \kep\ targeting for giants was also complex. The selection function therefore needs to be assessed carefully before using this data set for population studies \citep[e.g.,][]{Silvaaguirre2018}. For extended discussions of the targeting for the \kep-APOGEE sample see \citet{APOKASC1} and \citet{Simonian2019}. 

\subsection{The APOGEE Survey and Data Releases}
\label{sec:fullsample}

The Sloan Digital Sky Survey provides regular data releases; these include both new targets and new analysis techniques. Four different data releases (hereafter DR10, DR14, DR16, DR17) have been used in the APOKASC papers -- Data Release 10 \citep{DR10} for APOKASC-1; Data Release 14 \citep{DR14} for APOKASC-2, and Data Releases 16 and 17 \citep{DR16, DR17} for APOKASC-3. All of the data releases use the same underlying logic: spectra are fit in a multi-dimensional space to infer ``raw'' global parameters. The key ones for our purposes are \teffns, \loggns, \mh, and \amns. Here M is a metallicity index, closely correlated with Fe, and $\alpha$ is an index of elements associated with $\alpha$-capture species. 
Individual abundances are inferred from selected features in spectral windows. Of these, we use [C/Fe] and [N/Fe] to infer spectroscopic evolutionary states. The global parameters \teff and \logg are then placed on an absolute scale in a post-processing step. \teff is adjusted to corrected values tied to the Infrared Flux Method \citep{gonzalez-hernandez_bonifacio2009}, while \logg is tied to asteroseismic surface gravities \citep{Holtzman2018}. There are also small adjustments to the abundance scale for consistency. The calibration procedure for DR16 is discussed in \citet{Jonsson2020}, while that for DR17 can be found at \url{https://www.sdss4.org/dr17/irspec/aspcap/}.
We use the differences between DR16 and DR17 as a measure of systematic uncertainties in the spectroscopic properties of the sample. We use DR16 data for assigning spectroscopic evolutionary states (see below). Projected rotation velocities $v \sin i$, where $i$ is the inclination angle, are taken from \citet{Patton2024}.


The full APOGEE-\kep\ sample contains 23,363 unique targets. However, many of these are not in the domain where we expect asteroseismic detections from the long-cadence (30-minute) time series data obtained by \kep. These less evolved targets were observed either as part of other programs or as telluric standards \citep{Simonian2019}. 

Most, but not all, of our data is from DR17. The APOGEE survey has regular improvements and changes in the pipeline analysis, and as a result the stellar parameters change in each data release. There are 15,321 stars classified as giants in DR17. In 74 additional cases, we had valid solutions in DR16 but not DR17, and adopted the DR16 values.\footnote{Of the DR17 giants, 102 were not detected in DR16; this impacts only our spectroscopic evolutionary states for these stars, as discussed below.}
There were 175 additional targets classified as dwarfs in DR17 but as giants in DR16, so we searched for seismic signals in them as well.

In large samples there are rare but interesting objects, and they can have unusual spectra. We therefore have to be careful not to exclude these stars by definition from our sample. The APOGEE automated pipeline has quality-control checks, and if the fit is poor, the BAD STAR flag is triggered and calibrated values are not returned.  Some of these spectra simply have poor data, but others represent classes of stars, such as rapid rotators or binaries, that are poorly fit by the standard templates. We searched for asteroseismic signals in 174 targets without good spectral fits that were classified as giants in the KIC, but do not provide mass, radius, and age because they require spectroscopic information.

Our total giant sample was therefore 15,742. We also performed a background source search in the dwarfs, and found 66 such targets that we also included in our catalog. This gives a total catalog sample of 15,808.

\begin{figure}

\includegraphics[width=8cm]{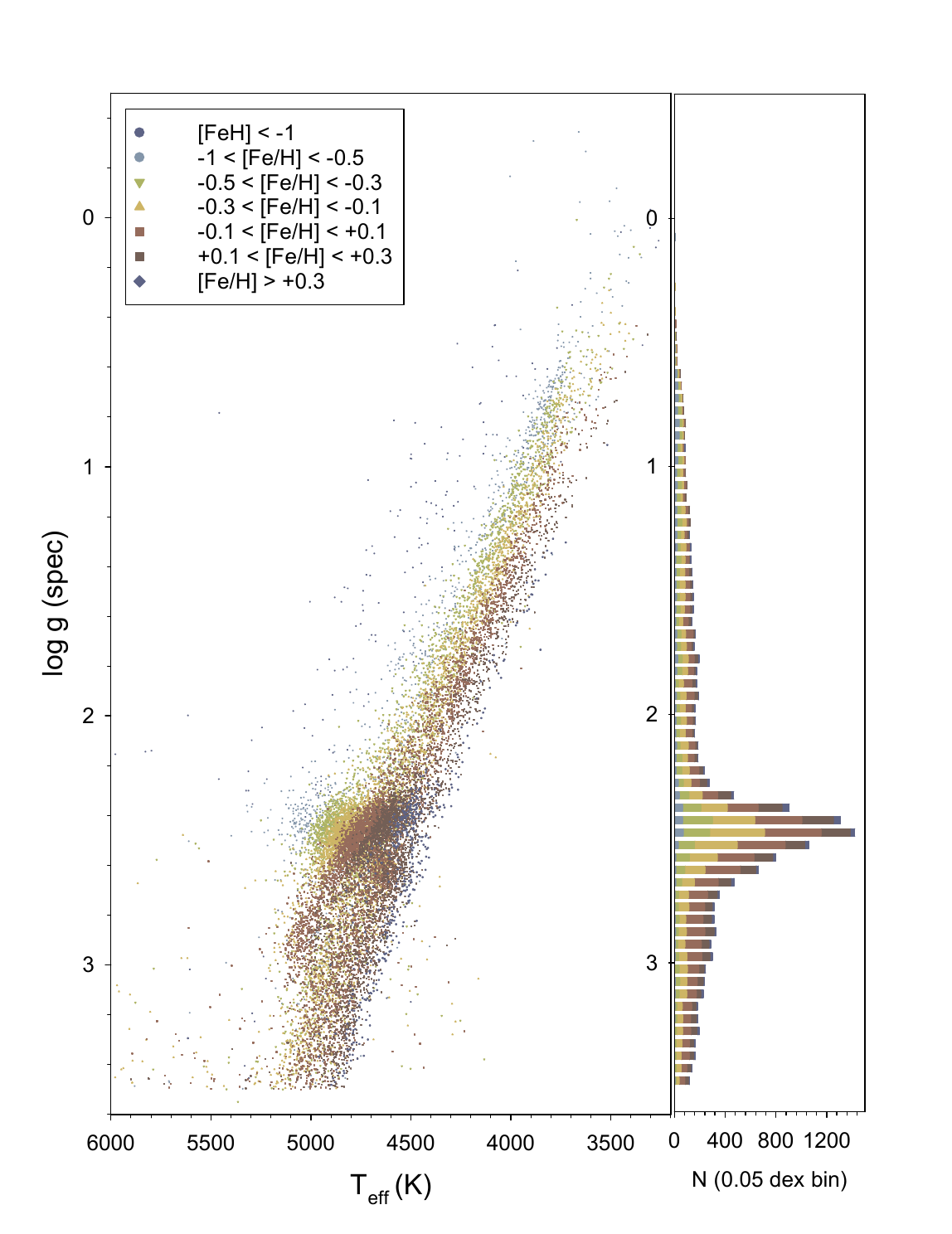}
\caption{The full APOKASC-3 evolved star sample. The histogram illustrates the number of targets, as a function of metallicity, in 0.05 dex \logg bins (right).}

\label{fig:allkiel}

\end{figure}

In Figure \ref{fig:allkiel} we show the 15,570 targets in the APOKASC-3 sample with spectroscopic data, along with a histogram of the \logg distribution.\footnote{In this paper, 65 of the 15,570 targets were discovered to be background asteroseismic detections in this paper, so the spectra do not correspond to the oscillations; these targets are included but not analyzed in detail.} A strong concentration of the sample around the location of the core He-burning RC and the red giant branch bump (RGBB) is apparent. Our sample range is 3000 K -- 6000 K in \teff and $-0.5$ to $3.5$ in \loggns. The lower \teff and \logg bounds are the limits of the APOGEE sample, while the higher \teff bound is set by the domain where solar-like oscillations are excited, and the higher \logg bound where \nmax\ exceeds the Nyquist sampling frequency.

\begin{figure}
\includegraphics[width=8cm]{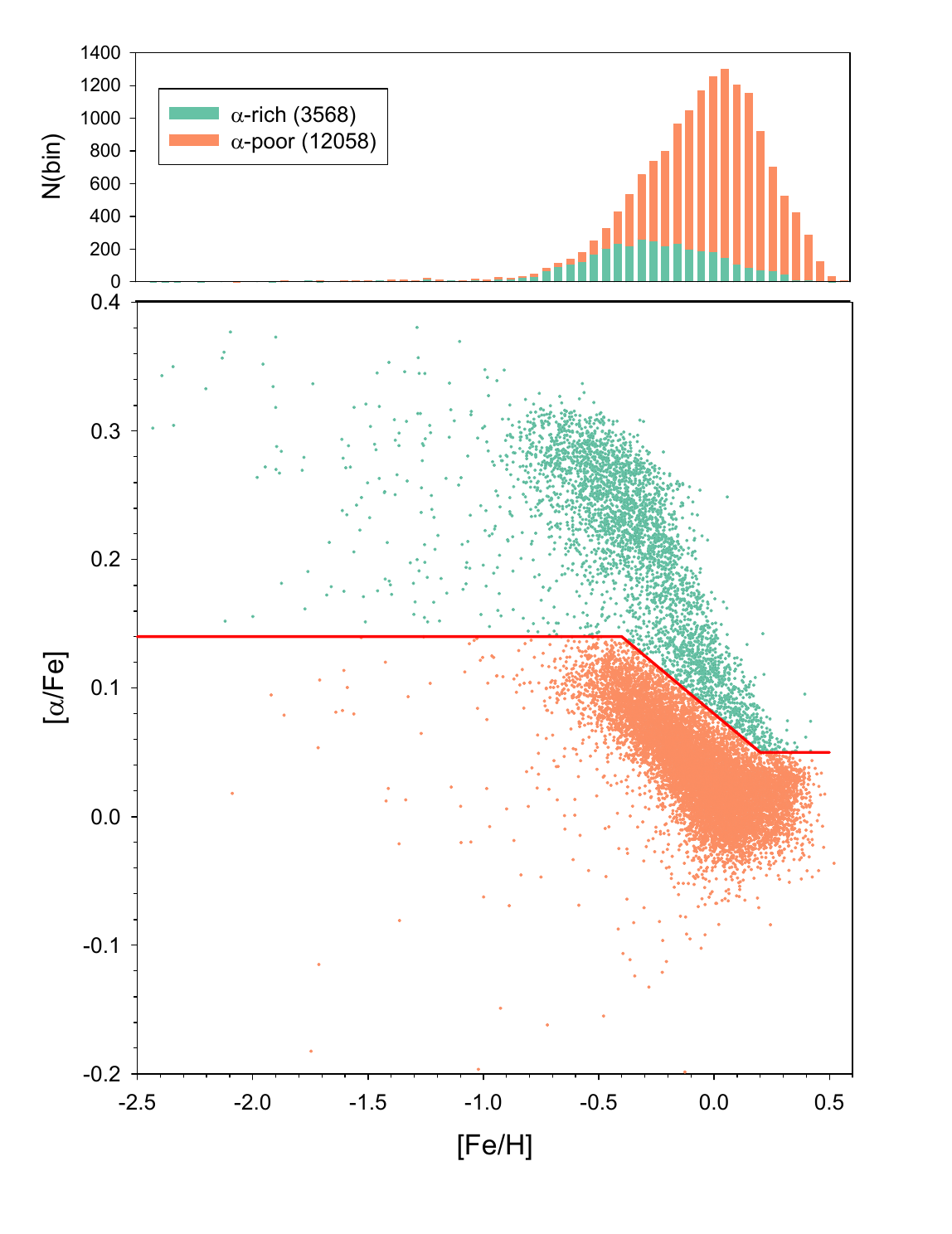}
\caption{The abundance distribution of the APOKASC-3 sample in the \afe -- \feh plane. The red line is the criterion that we use to distinguish high \afe from low \afe ones. For plotting purposes later in the catalog, we use a slight modification of the \citet{Roberts2024} criterion for distinguishing the two populations. A star was classified as $\alpha$-rich if \afe was above $(0.08 - 0.15 {\rm \fehns})$ in the range $-0.4< {\rm \fehns} <\ +0.2$; a threshold of $+0.14$ was assigned for \feh below $-0.4$; and a threshold of $+0.05$ was used for \feh above $+0.2$. Our criterion differs from the \citet{Roberts2024} one in that we use \afe and they used [Mg/Fe], including a slight zero-point shift.}
\label{fig:afefeh}
\end{figure}

Figure \ref{fig:afefeh} shows the heavy element mixtures for APOKASC-3 stars that we used for this study in the \afe vs. \feh\ plane. The bottom panel shows a scatter plot of the full sample, indicating the presence of a small sample of interesting low \afe stars; the top panel is a histogram  illustrating the distinctive metallicity distributions of our high and low \afe populations. The $\alpha$-poor population is predominant, with 12,058 members, compared to 3568 $\alpha$-rich stars. 

\footnote{For this, and subsequent plots, we use the \mh\ vector for metallicity, which closely tracks \feh in practice.} 

\subsection{The Asteroseismic Sample} \label{sec:seis_sample}

All of our targets have long-cadence \kep\ data, and are in the domain where we expect to be able to detect oscillations.  
We started with a broad spectroscopic prior for \nmax, which we used to identify background sources and reject outlier measurements. We describe the preparation of light curves and the 10 individual pipelines used for asteroseismic analysis in Appendix \ref{app:pipelines}. Our method for inferring \dnu\ and \nmax\ from the full set of measurements from individual pipelines is described in Appendix \ref{app:merging}.
From these data, we identified a sample with exceptionally high recovery. This was used to set relative \nmax\ zero-points and weights for individual pipelines. We then did a second outlier pass, rejecting measurements inconsistent with the ensemble data. In this round, some investigators also analyzed the outliers to weed out errors from the automated analysis, and some pipeline results were refined. 

We then defined four categories of targets: a Gold sample (5 or more valid measurements of both \dnu\ and \nmax); a Silver sample (2 -- 4 valid measurements of both \dnu\ and \nmax); a Detection sample (less than 2 valid \dnu\ measurements, but 2 or more \nmax\ ones); and a Non-Detection sample, with less than 2 \nmax\ detections. Errors and weights were inferred separately for each group. We returned median and weighted mean values for both \dnu\ and \nmax\ for the Gold and Silver samples, and \nmax\ alone for the Detections. 

The APOKASC-2 sample included only stars with $ 2 \;\rm{\mu Hz} < \nu_{\rm max} < 220 \;\rm{\mu Hz}$; here 
we measure oscillators across the full dynamic range ($ 0.1 \;\rm{\mu Hz} < \nu_{\rm max} < 280 \;\rm{\mu Hz}$) where they are detectable with \kep\ 30-minute cadence data, We report \dnu, \nmax, and the asteroseismic surface gravity for all detected stars. However, we only report masses and radii for $ 1 \;\rm{\mu Hz} < \nu_{\rm max} < 220 \;\rm{\mu Hz}$ (Section \ref{sec:low_high}).

However, our goal of recovering the full asteroseismic sample introduces new challenges relative to prior efforts: stars with incomplete data, unusual or suppressed oscillation patterns, and very high or low frequencies of maximum power. All of these populations are important for our study, and we begin by briefly describing the issues in turn. There is a small population of stars where oscillations are not detected even in excellent data. The primary culprit is thought to be stellar activity, which suppresses the amplitude of oscillations as observed in main-sequence stars by \citet{Chaplin2011} and \citet{Mathur2019}. \citet{Tayar2015} found that rapid rotators were far less common in asteroseismically detected giants than in the field. \citet{Gaulme_2020} reported that the vast majority of red giants without detected oscillations despite excellent data belong to close binary systems that are tidally locked. Such systems usually exhibit orbital periods shorter than 50 days. In addition, the fraction of binaries among the stars with partially suppressed oscillations remains significantly larger ($\approx 15\,\%$) than for RG oscillators that do not display detectable surface modulation \citep{Gaulme_2020}.

In some cases, there are also technical issues with light curves, or significant background contaminants. We therefore have examined our sample of non-detections individually, and note where stellar activity or backgrounds are responsible for the lack of a signal. We discuss sample completeness, recovery, and non-detections in Section \ref{sec:recovery}.

We then inferred masses, radii, and ages for our core sample. This required knowledge of evolutionary state, discussed in detail in a companion paper \citep{vrard2024}. Our evolutionary states are discussed in Section \ref{sec:evol_stat}. We discuss unusual targets (background sources and those without spectra) in Section \ref{sec:outliers}. Our mass, radius and age inference procedures, which require stellar interiors models and calibration to a fundamental system, are described in Section \ref{sec:massradiusage}.

\subsubsection{Low and High \nmax\ Targets} \label{sec:low_high}

Our asteroseismic approach breaks down in the low and high frequency domains.
On the high frequency side, we have to consider the Nyquist sampling limit for 29.4-minute long-cadence data, which is 283.447 $\rm{\mu Hz}$ \citep{Jenkins2010}. There is therefore an expected population with \nmax\ below this level. However, for stars close to this boundary, the true spectral power will extend above the Nyquist limit, leading to a distorted power spectrum and a reduced amplitude for \nmax\ close to the sampling limit.  As seen in Figure \ref{fig:nmaxtrends} (Appendix \ref{app:merging}), systematic differences between methods diverge above 220 $\rm{\mu Hz}$, which we take as the limit where different analysis techniques are required.  Near- and super-Nyquist asteroseismology is the subject of a separate paper (Liagre et al. in prep).

Along the RGB, the amplitude of the modes increases and \nmax\ decreases as surface gravity drops. Detecting oscillations is  therefore straightforward in luminous giants with a sufficiently long time series, which is typically true for the \kep\ sample. However, for these stars, precise asteroseismic characterization becomes progressively more difficult as \nmax\ decreases. There are fewer detected modes; 
although peaks in the power spectrum are narrower, their relative width, scaled to \dnu, increases when \dnu\  decreases \citep{Dreau2021}, making it hard to resolve individual mode frequencies. 
It is also difficult to define a consistent system relative to higher \nmax\ for a sparse set of modes, which is important for measuring \dnu. For the lowest frequencies, light curve systematics are also important. These challenges are reflected in measurement differences between methods, which grow as \nmax\ drops (Appendix \ref{app:merging}). 

The mapping of \nmax\ onto mean density also becomes more challenging in low \nmax\ stars, because we can observe only with low radial order $n$, where the asymptotic approximation in scaling relations becomes a poor one. This departure is accounted for in our \fdnu\ factor, but there are significant differences between different \fdnu\ approaches at low \nmax\ (Section \ref{sec:dnu}). Finally, there are known to be significant offsets between asteroseismic and fundamental radii for luminous giants \citep{Zinn2019Rtest,zinn+2023}.

For all of these reasons, we believe that the classical scaling relation approach is not the correct approach for the most luminous giants in our sample, and adopt \nmax\ of 1 $\mu$Hz as our threshold. This is the characteristic \nmax\ where \citet{Mosser2013} first detected a change in the oscillation frequency regime. \citet{Stello2014} also found that the low radial order modes ($n$ less than $5$ below 1 $\rm{\mu Hz}$) seen in these stars produce a highly non-uniform pattern even in theoretical models, further complicating analysis. We therefore provide ensemble averaged \nmax, \dnu, and asteroseismic \logg for giants with  $0.1 \;\rm{\mu Hz} < \nu_{\rm max} < 1 \;\rm{\mu Hz}$, but not masses and radii. 

\subsubsection{Overall Asteroseismic Recovery}
\label{sec:recovery}

\begin{figure}

\includegraphics[width=8cm]{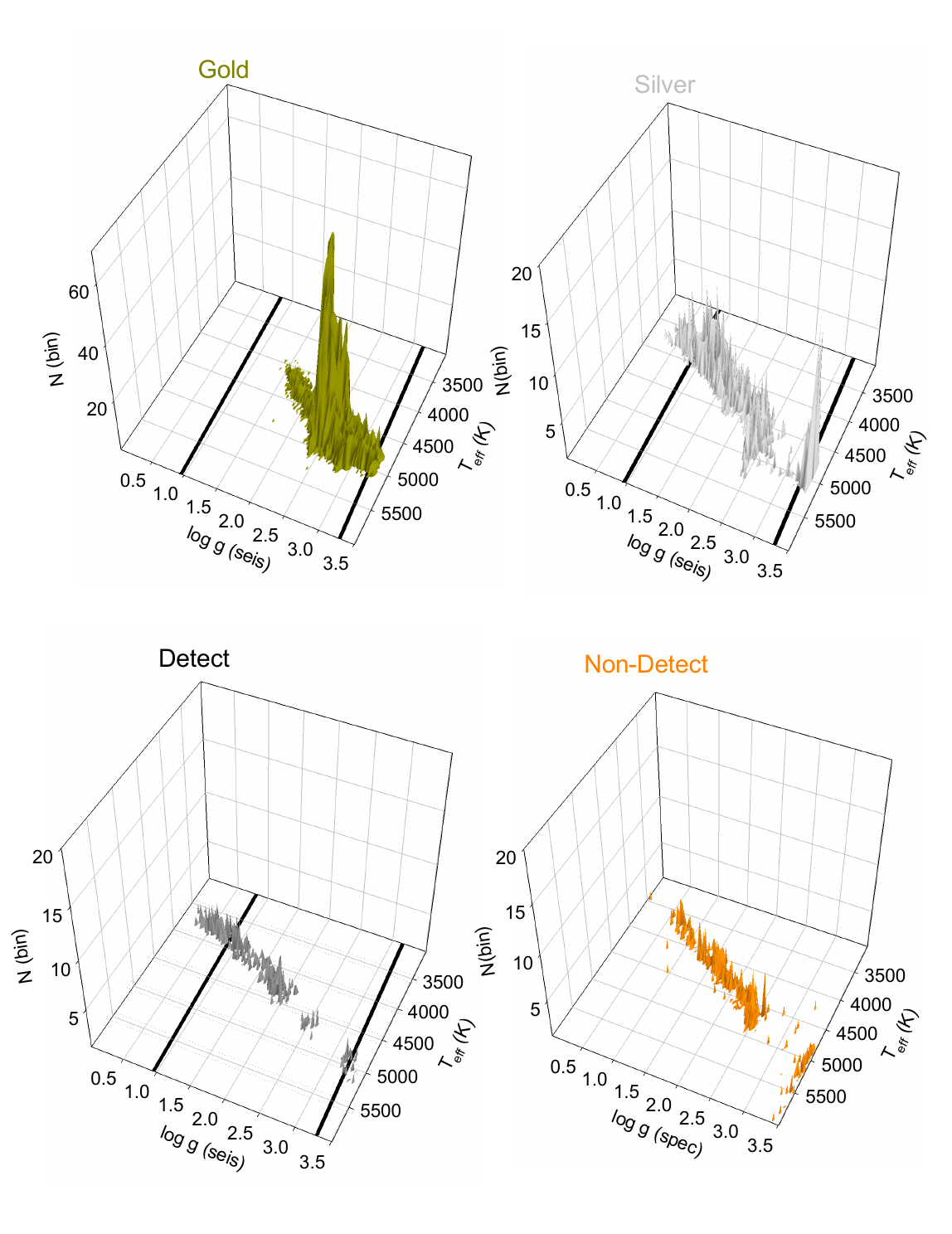}
\caption{Our four cohorts -- Gold (top left), Silver (top right), Detections (bottom left) and Non-Detections (bottom right) -- illustrated in three-D mesh plots as a function of \logg and spectroscopic \teffns. We use seismic \logg for all categories except Non-Detections, for which we use spectroscopic \loggns. The horizontal lines denote the approximate boundaries where we report masses and radii (between \logg of 1 and 3.3). Bin sizes of 40 K in \teff and 0.02 dex in \logg reflect typical measurement uncertainties. The vertical scale for the Gold bin is higher than for the other panels. The large majority of lower RGB and RC stars are robustly detected. In the detected groups we also note the number of stars with detections that are outside the domain where we provide masses, ages, and radii. In total, we have 12,448 targets with full data, 1,634 with partial data, and 1,423 without seismic data. Background sources (129) and stars without valid spectroscopic data (174) are not shown.}
\label{fig:kielcats}
\end{figure}

We present our 4 groups of targets -- Gold, Silver, Detection, and Non-Detection -- in Figure \ref{fig:kielcats}. 
The lines close to \logg of 1 and 3.3 denote characteristic \nmax\ of 1 and 220 $\rm{\mu Hz}$ respectively, which define the  domain where we infer mass, radius, and age. Stars without spectra and ones where the asteroseismic detection does not correspond to the spectroscopic source are not shown.

\begin{figure}

\includegraphics[width=8cm]{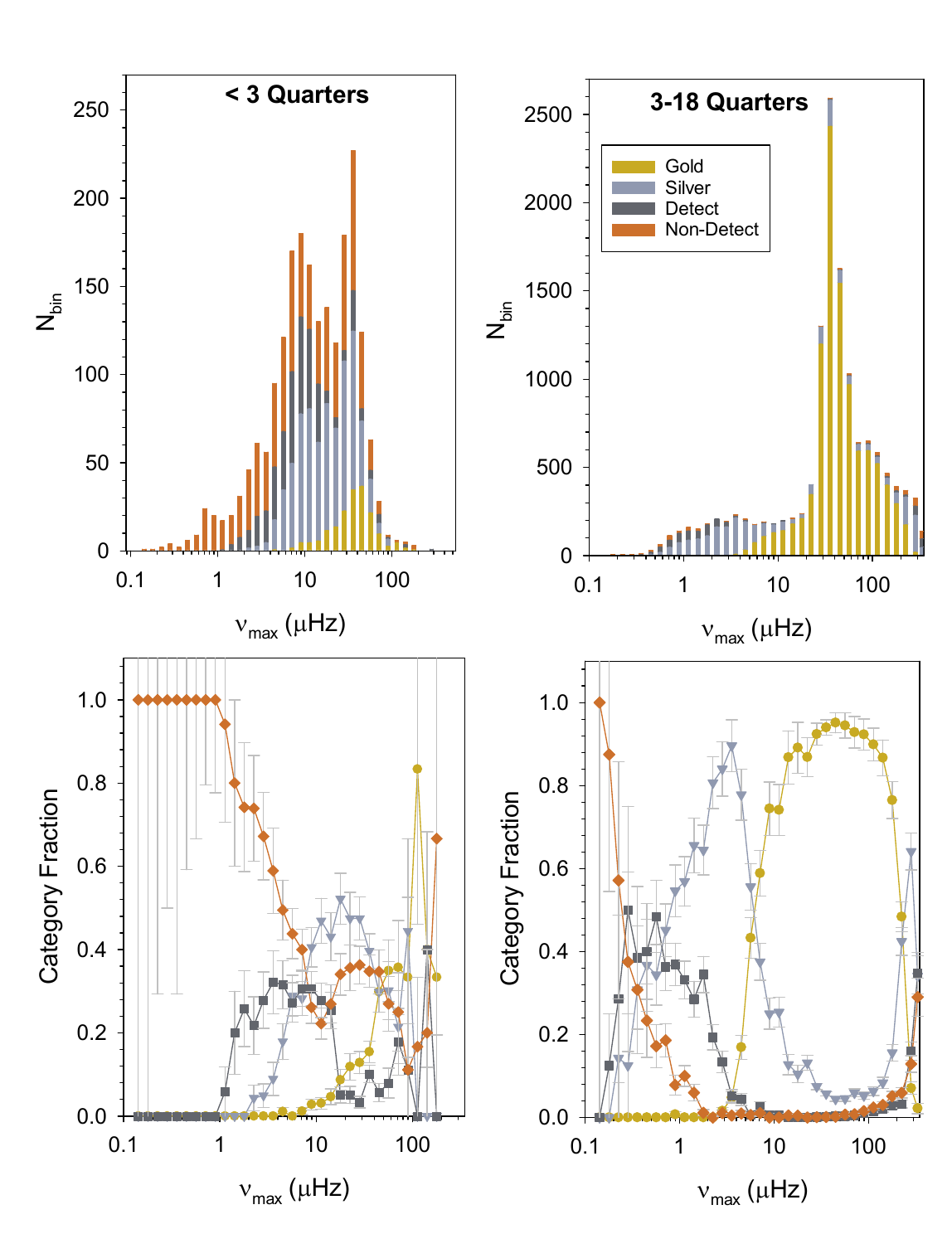}
\caption{Our four cohorts -- Gold (dark yellow, circles), Silver (light gray, triangles), Detections (dark gray, squares) and Non-Detections (orange, diamonds) -- illustrated as a function of spectroscopic \nmax\ for stars with less than 3 quarters of data (left) and for those with 3-18 quarters (right). Absolute numbers are in the top panels, and fractions in the bottom panels. The large majority of stars with good data are detected in our core domain of 1-220 $\rm{\mu Hz}$, while recovery is much more difficult for shorter time series.}
\label{fig:detcats}
\end{figure}

The non-detections and marginal detections may appear somewhat surprising, because the \kep\ sample is the highest quality asteroseismic data set currently in existence. \kep\ had a relatively large aperture, bright targets, and more than 4 years of continuous data, which was downloaded in 90 day segments referred to as quarters. The majority of our targets (13,543) had continuous data for 3 or more quarters of the mission, including 9,970 with 13 or more quarters. However, because some red giant stars were not explicitly included as planet search candidates, they were observed sporadically or infrequently; stars with less than 3 quarters of data are difficult to recover signals from. Figure \ref{fig:detcats} shows our recovery as a function of \nmax\ for targets with 1-2 quarters of data (left) and 3-18 quarters (right). For our core sample (1-220 $\rm{\mu Hz}$), 9,869 (80.4\%) of the stars with 3+ quarters of data were in the Gold sample, with only 123 (1\%) being non-detections. By contrast, of the stars with 1-2 quarters of data there were 826 non-detections (40.1\%) and 182 Gold sample stars (8.8\%). The APOKASC-2 sample was selected to emphasize stars with high quality light curves, and we obtained consistent results for almost all of those targets.

Stars without asteroseismic detections are nonetheless interesting objects, so we performed a manual analysis of targets.
We identified plausible reasons for non-detections in all cases. A total of 965 of the non-detections had less than 3 quarters of data; in many cases, this was actually less than one full quarter of data. There were 242 stars with predicted \nmax\ (from spectra) below 1 $\rm{\mu Hz}$, a domain where detection is challenging. An additional 147 stars had a predicted \nmax\ (from spectra) above 283 $\rm{\mu Hz}$, close to the Nyquist sampling rate. 262 stars not in these categories still had no detections, including 78 targets without good spectroscopic fits.

Of these, 29 stars had two oscillation power excesses, indicating either a true double giant system or a contaminating signal from a background source. In all cases, one of the peaks is close to the value predicted by spectra. These stars could be real binaries, but the majority of them are usually the result of a background star in the aperture, which contaminates the light curve of the main target. 

The amplitude of the asteroseismic modes is known to be reduced in rapidly rotating cool giants 
\citep[e.g.,][]{2010Sci...329.1032G,Chaplin2011,2018ApJS..237...17S,Mathur2019,Gaulme_2020}.
Rapid rotation is also associated with cool spots that can induce a photometric variability signal. For 69 stars, we do not detect modes, but can see spikes at low frequency that appear to correspond to the rotation of the star. We attribute these non-detections to mode suppression caused by rotation and activity.

We also have a sample of 87 stars where we can see signatures of transits or eclipses, preventing us from detecting the modes. Detailed work, outside of the scope of this paper, is required to interpret these light curves.

Finally, some power spectra present spurious spikes that can be due to instrumental issues or to the modes of classical pulsators from faint background sources. There are 79 stars in this category. We note that some stars were in more than one category. See Section \ref{sec:outliers} for a further discussion on outliers. We summarize the properties of our non-detections and outliers in Table \ref{table:outliers}.

\begin{table*}
    
    \centering
    \begin{tabular}{c c c c c c c c c c}
        KIC & STARCAT & SEISB & ROT & EB & BadTS & HighN & LowN & Nomodes & Short \\
          
 \hline
1434395 & SilverOl & 0 & 0 & 1 & 1 & 0 & 0 & 0 & 0 \\
1864258 & NoDetOl & 0 & 0 & 0 & 0 & 0 & 1 & 0 & 0 \\
1872749 & DetectOl & 0 & 0 & 0 & 0 & 1 & 0 & 1 & 0 \\
2010051 & LowNmax & 0 & 0 & 0 & 0 & 0 & 1 & 0 & 0 \\
2015616 & DetectOl & 0 & 1 & 0 & 0 & 0 & 0 & 1 & 0 \\
 
    \hline
    
    \end{tabular}
\caption{Outlier categories. We present only the first 5 rows; remaining entries can be found in the electronic version. KIC is the KIC ID, and STARCAT is the category the outlier is found in. The next 8 columns are toggles indicating the class of behaviors for which the light curve was flagged (1= yes, 0 = no). Objects could be flagged for more than 1 reason. SEISB = double-peaked power spectrum; ROT = strong stellar rotation signal; EB = Eclipsing Binary; BadTS = either pollution by a background source or a pathologicial time series; HighN = \nmax\ above or close to Nyquist; LowN = \nmax\ below $1 \;\rm{\mu Hz}$; Nomodes = no clear detection; Short = 1 or less quarters of data. Note that we do not list the non-detections with short time series; the Short category was only used for initial detections for which we identified contradictory or poor results.}
\label{table:outliers}
\end{table*}

\subsubsection{Evolutionary States} \label{sec:evol_stat}

He-core burning (RC) stars and stars with degenerate cores (RGB and AGB) exhibit distinct frequency patterns that can be clearly identified in the long duration time series data provided by space missions \citep{Bedding11,Mosser11-mixed}  After this was discovered, a number of automatic methods to distinguish RGB from clump targets with asteroseismology were developed. In this section, we present a summary of how we used those techniques to assess the evolutionary status of stars. The complete description of the method is presented in \citet{vrard2024}. 

We used 6 different methods that automatically determine 
the evolutionary status of red giant stars \citep{Elsworth19,Hon17,Kallinger12,Kuszlewicz20,Mosser19,vrard16}, one of them with two different codes. In some cases different approaches predicted different states for the same star. Some methods are based on direct inferences about core properties, which require higher quality data; others are based on parameters correlated to evolutionary state, which can be performed with moderate quality data. The indirect techniques were weighted less heavily in the final classification when their results conflicted with direct techniques, similar to the approach used in \citet{Elsworth19}. 
 
With those criteria, we manage to obtain a classification for 11,371 stars in the APOKASC sample (4,755 identified as RC and 6,616 identified as RGB or AGB). The RC category includes more massive core He-burning stars, both those fainter than the RC and sometimes referred to as the secondary RC, and the brighter intermediate-mass ones. Manual checks confirmed that the seismic evolutionary status identification was very reliable and robust when it disagreed with the spectroscopic state. We also had an agreement between the spectroscopic and seismic classification better than $94\%$ of the time. 

\citet{vrard2024} also investigated asteroseismic and spectroscopic methods for separating H-shell only (RGB) stars and double shell burning objects (AGB) stars. Asteroseismic techniques were in agreement with spectroscopic state inferences close to the RC, but were found to diverge for more luminous stars.  For stars with \dnu$\leq2 \;\rm{\mu Hz}$ the disagreement between the different techniques becomes close to $50\%$, thus the asteroseismic classification 
cannot be reliably assessed for luminous stars. 
In the APOKASC-3 catalog, we therefore treated shell-burning sources as potential AGB stars for \logg below 2.2; as AGB stars do not appear at higher \loggns, we classified all higher gravity shell-burning sources as RGB. 

\begin{figure}

\includegraphics[width=8cm]{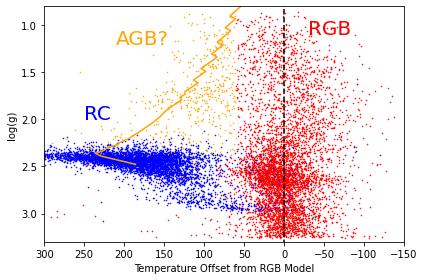}

\caption{Temperature offset, as defined in \citet{vrard2024}, relative to the mean RGB locus as a function of asteroseismic surface gravity for our upper RGB sample. The RC stars are shown in blue, RGB stars are red, and AGB stars are yellow. A one solar mass, solar metallicity MIST AGB track is shown for reference as a  yellow line.}
\label{fig:AGBRGB}

\end{figure}

However, AGB stars are hotter than RGB stars at the same surface gravity. The temperature difference is significant just above the RC and becomes small towards the tip of the RGB. We can therefore assign spectroscopic AGB or RGB evolutionary states to some stars above the RC, as shown in Figure \ref{fig:AGBRGB}. These assignments affect the inferred ages, as discussed in Section \ref{sec:massage}. For details of the approach used, see \citet{vrard2024}.

\section{Mass, Radius, and Age}
\label{sec:massradiusage}

The asteroseismic scaling relations (Eqs.~\ref{eqt-scalingM} and \ref{eqt-scalingR}) require corrections for precise work. The mean asteroseismic properties themselves are also subject to method-dependent offsets and trends. A calibration against fundamental data is therefore essential in our view.

We use stellar models and pulsation theory to define a star-by-star correction factor \fdnu\ to interpret the observed frequency spacings. The \nmax\ scaling has a strong underlying physical basis, tied to the relationship between the acoustic cutoff frequency and the surface gravity \citep{Belkacem2011}. A predictive theory, however, requires advances in our understanding of the excitation, damping, and reflection of modes. It is therefore not currently practical to define star-by-star corrections for \nmax. In practice, both the measurement systematics and the offsets between asteroseismic and fundamental radii are well-behaved functions of \nmax. Motivated by this result, we define an empirical \fnmax\ function that serves to calibrate our system.

In APOKASC-2, our primary calibrators were members of the open star clusters NGC 6791 and NGC 6819. There was limited overlap with eclipsing binary and interferometric samples, which remains true in the current sample, so these stars were not used as primary calibrators. A new tool is now a greatly expanded set of stars with reliable radii from \gaia. We use these data to define a correction function \fnmax\ that brings asteroseismic radii into statistical agreement with \gaia\ radii, inferred from knowledge of $L$ and \teffns.


\subsection{Mapping \dnu\ to Mean Density: Inferring \fdnu \label{sec:dnu}}
The observed oscillation pattern depends in detail on the internal structure of the star. In principle, it is possible to predict, identify, and fit individual frequencies. This approach, sometimes referred to as peak-bagging, has been employed for detailed studies of individual stars, and for stars in the open cluster NGC 6819 \citep{Handberg2017}. However, it is difficult to predict amplitudes from first principles, and real power spectra are complicated by mixed modes and rotational splittings. Even for the radial modes, there are known offsets between observed and predicted frequencies, even in the Sun \citep{CD1997}, that can be traced to poorly understood outer layer physics and near-surface reflection effects. See \citet{Li2023} for a recent discussion of this surface correction in the \kep\ context. Carefully chosen combinations of frequencies can mitigate these effects. 

Here, we collapse the frequency pattern down to a single figure of merit, \dnu, which is proportional to the square root of the mean density in the asymptotic limit of large radial order $n$ and a structure homologous to the Sun.  In real stars neither is strictly true. We correct for this by defining

\begin{equation}
\fdnu^2 = \frac {\langle \rho \rangle}{\dnu^2} \label{eq:dnu}
\end{equation}



\noindent and solving for it as described in the method sections below. We then took the inferred asteroseismic radii to define a \fnmax\ function, which perturbs the assigned asteroseismic gravities, and repeated the calibration exercise.  For all models we adopt the APOKASC-2 solar zero-point of 135.1416 $\rm{\mu Hz}$ for \dnu. We also adopt 3076 $\rm{\mu Hz}$ as the reference point for the solar \nmax; the \fnmax\ scaling is defined relative to this value.

To infer \fdnu, we need to choose stellar interiors models to produce theoretical frequency spectra. These model frequencies have to be converted into a \dnu\ that would have been observed from the model spectrum by choosing a subset of modes and assigning weights to spacings between individual modes. The difference between this predicted \dnu\ and the true mean density is then used to infer \fdnu. Following APOKASC-2, we adopted the Garstec models and the Mosser weighting approach.  However, we explored two different stellar evolution codes, and two different weighting schemes, to explore systematic uncertainties.

\begin{itemize}
 
\item{Sharma models and Weightings :}
The \textit{asfgrid}\footnote{The \textit{asfgrid} code is publicly available at \url{http://www.physics.usyd.edu.au/k2gap/Asfgrid/}}
\fdnu\ scheme \citep{sharma+2016} uses a grid of MESA v6950 models generated with the \textbf{1M\_pre\_ms\_to\_wd} test suite case \citep{paxton+2011,paxton+2013,paxton+2015}. 
Models were run without rotation, overshooting, diffusion, or mass loss. Convection was treated according to the \cite{cox_giuli1968} mixing length prescription. Opacities were generated from OPAL \citep{iglesias_rogers1993,iglesias_rogers1996} using \cite{grevesse_sauval1998} solar abundances, with C/O enriched abundance mixtures assumed for helium burning. In the low-temperature regime, molecular opacities are adopted from \cite{ferguson+2005}. Asteroseismic frequencies of the radial modes are then computed using GYRE v3.1 \citep{townsend_teitler2013}. A local measurement of \dnu\ in the vicinity of \nmax\ is performed as the slope of a linear fit to the radial mode frequencies versus their mode order using  a Gaussian weight function centered on \nmax, following \citet{white+2011}. The \fdnu\ term is defined as the ratio between the asymptotic and local \dnu, and tabulated as a function of evolutionary state, \fehns, \teffns, \nmax, and \dnu\ \citep{Stello2022}.

\item{Garstec models, White and Mosser Weightings :}
Our base case in APOKASC-2 used GARSTEC \citep{weiss:2008} models, which we also adopt here. GARSTEC models use OPAL radiative opacities \citep{iglesias_rogers1996} complemented at low temperature with molecular and dust opacities from the Wichita group \citep{ferguson+2005} and conductive opacities from Potekhin as updated in \citet{cassisi:2007}. The equation of state is FreeEOS \citep{freeeos:2012}. Hydrogen-burning reaction cross sections are taken from Solar Fusion II \citep{adelberger:2011}, and helium-burning reaction rates are from NACRE \citep{angulo:1999}. Stellar atmospheres are computed using the VAL-C T-$\tau$ relation from \citet{vernazza:1981}. Convection is modeled according to the mixing length theory, following the \citet{cox_giuli1968} implementation, and the mixing length parameter is fixed following the calibration of a standard solar model which, in combination to the T-$\tau$ relation, leads to $\alpha_{\rm MLT} = 2.012$. 

Models with masses $< 1.25~\msun$ include gravitational settling, following \citet{thoul:1994}. The efficiency of gravitational settling is artificially decreased for masses between $ 1.25~\msun$ and $ 1.35~\msun$, with a suppression factor increasing linearly from 0 to 1 in that range, and it is neglected for larger masses. The impact of this simplification for the structure of RGB stars is small. Turbulent mixing below the convective envelope is modeled diffusively, with the diffusion coefficient based on the parametrization from \citet{vandenberg:2012}. Convective over/undershooting is modeled using the diffusive approach \citep{freytag:1996}, and the free parameter fixed to $f=0.02$ in all convective boundaries. To avoid the well-known problem of large overshooting regions in very small convective cores, $f$ is decreased from its standard value down to 0 in the convective cores of models in the mass range $ 1.4~\msun$ to $<1~\msun$. The solar mixture is from \citet{gn:1993}, therefore (Z/X) = 0.02439 is used to define the spectroscopic solar reference abundance, \fehns\ = 0. The metallicity and helium enrichments are assumed to follow a linear relation such that $Y = Y_{\rm BBN} + (\Delta Y)/\Delta Z) Z$, and $\Delta Y /\Delta Z= 1.15$ as determined from $Y_{\rm BBN}=0.2485$ and our solar calibration. Mass loss is modeled using Reimers' prescription with the free parameter fixed to $\eta=0.2$. 

For each stellar model, version 0.3b of ADIPLS \citep{adipls:2008} is used to compute all radial modes with frequencies lower than the acoustic cutoff frequency. The $\dnu$ term is obtained as the slope of a weighed linear fit to the frequencies of the radial modes as a function of mode order. The weighting function represents a Gaussian distribution of power, centered in $\nmax$, and characterized by a FWHM value. Two possibilities have been considered for the latter: the prescription by \citet{white+2011} in which FWHM = 0.25$\nmax$, and the empirical determination obtained by \citet{mosser:2012}, for which FWHM = 0.66 $\nmax^{0.88}$ for $\nmax$ expressed in $\mu$Hz. The value of $\dnu$ is then used as in Eq.~\ref{eq:dnu} to compute $\fdnu$ for all the models in the grid. 

The grid of stellar models ranges from 0.6 to 5.0 \msun, with mass step of 0.02 \msun\ in the range 0.6-3 \msun, and 0.04 \msun\ for larger masses. \feh spans from $-2.5$ up to +0.6~dex, with steps of 0.05~dex, up to 3 \msun, and from $-1.0$ up to +0.6 for masses between 3 and 5 \msun. The chemical mixture is always solar-scaled. For this work, in BeSPP, $\alpha$-enhancement is taken into account by using a modified \feh value computed as \fehns$_{\rm corr}$ = \feh + 0.625 \afens. This is analogous, and quantitatively similar, to other transformations, such as the one from \citet{salaris:1993}. 

We then employ a Bayesian inference code, BeSPP, to interpret the model grid. BeSPP assumes a Salpeter mass function and a constant star formation rate as priors, and posterior distributions are obtained by marginalization as described in \citet{Serenelli2013}. The process for determining \fdnu\ is as follows: \dnu, \nmax, and \teff are used to obtain the seismic \logg, and an initial guess for the stellar mass using scaling relations and setting \fdnu\ $=1$. The input set of variables in BeSPP therefore is \loggns, \fehns$_{\rm corr}$, and the stellar mass along with their respective errors. These are used to construct the posterior distribution of \fdnu, which is used to refine the stellar mass and then used to start the new iteration. Convergence is defined as \fdnu\ varying less than one part in 10$^5$. Note that this is different from the traditional grid-based modeling in several respects. In particular, the \teff scale in stellar models is not used at all, as \teff is only used in scaling relations and not in the construction of any likelihood. 

\end{itemize}

\begin{figure}

\includegraphics[width=8cm]{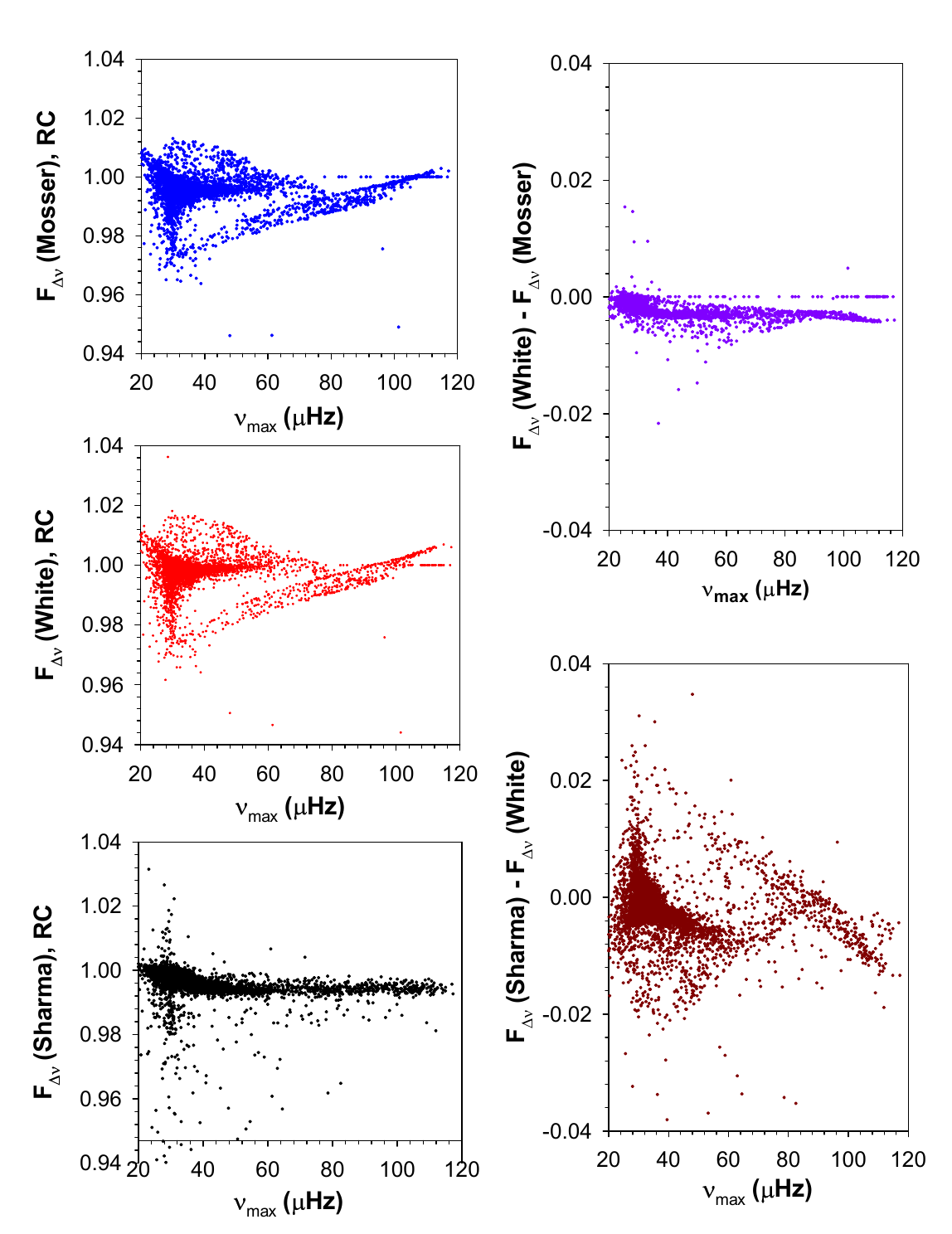}

\caption{Frequency spacing corrections for RC stars using the Garstec+Mosser (top left), Garstec+White (middle left) and Sharma+White (lower left) models and weights. Differences between Garstec+Mosser and Garstec+White (top right) measure the impact of the assigned weighting scheme. Differences between Garstec+White and Sharma+White measure the impact of the models used. For a minority of stars, the true \logg range of the sample did not correspond to the range in the model grid, which is responsible for some of the structures seen in the BeSPP results.}
\label{fig:rccor}

\end{figure}

\begin{figure}

\includegraphics[width=8cm]{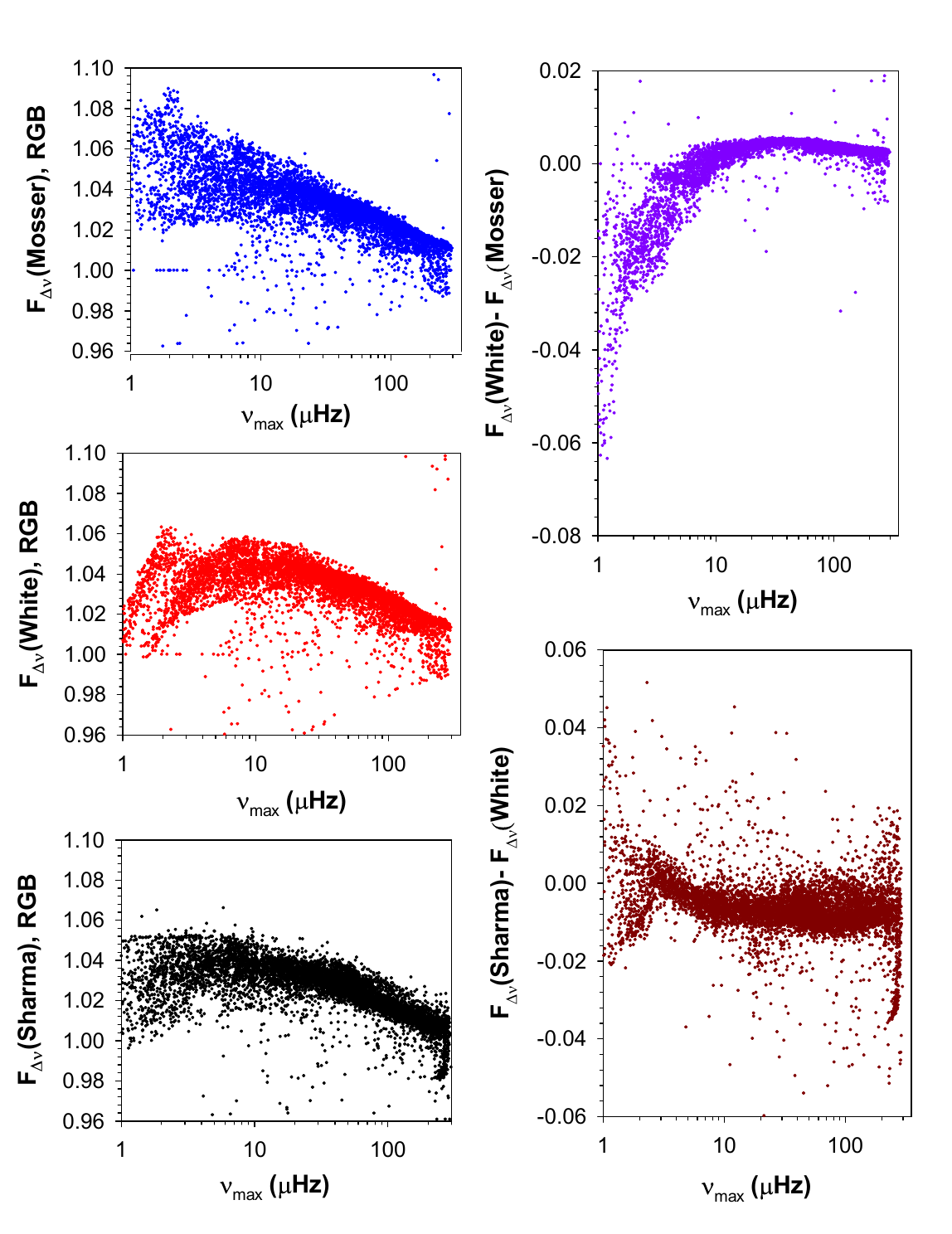}

\caption{Frequency spacing corrections for RGB stars using the Garstec+Mosser (top left), Garstec+White (middle left), and Sharma+White (lower left) models and weights. Differences between Garstec+Mosser and Garstec+White (top right) measure the impact of the assigned weighting scheme. Differences between Garstec+White and Sharma+White measure the impact of the models used.}
\label{fig:rgbcor}

\end{figure}

The derived correction factors for RC and RGB stars are illustrated in Figures \ref{fig:rccor} and \ref{fig:rgbcor} respectively. There is a significant difference between RC and RGB stars; the degenerate cores of AGB/RGB stars are more different from the Sun than the lower density RC stars, inducing a larger change in the relative scalings. Our applied corrections therefore require knowledge of the evolutionary states. In the absence of a well-defined state, we consider both solutions and adopt a larger uncertainty in the derived masses and ages. 

Relative offsets between methods become significant in low \nmax\ stars. If left unaddressed, this would induce method-dependent trends in the mass and radius scale as a function of \nmax. We address this by calibrating our absolute results to a fundamental reference system, which by construction places all three methods on a common zero-point and removes the relative trends seen in the left panels of Figures \ref{fig:rccor} and \ref{fig:rgbcor}. However, there is also a scatter in the inferred \fdnu\ for both populations that is apparent in the right panels of Figures \ref{fig:rccor} and \ref{fig:rgbcor}, even in the high \nmax\ domain for the RGB and in the main RC where the trends with \nmax\ are weak. For these domains, the scatter induced by the choice of weights ($\times$0.001) is less than that induced by the choice of models (0.005). We therefore take 0.005, as a minimum random error source for \fdnu. This is comparable to that induced by the uncertainty in the \dnu\ measurements themselves for high-quality data, which illustrates the importance of theoretical models for interpreting the pulsation frequencies.

\subsection{Asteroseismic and \gaia\ Radii: Inferring \fnmax} \label{sec:nmax}

At this point we are in a position to compute radii and masses. The zero-point of the \nmax\ scaling relation is formally the solar value. As discussed in APOKASC-2, however, the solar values differ between pipelines, and there is a poor correspondence between relative measurements in different pipelines and the relative solar values. In APOKASC-2 the zero-point of 3076 $\rm{\mu Hz}$ was set by calibrating on masses of open cluster stars in NGC 6819 and 6791. This calibration, however, does not address any potential systematics as a function of evolutionary state or surface gravity. 

\gaia\ parallaxes, combined with photometry, bolometric corrections, and an extinction map, can be used to infer luminosities. When combined with \teff from APOGEE, we can then solve for $R$, hereafter \gaia\ radii. Asteroseismic radii, which do not depend on parallax, can then be compared with these radii that do.  This combination was used as a test of the absolute \teff scale and the \gaia\ zero-point for DR2 \citep{zinn2019} and DR3 \citep{zinn2021}.



\begin{figure}

\includegraphics[width=8cm]{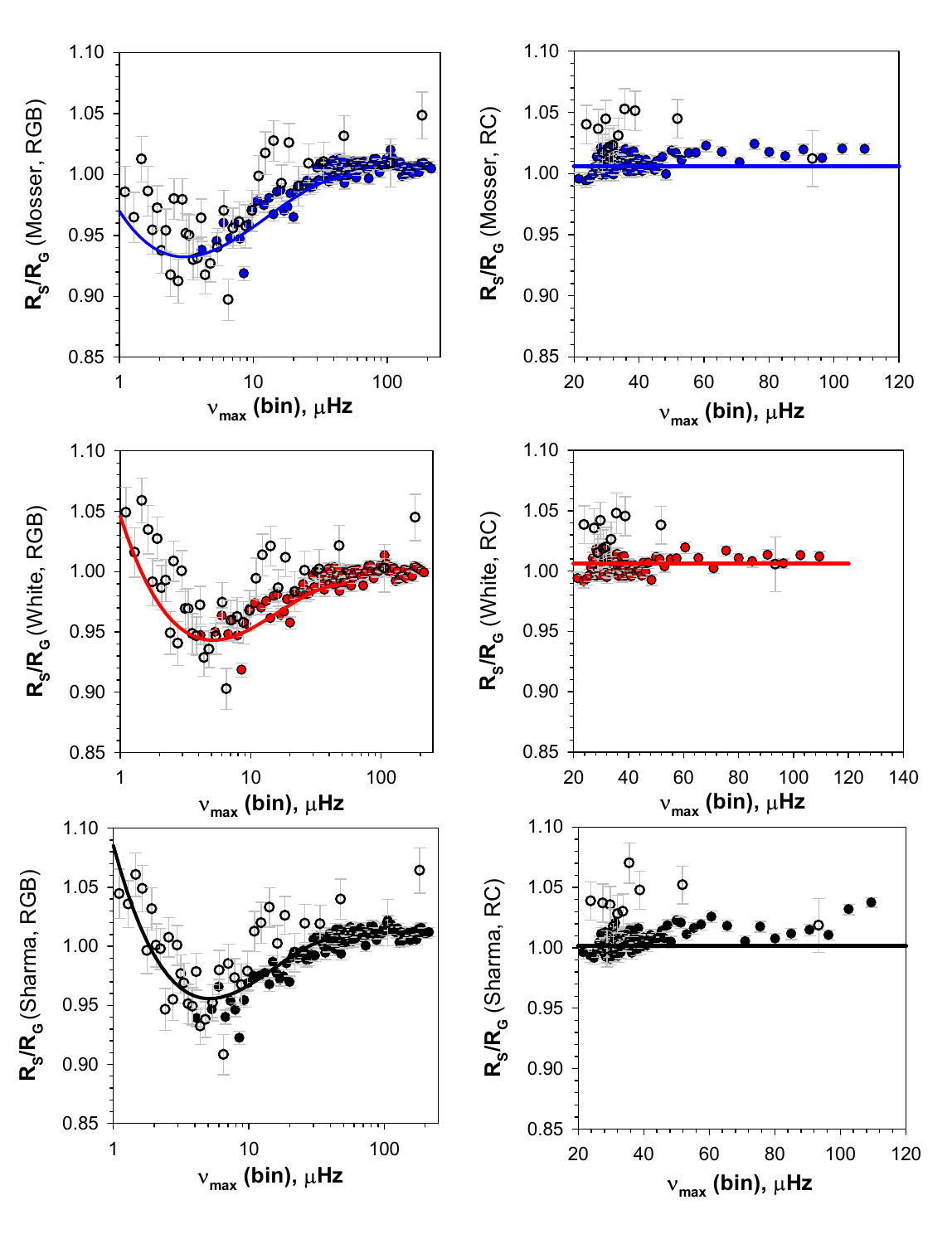}

\caption{The ratio of asteroseismic to \gaia\ radii for our sample. Rows denote results using different methods: the Garstec+Mosser (top), Garstec+White (middle), and Sharma+White (bottom) models and weights. RGB stars are shown on the left and RC on the right. Data points are 50 star bins, defined in a rank-ordered list in \nmax. Silver sample bins are open points, and Gold sample bins are closed points. The \nmax dependent fits to the RGB data are shown with the lines in the left panel.}

\label{fig:gaiar}

\end{figure}

Given the dependence of \fdnu\ on \nmax, it is plausible that a \nmax\ correction, \fnmax, also depends on \nmax. For RGB stars, which span several orders of magnitude in \nmax, we therefore implemented an empirical calibration. Binned means of the asteroseismic-\gaia\ radius ratio for RGB stars were taken as a function of \nmax, and are shown in Figure~\ref{fig:gaiar}. A third-degree polynomial was then fit using least squares as implemented in \textit{numpy}, giving each bin equal weight. The fit was anchored to the ratio of the radii at a \nmax\ of 50 $\rm{\mu Hz}$. The asteroseismic radii were then corrected according to this polynomial, with separate corrections performed for each \fdnu\ scale. This calibration can be thought of as a calibration of \nmax, since $R_{\rm seis} \propto \nmax$. The resulting calibration of \fnmax\ is: 
\begin{equation}
\fnmax = (1 + p)^{-1},
\end{equation}
where 
\begin{equation}
p = a (\ln \nmax)^3 + b (\ln \nmax)^2 + c (\ln \nmax) + d,
\end{equation}
with $\nmax$ in $\rm{\mu Hz}$, and is defined for $1.1 < \nmax < 50 \;\rm{\mu Hz}$. We adopt a fixed zero-point for $\nmax > 50\;\rm{\mu Hz}$ on the RGB and a separate one for the RC. The polynomial coefficients and zero points are provided in Table \ref{table:poly} for each of the $\fdnu$ scales. This calibration brings the asteroseismic radii and the \gaia\ radii into alignment for stars up to $R \approx 
50~R_{\odot}$. We adopt the GARSTEC+Mosser \fdnu\ values as our default case, which was the one used in APOKASC-2. The lower RGB and RC zero-points for the Mosser scale are the equivalent of an effective solar \nmax\ of 3063 and 3057 $\rm{\mu Hz}$ respectively.

\begin{table*}
\label{table:poly}
    \centering
    \begin{tabular}{c c c c c c c c}
MODEL & WEIGHT & $FNM_{RC}$ & $FNM_{RGB}$ & a & b & c & d
  \\
 \hline
SHARMA & WHITE & 0.9941 & 0.9893 & -0.00989448 &  0.08085812 & -0.18509446 &  0.08986743 \\
  GARSTEC &    MOSSER  &  0.9937 & 0.9959 & -0.00556347 &   0.04268084 & -0.07452754  & -0.02442568 \\
 GARSTEC  &   WHITE  &   0.9985 & 1.0032 & -0.00758535 &  0.06330463 & -0.14603078 &  0.05336784 \\

    \hline
    
    \end{tabular}
    
    \caption{Fit coefficients and zero-points for our three methods for inferring mean density. The underlying stellar models are in column 1, and the weighting scheme is in column 2. Zero-points for the RC (column 3) and RGB (column 4) are the total scale factors relative to the APOKASC-2 value of 3076 $\rm \mu Hz$. Polynomial coefficients a, b, c, and d, as described in the text, are in columns 5-8.} \label{table:fit_coeffs}
\end{table*}


Figure~\ref{fig:gaiar} compares \gaia\ radii to asteroseismic radii. We rank order stars in \nmax\ and show averages for 50 star bins. The three left panels show RGB data adopting different \fdnu\ methods (from top to bottom, Garstec+Mosser, Garstec+White, and Sharma+White); the three panels on the right show RC stars using the same methods. Open symbols are the Silver sample, and closed ones are the Gold sample. 

Asteroseismic radii are in excellent agreement with fundamental data in the RC and the lower RGB. However, there are clear, and highly significant, deviations between asteroseismic radii and fundamenta data in lower \nmax\ RGB stars; furthermore, the magnitude of the offsets depends on the method used to infer \fdnu. The lines in our left panels represent the fits, using the coefficients in Table \ref{table:fit_coeffs}. 

There are also offsets between the Silver and Gold RC samples, but they are of marginal statistical significance.  The Silver RC sample is quite small, with large uncertainties, as most \kep\ RC stars have high quality data. Given the lack of a significant trend in \nmax\ for the RC, a constant \nmax\ zero point was used for the RC.

Our fit differs from that used in APOKASC-2, which was based on masses in the open clusters NGC 6791 and NGC 6819. However, our system does produces results in good agreement with open cluster data. In a companion paper focused on luminous giants, \citep{Ash2024} found mean lower RGB and RC masses in NGC 6791 of $1.15 \pm 0.01$ and $1.12\pm0.01$ \msun, respectively, in excellent agreement with the \citet{APOKASC2} fundamental mass of $1.15 \pm 0.02$. For NGC 6819, average lower RGB and RC masses were $1.65 \pm 0.02$ and $1.64 \pm 0.02$
\msun, respectively, roughly $2 \sigma$ higher than the $1.55 \pm 0.04$ \msun\ value adopted by APOKASC-2. However, we note that our values are close to those obtained by \citet{Handberg2017} ($1.61 \pm 0.02$ and $1.64 \pm 0.02$ \msun\ for the RGB and RC respectively.) We discuss our systematic and random error models in Section \ref{sec:uncertainties}.

\subsection{Masses and Ages}
\label{sec:massage}

\subsubsection{Masses}
\begin{figure}
\includegraphics[width=8cm]{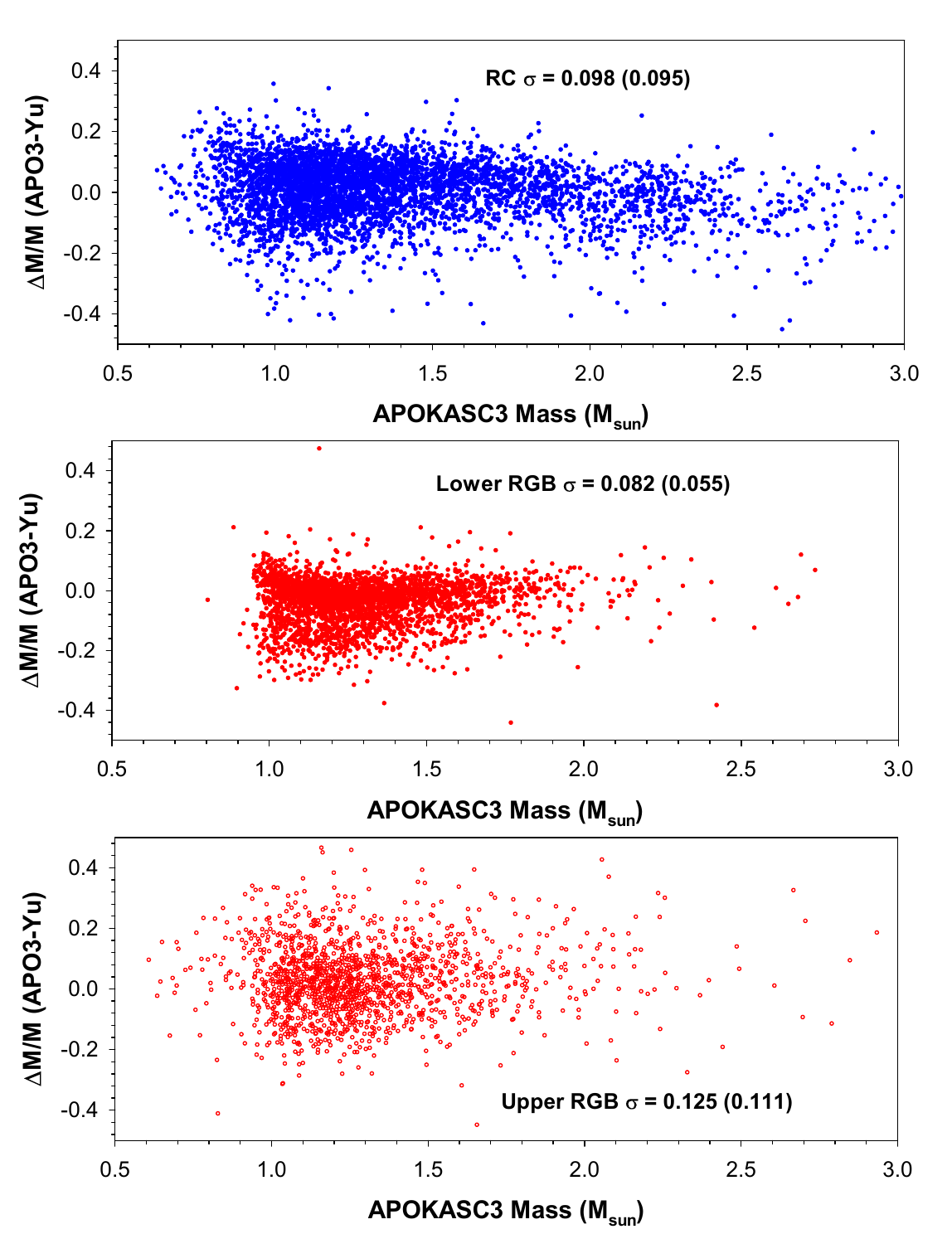}
\caption{APOKASC-3 and Yu masses compared for Gold sample stars in the RC (top), lower RGB (middle) and upper RGB (bottom). Errors in each panel are the standard deviation and the MAD converted to $\sigma$ (in parens). (in brackets)}
\label{fig:massoffset}
\end{figure}

For masses, we adopt the \fdnu\ and \fnmax\ data from above for all three of our \fdnu\ methods, which gives us three different mass scales. For consistency with APOKASC-2 we choose to use the Garstec+Mosser corrected mass scale as our base case, and provide the alternative options in Table \ref{table:altvalue}. We discuss population tests of our mass scale and uncertainties in Section \ref{sec:populations}. The \citet{Yu18} data set is the largest homogeneous comparison set for asteroseismic masses, and provides a good check on our results. They focused on stars with $\nmax > 5 \;\rm{\mu Hz}$ and used the SYD pipeline (Appendix \ref{app:syd}), which is one of the ones used in our study. In order to make a comparison with the \citet{Yu18}, we focus on the Gold sample, which describes the large majority of stars in that domain. We compare our data with Yu et al. in Figure  \ref{fig:massoffset}. To quantify the comparison, it is useful to divide the sample into 3 cohorts: RC, lower RGB (\logg $>$ 2.5) and upper RGB (\logg $<$ 2.5). The median fractional mass differences for these three groups are modest (+0.016, $-0.023$, and +0.022 for the RC, LRGB, and URGB, respectively); the dispersion is larger (0.098, 0.082, and 0.125 respectively). 

The \citet{Yu18} masses are in good average agreement with ours, which is expected because the underlying asteroseismic data is on a similar scale. The derived asteroseismic parameters from SYD are in excellent mean agreement with our system (Table \ref{table:zeropt} in Appendix \ref{app:merging}), with zero-points in \nmax\ and \dnu\ within 0.1\% of our mean. \citet{Yu18} used the Sharma+White model for inferring \fdnu. This method, as shown in Figures \ref{fig:rccor} and \ref{fig:rgbcor}, is close to our central value for the Gold sample \nmax\ range. 

We can trace the mass differences to larger measurement uncertainties in a single method relative to an ensemble average. The scatter of the SYD pipeline around our ensemble mean is $\sim 0.015$ in \nmax\ and \dnu\ for the Gold sample, which by itself would produce a predicted scatter in mass around our relationship of 0.075 \msun. The remainder of the mass scatter can be explained by the non-seismic data. \citet{Yu18} adopted a heterogeneous sample of \teff and \feh literature measurements, and only had a modest subset of 5,678 APOGEE measurements available; the data was also from DR12, and there have been substantive changes to APOGEE since that time. Photometric \teff measurements have large random uncertainties \citep{Pinsonneault2012}, and are subject to systematics from the adopted extinction map, and \teff is a direct ingredient of the scaling relations. \feh has an indirect effect on photometric \teff and on \fdnu. We therefore believe that the differences shown here can be traced to known effects.

Overall, we believe that our ensemble averaging method produces more precise data, and should be used in preference to single pipeline results when available. We explore the stellar population results more fully in Section \ref{sec:populations}. 
We also note that the Yu sample did not report masses for 695 of the Gold sample stars; we report results for these stars because our ensemble method allows us to recover signals in targets missed by individual pipelines. Conversely, the Yu sample includes stars without APOGEE spectra, which makes it of comparable size overall (16,094) to APOKASC-3. 




\vskip 1cm
\subsubsection{Ages}
We use stellar evolution models to infer ages from the measured mass, surface gravity, and composition. However, stellar ages are intrinsically model-dependent. We also explore several different methods for doing so. The first option for estimating the ages, and our reported base case, is most similar to the method used for APOKASC-2. In this case, the seismically inferred stellar mass and surface gravity are used, and the metallicity and $\alpha$-enhancement are combined into the corrected \feh as described in Section \ref{sec:dnu}. 



 


The corrected metallicity, the seismic mass, and \logg are then used as inputs in our Bayesisan code BeSPP, discussed in Section \ref{sec:nmax}, to determine the stellar age. No prior was used with respect to the initial mass function when inferring ages, it being used before to determine the seismic mass. We use the stellar models described in Section \ref{sec:dnu}. A priori, the stellar age and uncertainties could then be determined directly from the posterior distribution function. However, a typical problem that may arise is a lack of consistency between age and mass estimates originating in the non-linear relation between mass and age. To avoid this, for any given star, we have carried out three runs in BeSPP: one using the seismic mass as input, and two others using the seismic mass increased and decreased by its uncertainty respectively. In each case, very small mass errors are used in BeSPP to avoid non-linear effects in the age posterior distribution. The age determined from the posterior distribution function in the first case is taken as the central value of the stellar age, and the standard deviation is adopted as the statistical uncertainty linked to \feh and \logg errors. The age values determined in the two other runs are defined as the lower and upper 1-$\sigma$ uncertainties after quadratically combining them with the induced \feh and \logg errors.

\begin{figure}
\includegraphics[width=8cm]{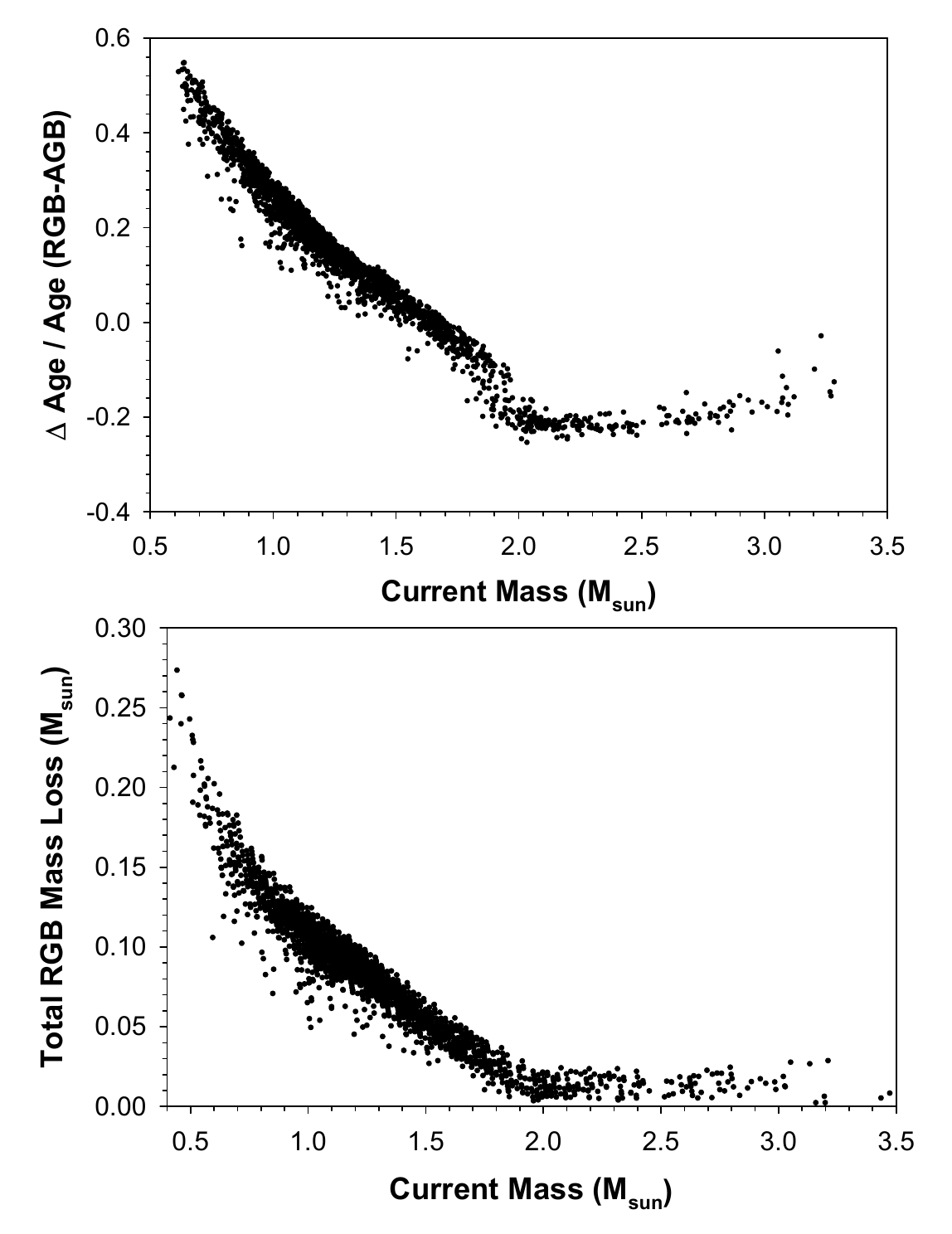}

\caption{Fractional difference in age between AGB and RGB stars (top) and the total mass loss on the RGB (bottom) as a function of the current mass of the star. Low mass stars experience much more mass loss.}
\label{fig:mdot}
\end{figure}

We measure only the current mass in our sample; stars that have experienced significant mass loss had a higher birth mass, and therefore a younger age than the one we would infer in the absence of mass loss. The standard Reimers mass loss rate we have used is defined by assuming that a fraction $\eta$ of the luminosity is used to provide the lost gravitational potential energy associated with a stellar wind;  $\eta L = -\frac{GM}{R} \frac{dM}{dt}$. We adopted $\eta = 0.2$ for our initial calculations. Low mass stars experience much more mass loss than high mass stars, as seen in the bottom panel of  Figure \ref{fig:mdot}. This effect means that low mass AGB stars had a higher birth mass, and therefore a younger age, at a given current mass (top panel, Figure \ref{fig:mdot}). Higher mass stars have little mass loss, but are actually \textit{older} at fixed mass in the AGB phase as opposed to the RGB phase. In this case, there is minimal mass loss, and the AGB star is older because it has lived through both the core-H and core-He burning phases.

We use the $\alpha$-rich population to validate our mass loss model (see Section \ref{sec:populations}). In brief, the large majority of the stars in that population have similar ages, and therefore the difference in mass between RC and RGB stars is a reasonable diagnostic of mass loss, which is 0.10 \msun\ for our sample. This corresponds to an effective Reimers $\eta = 0.17$, which we adopt as a central value. The RC population has a larger dispersion in mass and age than the RGB precursors, which is consistent with a dispersion in mass loss of 0.03 \msun. This is equivalent to a dispersion in $\eta$ of 0.05, which we propagate into an enhanced age uncertainty for the RC population.

We therefore define our ages in three domains. On the lower RGB, non-interacting stars are not expected to experience significant mass loss. On the RC, stars have in general experienced mass loss in the prior RGB phase. Upper RGB stars are a mixture of first ascent giants and AGB stars; the latter have, to a first approximation, the same degree of mass loss as RC stars. We discuss population tests of our mass loss model in Section \ref{sec:populations}.

The majority of luminous giants are on the first ascent of the RGB because hydrogen burning is much more efficient than helium burning, but a significant fraction \textbf{$\sim 1/6$} are AGB stars;
in most cases, the RGB age will therefore be applicable, but the two populations have significant overlap in the HR diagram. However, we can use \teff to distinguish between the two in some domains, as shown in Figure \ref{fig:AGBRGB}. We provide two sets of ages for shell-burning stars above the RC that cannot be reliably sorted into AGB or RGB alone. 

Stellar ages are subject to significant systematic uncertainties. In low mass stars, assumptions about mass loss are important, as shown in Figure \ref{fig:mdot}.  In higher mass stars, the model-dependent treatment of convective cores has a strong impact on the main sequence lifetime, and therefore on the age on the RGB or RC. To quantify this, we have computed independent models for ages, following the method described in \citet{Tayar2017a}. Here the mass, surface gravity, [M/H], and [$\alpha$/M] are interpolated in a grid of models build using the YREC code \citep{Demarque2008}. These models assume no core overshoot, a \citet{grevesse_sauval1998} mixture, no gravitational settling, and a gray atmosphere.For these stars, an empirical calibration of the metallicity dependent mass-loss is assumed for clump stars with masses below 1.3 \msun\ using the same formulation as \citet{Tayar2023RNAAS}, namely  $\Delta M=(0.12([M/H]+1.0)^2+0.95)-(0.3[M/H]+1.0)$ M$_{\odot}$.
Mass loss is not included for age inferences in higher mass RC stars or any RGB stars.
For this interpolation, the ages are estimated as if all of the stars are on the first-ascent RGB. Because we included neither mass loss nor core overshooting in these models, the differences are a measure of the systematics from these effects. 

\begin{figure}
\includegraphics[width=8cm]{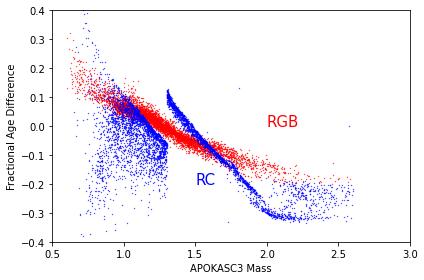}
\caption{Difference between our main (GARSTEC) and alternative (YREC) ages for both RGB stars (red) and clump stars (blue), in the sense GARSTEC-YREC. Mass loss becomes important in low mass stars, while differences in convective overshoot treatment are significant for higher masses. Systematics are at the 10 \% level for the majority of the RGB stars, and somewhat larger for more massive RC stars. The sharp edge in the clump offsets comes from the 
assumptions about mass loss in the alternative (YREC) ages.}
\label{fig:ageoffset}
\end{figure}

Even with these simplifications, as shown in Figure \ref{fig:ageoffset}, we find surprisingly good agreement.  For RGB stars there are trends at the $\pm$ 10\% level, with YREC ages being higher at low mass and lower at high mass). Typical masses on the lower RGB are  $\sim$ 1.4 \msun, so the progenitors of these stars had a small or absent main sequence convective core, and little mass loss is expected for them. These insensitivities explain our modest differences. At higher masses and younger ages, the differences in overshoot are important, but little mass is expected to be lost. At lower masses and older ages, choices about gravitational settling are important, as are the assumptions about mass loss for RC and upper RGB stars. 


\vskip 1cm
\subsection{Uncertainties}
\label{sec:uncertainties}

Our error model begins with a formal error based on measurement uncertainties, such as \nmax, \dnu, \teffns, and \fdnu. We confirm these by population and external checks on mass, radius, and age. Systematic uncertainties are tied to those in the calibration system and model dependent inferences (such as \fdnu\ and the stellar interiors models used to infer age from mass, composition, and evolutionary state). We discuss these effects and their impact next, and summarize the outcome of external tests at the end. 

\subsubsection{Uncertainties in Asteroseismic Properties}
\label{sec:astunc}

Our error model is similar to that used in APOKASC-2. We take the scatter of individual pipeline measurements of asteroseismic properties around the mean as a measure of random error. We also assume that the uncertainties are uncorrelated, taking the standard error of the mean as our measurement error for \dnu\ and \nmax. We provide alternative uncertainties in Appendix \ref{app:merging} for those who wish to use other measures. 

The dispersion between calibrated \fdnu\ values is another measure of random uncertainty. From Figures \ref{fig:rccor} and \ref{fig:rgbcor}, it may appear that these could be quite large; however, our calibration ties these to a common value and the dispersion in calibrated mass and radius values induced by the method for inferring \fdnu\ is much smaller than the pre-calibration scatter.  When combined with uncertainties in \teff described below, these error sources are propagated in quadrature to infer an overall uncertainty in mass and radius.

\subsubsection{Systematic and Random Spectroscopic Uncertainties}
\label{sec:specunc}

For this catalog, we adopt APOGEE DR17 central values \citep{DR17} where available. Asteroseismic measurements are independent of the spectroscopic ones. However, uncertainties in \teff propagate directly into ones for radius and mass. In addition, we need to know the stellar composition to properly map the observed \dnu\ onto mean density and to infer ages.

Changes between data releases are a good check on systematic and random errors. \citet{APOKASC2} used results from DR14, while the current effort uses results from DR17. The overlap samples, for DR17 giants, are 12,888 (DR14) and 15,422 (DR16). In order not to be biased by a small number of large changes, we used median statistics, and converted the median absolute deviation (MAD) to an effective sigma (for a normal distribution, $MAD = 1.4826 \sigma$). 

We present our results in Table \ref{table:uncertainties}. These properties for our sample are comparable to those derived from the full data releases \citep{Spoo2022}.

\begin{table*}
    \centering
    \begin{tabular}{c c c c c c c c c}
         Category & Units & Adopted & DR17 & DR17-DR14 & DR17-DR14 & DR16 & DR17-DR16 & DR17-DR16  \\
                -  & & $\sigma$   & $\sigma$ & median & $\sigma$ & $\sigma$ & median  & $\sigma$  \\
 \hline
\teff & K & 0.53 DR16 & 8.6 & $-$7.7  & 23.4  & 85.6 & $-$11.4  & 21.7\\
\logg (spec) & dex & 0.065 & 0.024  & 0.031 & 0.069 & 0.053 & 0.023 & 0.053 \\
 \feh & dex & 0.05 & 0.007 & $-$0.075 & 0.019 & 0.007 & $-$0.023 & 0.020 \\
\afe & dex & 0.02 & 0.006 & 0.021 & 0.028 & 0.007 & 0.025 & 0.022 \\

    \hline
    
    \end{tabular}
    
    \caption{Adopted uncertainties (third column) for our key spectroscopic observables (first column). We also include the median random error in DR17 (4th column), median difference between DR17 and DR14 (5th column), and the dispersion between them 6th column). The 7th, 8th, and 9th columns are the median random error in DR16, median difference between DR16 and DR14, and the dispersion between them. Note that we do not adopt the DR17 \teff errors where available; instead we adopt random errors of 0.53 times the DR16 errors. Stars without DR16 errors are treated as discussed in the text.} \label{table:uncertainties}
\end{table*}

Systematic scale differences are small for \teff and \logg because they are tied to absolute reference systems. Random errors for both are important, and provided by APOGEE for all spectroscopic measurements.
The notional DR17 uncertainties are much smaller than those in prior releases. Based on the dispersion in results between data releases alone, we argue that the DR17 uncertainties are underestimated. For more realistic errors we use external comparisons.

For \teffns, we take advantage of a robust feature of our sample: the temperature offset between RC and RGB stars \citep{Holtzman2018}; 
see also \citet{vrard2024}. The RGB locus has well-defined \teff trends with \loggns, mass, and \fehns. When these trends are removed, there is a clear separation between RC and RGB stars that is a weak function of stellar observables. We also have independent asteroseismic techniques for evolutionary classification, which can in turn be used to train spectroscopic state diagnostics. If the errors were as large as the quoted DR16 ones, we would expect a much higher rate of incorrect spectroscopic states than what we see. The observed false positive rate is consistent with a median \teff error of 45\,K, 
corresponding to 0.53 of the DR16 errors. We adopt this scaled version of DR16 uncertainties for our \teff uncertainty. For stars without DR16 data, we inflate the DR17 errors by a factor of 5, which is the ratio of the mean \teff uncertainties in the two data releases. 

Spectroscopic \logg errors are important for detecting background sources not associated with the APOGEE target and rejecting outlier measurements. We therefore discuss them more fully in Appendix \ref{app:merging}. Our adopted error in Table \ref{table:outliers} is inferred from the dispersion derived from the median absolute deviation between APOKASC-2 asteroseismic data and DR17 spectroscopic values.

For abundance errors, both systematic and random uncertainties are important, as we are comparing observed abundances to absolute abundances in theoretical models. Revisions between data releases induce larger shifts than the quoted random errors in DR17.  We adopt 0.05 dex for \feh and 0.02 dex for \afe as minimum uncertainties for the purpose of comparing models to data, which are characteristic of the offsets seen between methods. For [C/N], used for the evolutionary state specification, we adopt the DR17 values, as the \teff errors are the largest component of the error budget for state inference.


\subsubsection{Systematic Radius and \teff Uncertainties}
\label{sec:sysrteff}

Our comparisons between \gaia\ radii and asteroseismic radii are a test of the error model. We focus on RGB stars with \nmax\ between 50 and 200\;$\mu$Hz. The Gold sample stars have a scatter not much larger than that expected from errors in the \gaia\ radii alone; this indicates that our predicted uncertainties in the asteroseismic radii are reasonable.  The scatter in the Silver sample is larger than that expected from the \gaia\ radii, consistent with the larger scatter for these that that we expect from the error model. We attempted to use the error budget to infer a scale factor for the radius, and by extension mass, estimates; however, the derived results are not robust to outliers, and are sensitive to the treatment of large astrometric errors.  As a consequence, we can claim that our errors are consistent with this check, but we do not attempt to calibrate them on the joint data.

External comparisons are also useful for quantifying the systematic uncertainties in the APOGEE temperature scale. Given that APOGEE is calibrated to the fundamental Infra-Red Flux Method (IRFM) scale \citep{gonzalez-hernandez_bonifacio2009}, the absolute accuracy of the temperatures is set by that of the IRFM system. \citet{gonzalez-hernandez_bonifacio2009} found that their giant temperature scale agreed to within 40\,K when compared to the independent \citet{alonso_arribas_martinez-roger1994} temperature scale for stars with \feh $> -0.4$, which we take as a 2$\sigma$ systematic uncertainty on temperature. In practice, we adjust the zero-point of our seismic radius scale to agree with \gaia\ radii, making our results insensitive to the absolute \teff zero-point.



\subsubsection{Systematic Uncertainties Arising from the \gaia\ Radius Calibration}
\label{sec:sysgaia}

The asteroseismic radius systematics are best considered in three separate regimes: stars with $R < 30$ \rsun, stars with $R > 30$ \rsun, and stars with [Fe/H]$< -1$. The majority of stars in our sample fall into the first category. These stars have systematics set by the \gaia\ calibration we perform and the temperature systematics discussed above. The \gaia\ systematics arise from zero-point offsets in the parallax scale, which are documented to be position-, color-, and magnitude-dependent \citep{lindegren+2021,khan+2023}. The DR3 parallax scale appears to have a global offset of $15 \;\mu\mathrm{as}$ with respect to asteroseismic parallaxes \citep{zinn2021}, which would translate to a 2\% systematic uncertainty in radius. Temperature systematics are smaller than this, and so the parallax uncertainty dominates the systematic uncertainty in asteroseismic radii via their calibration to \gaia. We therefore estimate that our calibration of \fnmax\ is accurate to $2\%$ for the large majority of the sample.

Radius systematics are larger than 2\% for the minority of stars with large radii or low metallicities. Stars with $50\;\rsun > R > 30\;\rsun$, corresponding to $1 \;\rm{\mu Hz} < \nmax < 4 \;\rm{\mu Hz}$, have radii that disagree with \gaia\ radii at up to the 10\% level before calibration. Although our fit corrects for these deviations, it is important to understand the physical origin of these differences \citep{zinn+2023}.
Stars with [Fe/H] $< -1$ are in a domain where the IRFM has limited data, and there is the possibility of systematic offsets in the underlying \teff\ scale, as discussed in \citet{schonhut-stasik+2023} for the APO-K2 catalog. Although metallicity is accounted for in the \fdnu\ factor, it is not in the \nmax\ scaling relationship. Theoretical considerations would also predict that the asteroseismic \nmax\ scaling relations could be sensitive to mean molecular weight, and thus metallicity \citep{viani+2017}; however, in APOKASC-2, the data did not show evidence for a 
metallicity-dependent systematic offset in the close to solar metallicity domain. We therefore caution that our results may be subject to larger systematic errors in the large radius and low metallicity domain than for the bulk of the sample.

\subsubsection{Systematic Mass and Age Uncertainties}
\label{sec:sysmage}

Our calibration method places asteroseismic and Gaia radii on a common absolute scale. Formally, however, this is only a constraint on the product of $\fnmax T_{\rm eff}^{1/2} \fdnu^{-2}$.
It is therefore possible that some of the offsets captured in this term could arise from errors in \teffns or \fdnu, rather than being an offset in the \nmax\ scaling relation. In the limit where the \nmax\ relation is correct, and the error is in the \dnu relationship alone, the mass correction would scale with $\fnmax^2$; our approach, assigning the full error to \nmax, scales with $\fnmax^3$. There is therefore a systematic scale factor error of order \fnmax in our masses, which is related to the origin of the radius offset. For the vast majority of our sample this effect is small, but it becomes important in more luminous RGB stars (see below). 

Our mass and radius uncertainties are taken from the quadratic sum of the uncertainties in \dnu, \nmax, \teff and \fdnu. Similarly, the age uncertainties are taken from the uncertainties in mass and composition.  However, there are domains where systematic errors can be significant, and where the assumptions used in our error model can break down. 
We therefore test our results in several different ways (Section \ref{sec:populations}). The mean masses and mass scatter in star clusters are a test of whether our radius-based reference system produces sensible masses, and whether our error model produces sensible uncertainties.
The properties of the $\alpha$-rich population, which can be treated as a pseudo-cluster in some respects, also provide interesting tests. In this case, the small observed age dispersion in the lower RGB, the relative ages of the RC and RGB, and the mass difference between them are all consistent with our overall model \citep{Ash2024}. 

For the more luminous stars, \fnmax\ deviates significantly from unity, so there are larger systematic uncertainties. 
We see evidence of differences in age relative to lower RGB stars of 17\% - 37\% for moderate and high luminosity $\alpha$-rich stars, corresponding roughly to mass offsets of 6\% - 12\%. These stars are a mix of AGB and RGB stars, and population effects could be important, so we do not believe that this is sufficient grounds to revise the underlying system. 

\subsection{Outliers and Rotation}
\label{sec:outliers}

Any large astrophysical sample will contain outliers, and ours is no exception. Some of these objects are of genuine astrophysical interest, but many are simply the result of automated analysis of large data sets. We therefore individually examined the light curves and analysis for a subset of 201 stars in unusual regimes of phase space or with unusual ensemble asteroseismic properties. Common issues included more than one asteroseismic target in the aperture, typically from chance alignments; eclipsing binaries; artifacts in the light curves; power spectra polluted by background classical pulsators, typically producing high spikes in Fourier space; and rotation. These stars were classified as outliers. Some targets also had very low or high \nmax\ outside of our calibration domain. In all of these cases, we did not provide mass, radius or age. However, 37 of these stars were confirmed to be valid measurements with interesting properties. A total of 114 Silver sample stars and 12 Gold sample stars were classified as outliers, with the remainder as detections. We summarize the categories below. Table \ref{table:outliers}, presented earlier, includes these results.

\begin{itemize}
    \item \textbf{Super Nyquist Stars} 49 of our candidates were confirmed in Liagre et al. (in prep) as detections above the Nyquist limit.  Of these stars, 39 were in our Silver sample. We treat all as outliers for this paper, classify them as HighN stars, and do not provide asteroseismic properties for them in this paper.
    \item \textbf{High Scatter}. 60 of our candidates had unusually high measurement scatter in \nmax. Of these, 14 were valid measurements. Many of the others were double or multiple sources (14) and eclipsing binaries (8). The remainder were classified as non-detections.
    \item \textbf{Stars Below the RC}. Red Clump stars cluster in a narrow range of radii, and smaller stars are unexpected and interesting. The large majority of RC stars had asteroseismic states, but a minority did not. Stars with RC state assignments based only on spectra were more likely to have spurious measurements, and we discuss them separately from seismic states.
    We had 11 spectroscopically classified RC stars below 6.5 \rsun. Of these, none were true RC stars, and none were confirmed as true asteroseismic detections at all. Of 32 stars with RC spectroscopic states, mass below 1 \msun, and radius between 6.5 and 10.5 \rsun, 5 were valid. However, 6 out of 11 small RC stars with asteroseismic evolutionary states are confirmed as true RC stars.
    \item \textbf{Unusually Massive Stars}. Very massive stars are rare, but present. We looked at 28 RC stars with masses above 3.5 \msun\ and 34 RGB stars with masses above 3 \msun. Of these stars, 26 were rejected as false positives; however, we do find truly massive stars in our data. In the catalog there are 97 stars with masses above 3 \msun, roughly evenly split between RC and RGB, 28 of which are in our Gold sample. The highest mass cohort (above 3.5 \msun) has 36 stars. An interesting number of these stars may be post-main sequence mergers \citep{Deheuvels2022}. Our data set has a number of targets in common with the recent \citet{Crawford2024} study of 48 high mass stars, although our mass estimates are systematically lower than those in that study for stars in common.  
    \item \textbf{Low Mass RGB Stars.} Below the RC, there is little mass loss expected in RGB stars, and the finite age of the Galaxy imposes a maximum age and minimum mass. We therefore expect a metallicity-dependent, but sharp, lower bound on the mass distribution. We checked 17 stars with mass below 0.9 \msun\ and radius below 10 \rsun; of these, 2 were valid, including one metal-poor stars for which a lower mass is reasonable. Very low stellar masses are also suprising, even on the upper RGB; of 52 stars with formal masses below 0.5 \msun\ at any \loggns, none were found to be valid asteroseismic measurements.
\end{itemize}


There is another unique population of stars which makes up 
4.0\% of the APOKASC-3 sample.
\footnote{Note that the number of stars reported here is less than what originally appeared in \citet{Patton2024}. This is due to differences in selection criteria for giants (log \textit{g} < 3.85 in \citet{Patton2024} and log \textit{g} < 3.5 in this work) and which APOGEE data release the spectra came from. Note also that in \citet{Patton2024} 15,220 giants were identified, whereas the giant catalog presented here contains 15,808 stars. This is again due to differences in selection criteria. Giants only made the APOKASC-3 main catalog if they had complete spectroscopic solutions in DR17, whereas \citet{Patton2024} used spectroscopic stellar parameters from DR16.}
These 
631 rotationally enhanced giants were originally reported in \citet{Patton2024} and have $v \sin i$ values between 5 and 75 km s$^{-1}$, 5+ times faster than the typical rotation speed expected for a giant \citep[e.g.,][]{Carlberg2011}. We note that 113 stars have $v \sin i >10$ km s$^{-1}$, the typical literature threshold for defining rapid rotation in giants; this corresponds to 0.7\% of our sample. There is significant evidence that the intermediate and rapid rotator cohorts have distinct properties \citep{Patton2024}. 

The spectra used to estimate $v \sin i$ came from APOGEE DR16, and we included estimates for giants whose spectra made it past one or both rounds of spectral fitting in ASPCAP. We cannot estimate $v \sin i$ for giants which were rejected by the pipeline from the outset. Further details on how the giants were selected and how $v \sin i$ was estimated are in \citet{Patton2024}.

\begin{figure}
    \centering
    \includegraphics[width=8cm]{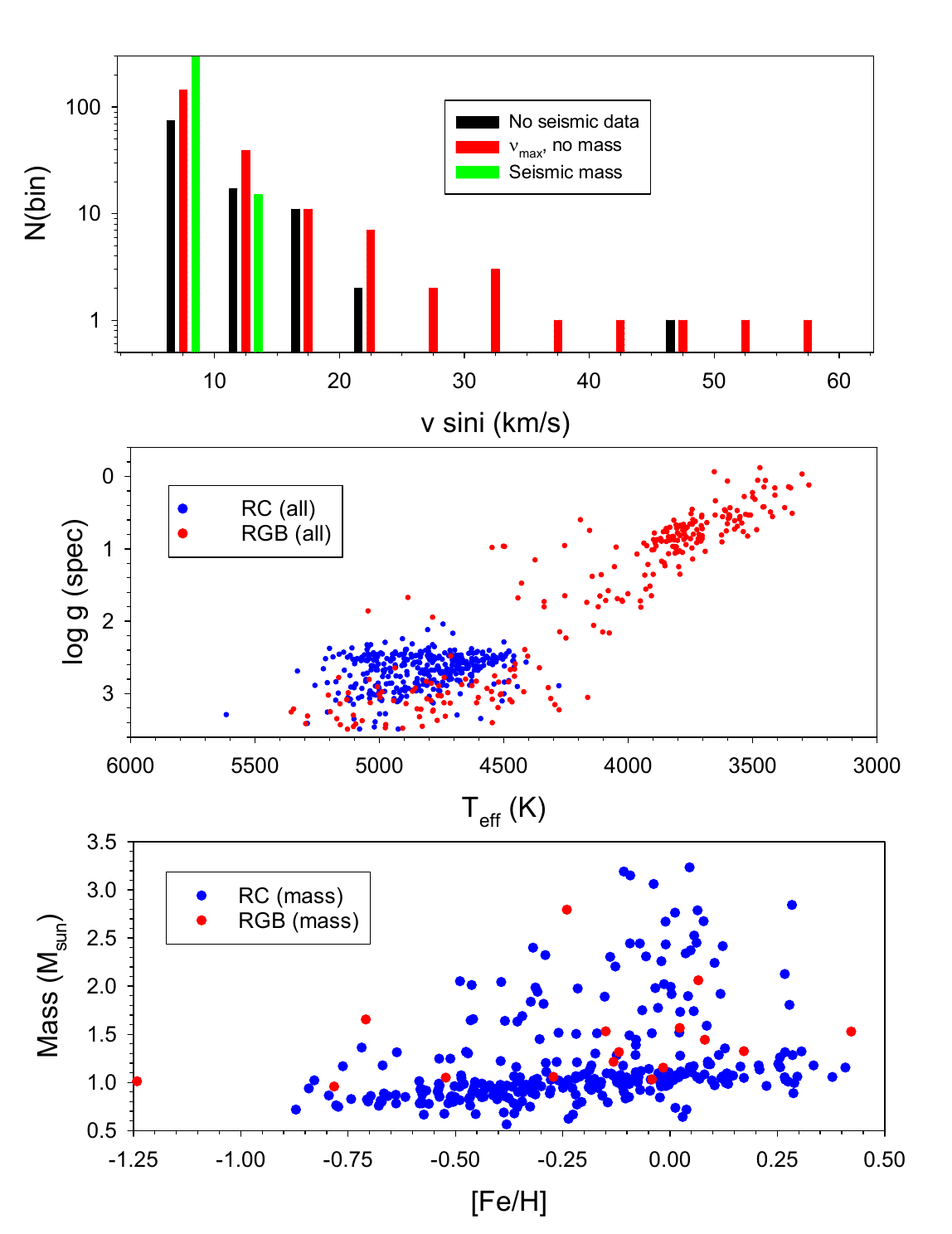}
    \label{fig:vdist}

    \caption{The distribution of $v \sin i$ for giants without a seismic detection (black), with partial data (red), and with full seismic data (green) are shown in bins of 5 km s$^{-1}$ in v sini in the top panel. Stars of different type are slightly offset at the same bin location for clarity. In the middle panel, we present spectroscopic log \textit{g} versus $T_\mathrm{eff}$ taken from APOGEE DR17 for the rotationally enhanced giants. The RC stars are plotted as blue circles and first-ascent giants are plotted as red ones. In the bottom panel we show mass as a function of [Fe/H] for the sample with full asteroseismic data.}
\end{figure}

It is well-established that rapid rotation can suppress seismic signals \citep{Gaulme_2020}, which we see in our population of rotationally enhanced giants. Figure \ref{fig:vdist} shows the distribution of $v \sin i$ for giants with and without even a partial seismic detection. Of the 
631 rotationally enhanced giants, 
316 have complete seismic detections (derived mass and radii) and 
184 have partial detections. Unsurprisingly, the vast majority of the rotators with seismic detections rotate more slowly, peaking in the 5-10 km s$^{-1}$ range and extending up to 20 km s$^{-1}$, except for a few outliers. Of the 91 stars with $v \sin i >$ 10 km s$^{-1}$ and good spectral fits, only 15 have derived masses, including no RGB stars and no stars with $v \sin i >$ 15 km s$^{-1}$.

Rapid rotation is not expected in red giants because of spin down caused by mass loss and expansion. Rotationally enhanced giants either did not spin down as much during the main sequence or were spun up. Spin-up can occur in binaries either through merging with a companion or by tidal synchronization. \citet{Patton2024} found a high binary fraction for rotationally enhanced red giants in the field, but were limited in their assessment of binarity in the APOKASC-3 sample due to many targets having only one visit from APOGEE. 

Nevertheless, many of these stars likely have experienced binary interaction. This unique population spans the giant branch and clump, as shown in Figure \ref{fig:vdist}. The majority of the sample is in the RC, consistent with their being upper RGB interaction products. Interestingly, most of these RC stars are low mass, as seen in the lower panel. As another intriguing data point, the majority of the upper RGB detections do not show evidence of a current binary companion, suggesting that they could be in post-merger systems. The rotationally enhanced giants with seismology provide a unique opportunity to probe the internal structure of stars, in various evolutionary states, with unusual histories.

\section{The APOKASC-3 Catalog} \label{sec:catalog}

We now turn to presenting the full set of catalog data. For ease of use, we separate our results into three full tables. Table \ref{table:main} contains our recommended values for key parameters. Our algorithm for doing so is as follows:
\begin{itemize}
    \item \textbf{Identifiers}. We include the KIC ID, \gaia\ DR3 ID, and 2MASS ID for each target. The latter is used as a unique identifier for the APOGEE survey.
     \item \textbf{Evolutionary State}. We adopt asteroseismic states when available, and spectroscopic states when they are not. Stars without valid spectra do not have assigned states. For a small minority of stars, we had only DR17 data and not DR16 (the latter was used to assign states); in this case some stars were assigned RC/RGB states if they were in the \logg domain where both populations are present. Stars classified as dwarfs or subgiants are not included in the table. We also specify whether the state was derived from spectra or asteroseismology.
    \item \textbf{Category}. \textit{We only present masses, radii, and ages for stars in the Silver and Gold samples.} To be included in either sample, \nmax\ must be between 1 and 220 $\rm{\mu Hz}$, the source must not be flagged as a background source or one with poor asteroseismic measurements, and it must have a valid spectroscopic solution. Silver sample stars have between 2 and 4 valid measurements of \nmax\ and \dnu. Gold sample stars have 5 or more measurements of both. A minority of both classes were classified as outliers, with anomalies in the measurements from automated methods; we do not present masses and radii for these targets, and they are labeled SilverOL and GoldOL respectively. Stars with a \nmax\ inconsistent with the spectroscopic value are categorized as Background. Stars with \nmax\ below 1 $\rm{\mu Hz}$ and 2 or more valid asteroseismic measurements are classified as LowNMax. Stars with \nmax\ above 220 $\rm{\mu Hz}$ and 2 or more valid measurements are classified as HighNMax (Liagre et al. in prep). Stars with less than 2 valid \dnu\ measurements, but at least 2 \nmax\ measurements, were classified as detections (Detect); only \nmax\ entries are included. Entries with poor data are classified as DetectOL. Non-Detections are split into 2 groups: NoDetSh are ones with less than 3 quarters of data, while NoDetOL are ones with other recognized issues. Outlier categories are described in Table \ref{table:outliers}.
    \item \textbf{Spectroscopic Data}. We adopted DR17 spectroscopic parameters when available, and DR16 if not; a table column indicates which data release is the source.  A small number of stars lacked valid spectra and were labeled NoSpec. The central values for spectroscopic \teffns, \loggns, \fehns, and \afe were taken directly from APOGEE. Their errors were inferred as described in the text. Values and errors for [C/Fe] and [N/Fe] were taken directly from APOGEE. Projected rotation velocities ($v \sin i$) were taken from \citet{Patton2024}, and are listed only if they are detections above 5 km/s. Stars with high \afe at a fixed \feh are a distinct stellar population from sun-like stars with lower \afens. We classify stars as $\alpha$-rich or $\alpha$-poor as described in Section \ref{sec:fullsample}.
    \item \textbf{Asteroseismic Parameters}. We used the weighted mean averages for \dnu\ and \nmax\ for the Gold and 
    Silver samples. Simple averages were used for detections. We adopted \fdnu\ (Section \ref{sec:dnu}) and \fnmax\ (Section \ref{sec:nmax}) values from the Garstec models with Mosser frequency spacing weights discussed in Section \ref{sec:dnu}.
    \item \textbf{Global Stellar Properties}. Asteroseismic mass, radius, and \logg  \footnote{Our asteroseismic gravities were derived without including the \fnmax term, as discussed in \ref{sec:nmax}.}
    were taken from the Mosser \fdnu\ scaling relation and the spectroscopic \teffns. Ages were derived using the GARSTEC models. 
       For lower RGB and RC stars we provide ages inferred from models for the appropriate state only. For luminous giants, we distinguish between AGB, RGB, or AGB/RGB using spectroscopic criteria. Both AGB and RGB ages are provided for stars with ambiguous states. \gaia\ radii were inferred as described in the text.
\end{itemize}

\begin{table}
    \centering
    \begin{tabular}{c c}
         Label & Contents  \\
          
 \hline
 KIC ID & Number in the \kep\ Input Catalog \\
EvolState, ESSource & Evolutionary State and Source\\
CatTab & Category \\
SeisSource & Seismic Weighting Scheme \\
SpecSource & Spectroscopic Data Release \\
NNumax & Number of Filtered \nmax\ values \\
NDnu & Number of Filtered \dnu\ values \\
NQuar & Quarters of \kep\ data \\
Numax, SNumax & \nmax\ ($\mu$Hz) and $\sigma$ \\
Dnu, SDnu & \dnu\ ($\mu$Hz) and $\sigma$ \\
FDnu, SFDnu & Mosser \fdnu\ and $\sigma$ \\
Fnumax & Mosser \fnmax\ \\
Mass, SMass & Mosser Mass (\msun) and $\sigma$ \\
Radius, SRadius & Mosser Radius (\rsun) and $\sigma$ \\
Loggseis, Sloggseis & Mosser Seismic \logg (cgs) and $\sigma$ \\
Teff, STeff & \teff and $\sigma$ (K)\\
Loggspec, Sloggspec & Spectroscopic \logg (cgs) and $\sigma$ \\
FeH, SFeH & \feh ([M/H]) and $\sigma$ \\
AlpFe, SAlpFe & \afe and $\sigma$ \\
CFe, SCFe & [C/Fe] and $\sigma$ \\
NFe, SNFe & [N/Fe] and $\sigma$ \\
InvRGaia, SInvRGaia & MIST K $\frac{1}{R_{\gaia}}$ and $\sigma$ \\
RecAge & Recommended Age Class \\
AgeRGB & Garstec Age (Gyr), RGB \\
SAgeRGB+,SageRGB- & $\pm $Garstec Age $\sigma$ (Gyr), RGB \\
AgeRC & Garstec Age (Gyr), RC or AGB \\
SAgeRC+, SAgeRC- & $\pm$ Garstec Age $\sigma$ (Gyr), RC or AGB \\Vsini & $v \sin i$ (km/s) \\
$\alpha$ Class & As defined in the Figure \ref{fig:afefeh} caption\\
\gaia\ ID & Number in the 
\gaia\ Catalog \\
 2MASS ID & Number in the 2MASS Input Catalog \\

    \hline
    
    \end{tabular}
    
    \caption{Catalog of Recommended Stellar Properties. See text for details.}
    \label{table:main}
\end{table}

We also present two additional tables. Table \ref{table:altvalue} includes alternative measurements of key properties. This includes DR16 spectroscopic data; \fdnu, \fnmax, mass, radius, and \logg values from the Sharma and White weighting schemes; ages from YREC; and \gaia\ radii derived using the \citet{gonzalez-hernandez_bonifacio2009} version of the IRFM. Alternative weightings for the mean asteroseismic parameters themselves, and the raw measurements used for them, are presented in the appendices.

\begin{table*}
    
    \centering
    \begin{tabular}{c c}
         Label & Contents  \\
          
 \hline
 KIC & Number in the \kep\ Input Catalog \\
 Nquar & Number of quarters of \kep\ data \\
ESSeis, ESSpec & Asteroseismic and Spectroscopic Evolutionary States\\
Teff16, STeff16 & DR16 \teff and $\sigma$ (K)\\
Logg16, Slogg16 & DR16 Spectroscopic \logg (cgs) and $\sigma$ \\
FeH16, SFeH16 & DR16 \feh ([m/h]) and $\sigma$ \\
AlpFe16, SAlpFe16 & DR16 \afe and $\sigma$ \\
CFe16, SCFe16 & DR16 [C/Fe] and $\sigma$ \\
NFe16, SNFe16 & DR16 [N/Fe] and $\sigma$ \\
Teff17, STeff17 & DR17 \teff and $\sigma$ (K)\\
Logg17, Slogg17 & DR17 Spectroscopic \logg (cgs) and $\sigma$ \\
FeH17, SFeH17 & DR17 \feh ([m/h]) and $\sigma$ \\
AlpFe17, SAlpFe17 & DR17 \afe and $\sigma$ \\
CFe17, SCFe17 & DR17 [C/Fe] and $\sigma$ \\
NFe17, SNFe17 & DR17 [N/Fe] and $\sigma$ \\
PI, SPI & Gaia parallax $\pi$ (mas) and $\sigma$ \\
KS, SKS & 2MASS K magnitude and $\sigma$ \\
InvRG16, SInvRG16 & DR16 MIST K-band $\frac{1}{R_{\gaia}}$ and $\sigma$ \\
InvRG17, SInvRG17 & DR17 MIST K-band  $\frac{1}{R_{\gaia}}$ and $\sigma$ \\
InvRGaia, SInvRGaia & GHB09 $\frac{1}{R_{\gaia}}$ and $\sigma$ \\
FDnuWh, SFDnuWh & \fdnu\ and $\sigma$, White weighting \\
FnumaxWh & \fnmax, White weighting\\
MassWh, SMassWh & Mass (\msun) and $\sigma$, White weighting \\
RWh, SRWh & Radius (\rsun) and $\sigma$, White weighting \\
LoggseisWh, SloggseisWh & Seismic \logg (cgs) and $\sigma$, White weighting \\
FDnuSh, SFDnuSh & \fdnu\ and $\sigma$, Sharma models \\
FnumaxSh & \fnmax, Sharma models\\
MassSh, SMassSh & Mass (\msun) and $\sigma$, Sharma models \\
RSh, SRSh & Radius (\rsun) and $\sigma$, Sharma models \\
LoggseisSh, SloggseisSh & Seismic \logg (cgs) and $\sigma$, Sharma models \\
RGBAgeM, SRageMP, SRageMM & Garstec RGB Age (Gyr) including mass loss and $\pm \sigma$ \\
RCAgeM, SaageMP, SaageMM & Garstec RC/AGB Age (Gyr) including mass loss and $\pm \sigma$ \\
RGBAgeNoM, SRageNoMP, SRageNoMM & Garstec RGB Age (Gyr) not including mass loss and $\pm \sigma$ \\
RCAgeNoM, SaageNoMP, SaageNoMM & Garstec RC/AGB Age (Gyr) not including mass loss and $\pm \sigma$ \\
YRECAge & YREC Model Age (Gyr) (see text) \\
Fagecor, Mtrial & Age correction factor between trial and final mass, trial mass (\msun)\\
Fagemdrgb,Fagemdrc & Age correction factor for mass loss ($\eta = 0.2$) for RGB and RC stars.\\

    \hline
    
    \end{tabular}
    
    \caption{Catalog of Alternate Stellar Properties. We include evolutionary states derived from asteroseismology as well as those derived from spectroscopy alone. Spectroscopic properties are from DR16, as compared with the DR17 values adopted for the large majority of the sample. We also present here asteroseismic data derived using the Sharma and White values, including \fdnu, \fnmax, mass, radius, and asteroseismic surface gravity. Alternate ages measurements and inferred \gaia\ radii are also included. For details see text.}
    \label{table:altvalue}
\end{table*}

We also make available on Zenodo \footnote{https://zenodo.org/records/13308665} an additional table that contains an extensive list of other key properties of interest, collated from other catalogs and sources.  This includes a variety of photometric measurements, literature data from other large catalogs, and notes about special stars. Because this table is so large, over 500 columns, and contains data from many sources, it is not as carefully vetted as the rest of the results that we present here. However, over the past decade, we have found it advantageous to have a pre-collated dataset including magnitudes, spectroscopic parameters from various surveys, seismic results from our previous work and the literature, annotations about binaries and other stars of interest, and individual spectroscopic state indicators over time. We therefore make this previously internal and exploratory table accessible to everyone with this publication.
We also refer to \citet{godoy2024} for a complementary characterization of the color-magnitude diagram and binary systems.



We now present key properties of our sample, starting with stellar physics and following up with stellar populations.

\subsection{Stellar Physics}
\label{sec:physics}

Stellar theory makes strong predictions about the properties of evolved red giant stars. Our asteroseismic sample permits stringent tests of stellar theory; in our view, it should be used as a standard calibration set for theoretical 
stellar interiors models. Here we present some examples of asteroseismology as a test of stellar physics. 

\subsubsection{Composition Trends} We begin by showing composition trends in a sample of stars with masses between 1.1 and 1.2 \msun\ (Figure \ref{fig:comptrend}), restricting our data set to stars classified as $\alpha$-poor.  In this and the following figures, Gold sample 
shell-burning stars are shown with filled symbols; Gold sample core He-burning stars with open ones, and the Silver sample with crosses. The latter group consists predominantly of more luminous shell-burning stars. The position of the RGB and RC is strongly composition dependent, with a nearly constant \teff offset between the two. The narrow width of the RC in \logg is also striking, and consistent with stellar theory. Double shell burning, or AGB, stars are seen above the RC and hotter than the main RGB, producing a broader upper RGB relative to the lower RGB. The red giant branch bump (RGBB) is clearly seen below and to the right of the RC; this phenomenon occurs in theoretical models because there is a composition discontinuity associated with the lower boundary of the surface convection zone. When the H-burning shell at the top of the growing core reaches this boundary, the star temporarily becomes fainter and then brighter, producing about a factor of three local increase in the density of stars. 

\begin{figure}

\includegraphics[width=8cm]{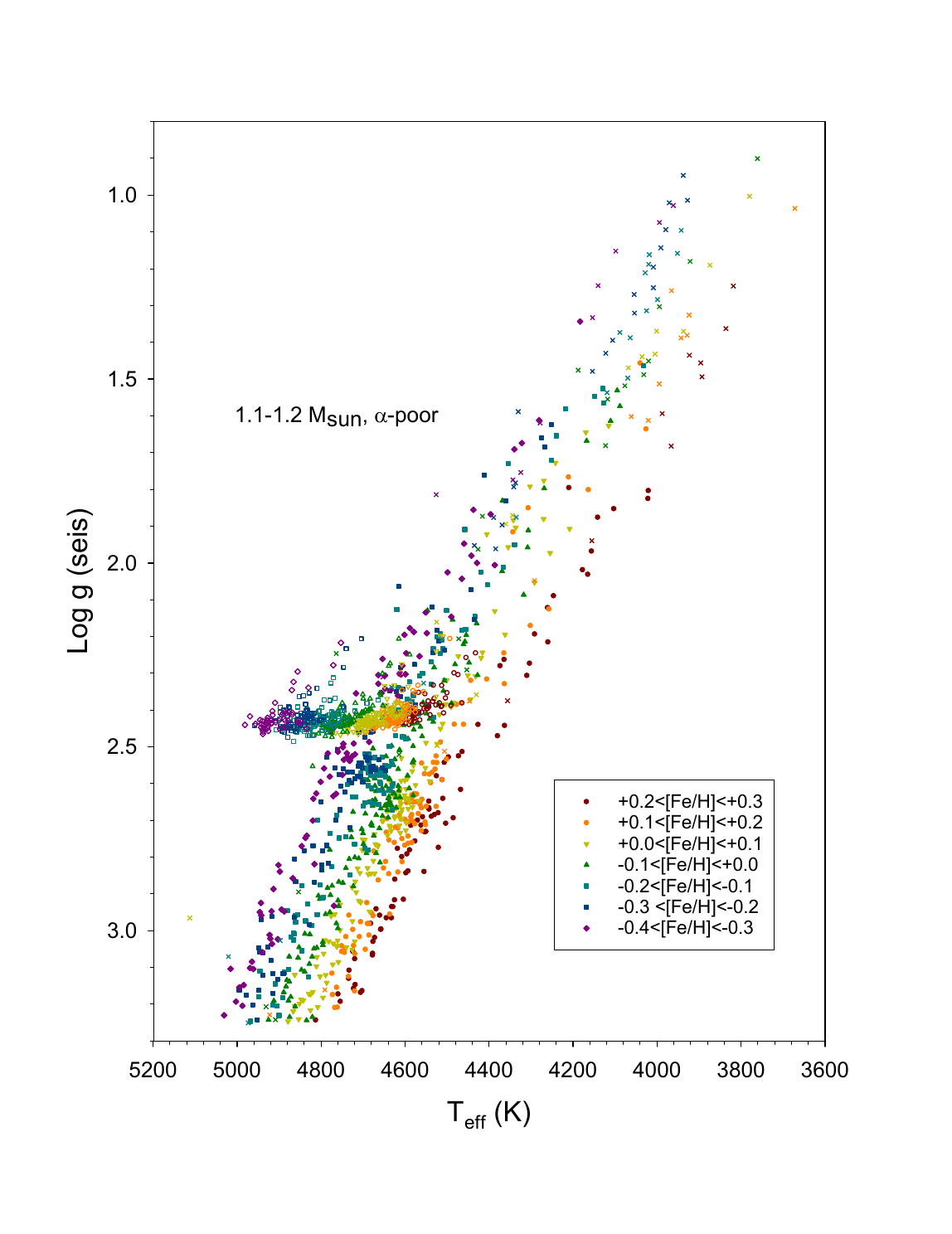}

\caption{Stellar properties at fixed mass and variable metallicity. APOKASC-3 stars classified as $\alpha$-poor, with masses between 1.1 and 1.2 \msun, shown in the asteroseismic \logg -- spectroscopic \teff plane. Filled symbols are Gold sample RGB, open symbols are Gold sample RC, and crosses are Silver sample stars. Symbol types and colors reflect metallicity; stars shown are in the range $-0.5 <$ [Fe/H] $<+0.3$.}

\label{fig:comptrend}

\end{figure}

\subsubsection{Mass Trends} 
\begin{figure}

\includegraphics[width=8cm]{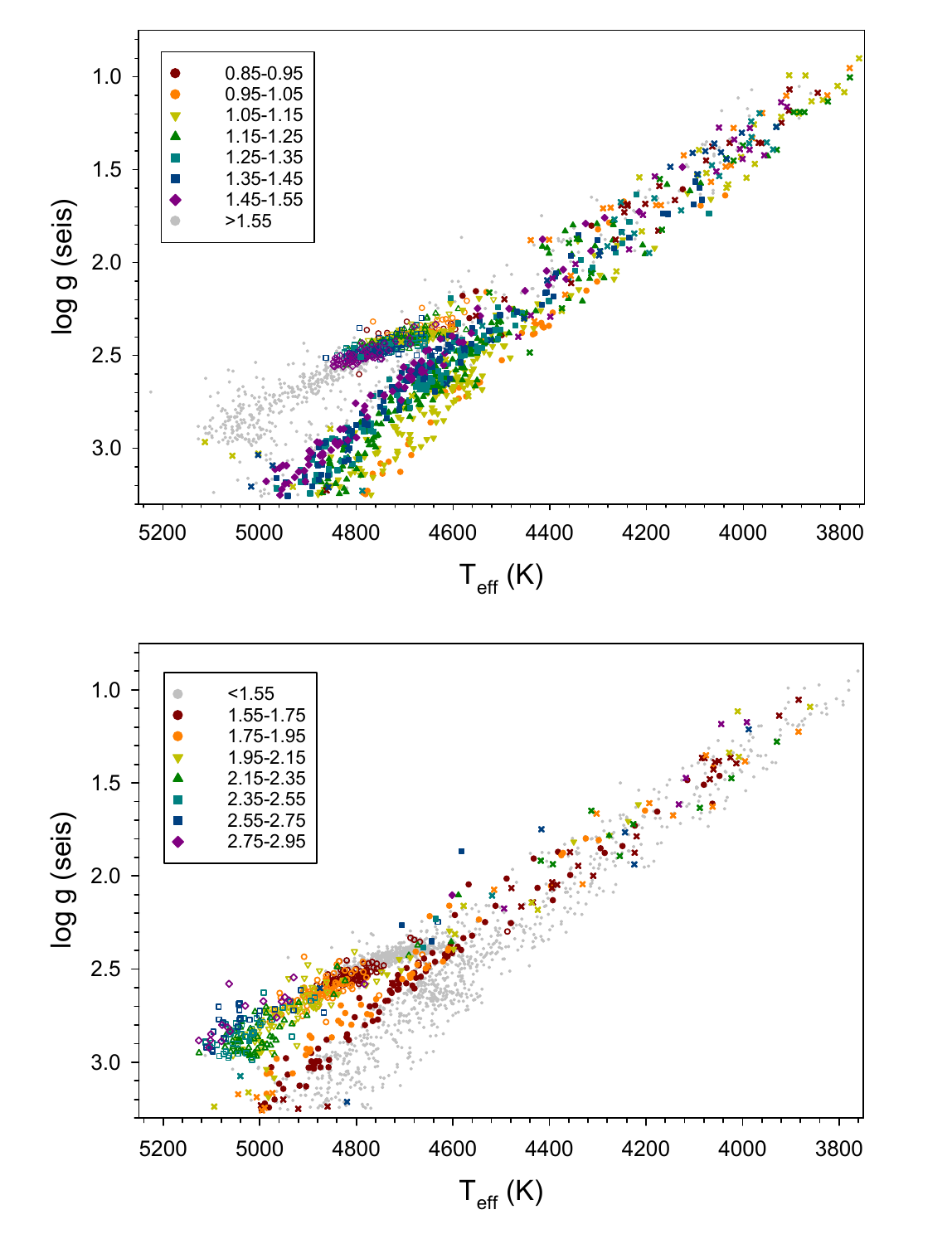}

\caption{Stellar properties at fixed metallicity and variable mass. APOKASC-3 stars with $-0.05<$ [Fe/H] $<+0.05$, shown in the asteroseismic \logg – spectroscopic \teff plane. Filled symbols are 
Gold sample RGB stars, open symbols are Gold sample RC stars, and crosses are Silver sample stars. Symbol types and colors reflect mass. The upper panel shows stars between 0.85-1.55 \msun, while the lower panel shows stars between 1.55-2.95 \msun. Higher mass stars (upper panel) and lower mass ones (lower panel) are shown in gray.}

\label{fig:masstrend}

\end{figure}

We then isolate stars with close to solar metallicity ($-0.05 <$ [Fe/H] $< +0.05$), and show mass trends in Figure \ref{fig:masstrend}. Higher mass corresponds to hotter \teff, on average, but the effects are much smaller than those due to composition; furthermore, there is a strong relationship between surface gravity and mass in the RC. The latter is expected from theory, because low-mass RC stars have similar radii.  The AGB / RGB contrast is more visible in this plane, where the strong dependence of the RGB locus on composition is absent; for example, the orange points in the upper panel are confined to the cool edge below the RC, but are observed at the hot and cool edges on the upper RGB. The mass dependence of the RGBB is also clearly seen here. At higher masses (lower panel) the RC population dominates, and on the lower RGB massive stars are only seen on the hot edge.

\subsubsection{The First Dredge-Up} 
\begin{figure}

\includegraphics[width=8cm]{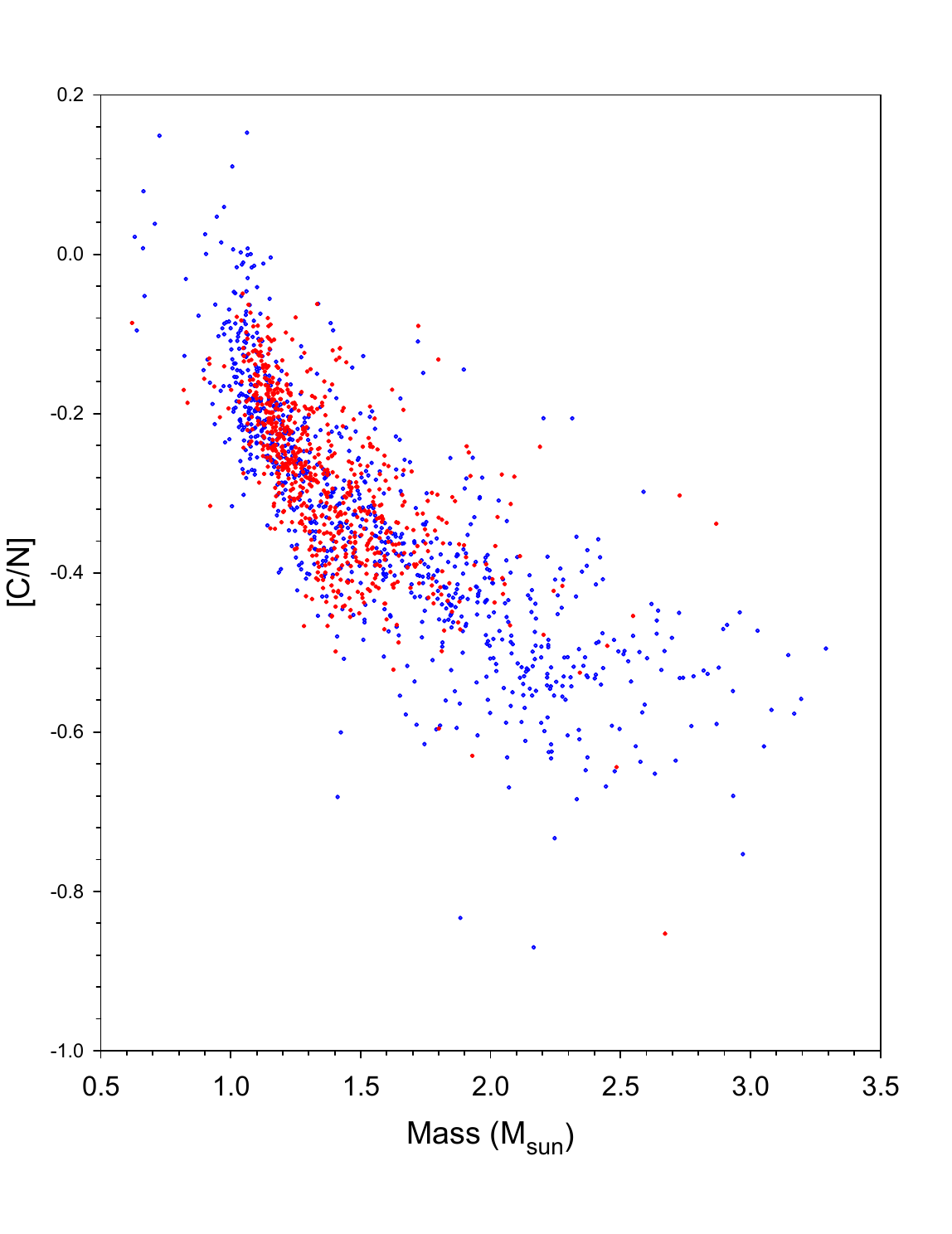}

\caption{The first dredge-up at solar metallicity. APOKASC-3 stars classified as $\alpha$-poor, with $-0.05<$ [Fe/H] $<+0.05$, are shown in the [C/N]-mass plane. Red symbols are Gold sample RGB stars, and blue symbols are Gold sample RC stars. Higher mass stars are preferentially seen in the RC due to lifetime effects.}

\label{fig:FDU}

\end{figure}

The development of a deep surface convection zone in giants also leads to a dredge-up of CN-processed material
(see \citealt{Roberts2024} for an empirical discussion in the APOKASC-3 context). The mass dependence of the first dredge-up at solar metallicity ($-0.05 <$ \feh $< +0.05$) is shown in Figure \ref{fig:FDU}. There is a strong mass dependence at low mass that flattens out at higher mass, with a significant scatter at fixed mass seen even in this sample with a narrow \feh range. The RC and RGB stars are similar, with some evidence of mass loss in the RC relative to the RGB in the lowest-mass stars.

\subsubsection{The Red Clump} 
\begin{figure}

\includegraphics[width=8cm]{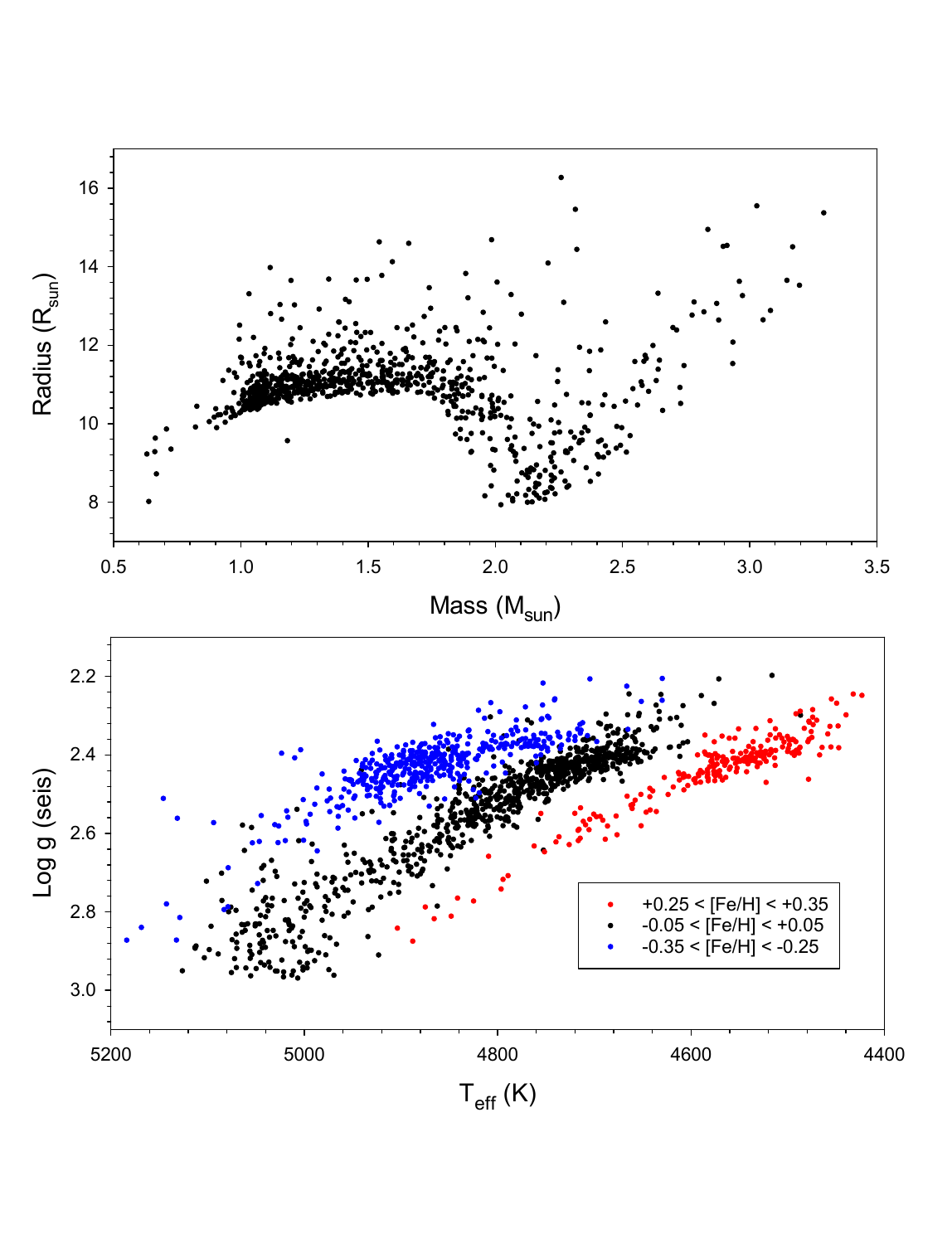}

\caption{In the upper panels we show the solar \feh sample in the mass-radius plane. The RC in the \logg-\teff plane for three metallicity domains ($-0.3$, blue; 0, black; and +0.3, red; with a range of $\pm 0.05$ dex for each) is shown in the bottom panel. In all cases, we choose stars from the $\alpha$-poor sample. }

\label{fig:RCpop}

\end{figure}

Reproducing the properties of the RC in detail is particularly challenging for stellar models. Core He-burning stars have convective cores, leading to significant uncertainties in the lifetime of the RC phase depending on the adopted model, especially the treatment of mixing at the outer boundary \citep{Bossini2015,Constantino2015}. G-mode period spacings (not discussed here) are difficult to reproduce with existing models, and they tend to favor longer lifetimes than the ones typically predicted by isochrones \citep{Montalban2013}. Furthermore, the starting conditions (core mass as a function of initial mass and composition) are contingent on the prior evolution. Our sample provides an extraordinarily precise characterization of the RC. In Figure \ref{fig:RCpop} we show some key properties of the RC. The bottom panel shows the location (in \logg and \teffns) of three cohorts of $\alpha$-poor stars (metallicity close to $-0.3$, 0 and +0.3, respectively). Metallicity induces significant \teff offsets, while there are strong mass-dependent \logg trends.  The main (upper) portion of the RC shifts to lower \logg for the most metal-rich stars. 
The lower boundary of the RC in mass-radius space is a distinctive feature clearly seen in earlier catalogs \citep{APOKASC2, Yu18}.
However, in the APOKASC-3 sample (upper panel), we see a strikingly sharp lower boundary for the RC phase, shown here for the solar metallicity cohort. The drop in $R$ close to 2 \msun\ can be traced to the transition between degenerate and non-degenerate He ignition; the rise in $R$ at higher masses can be traced to larger core masses in stars with higher birth mass.

\vskip 1cm
\subsection{Stellar Populations}
\label{sec:populations}

Galactic Archeology, or the study of the formation history of the Milky Way Galaxy, is a flourishing topic. Massive spectroscopic surveys, such as APOGEE and GALAH, have yielded detailed information on stellar abundances. Detailed kinematic, photometric, and distance information from the \gaia\ mission allow us to map out these trends across the Galaxy. Asteroseismology adds mass and age data for luminous evolved stars. The combination of all three is extremely powerful, and interesting insights have emerged from the \kep\ fields \citep{Silvaaguirre2018, Miglio2021,Montalban2021, Huber2024}.
Here we highlight how our precise and accurate data allows us to see population features with high fidelity.

Before proceeding, it is important to acknowledge that there are significant selection effects in the \kep\ data \citep{APOKASC1} that must be accounted for in detailed population studies. We therefore focus on clear global features, including sharp population boundaries and large trends, that are insensitive to details of the selection function.

\begin{figure}

\includegraphics[width=8cm]{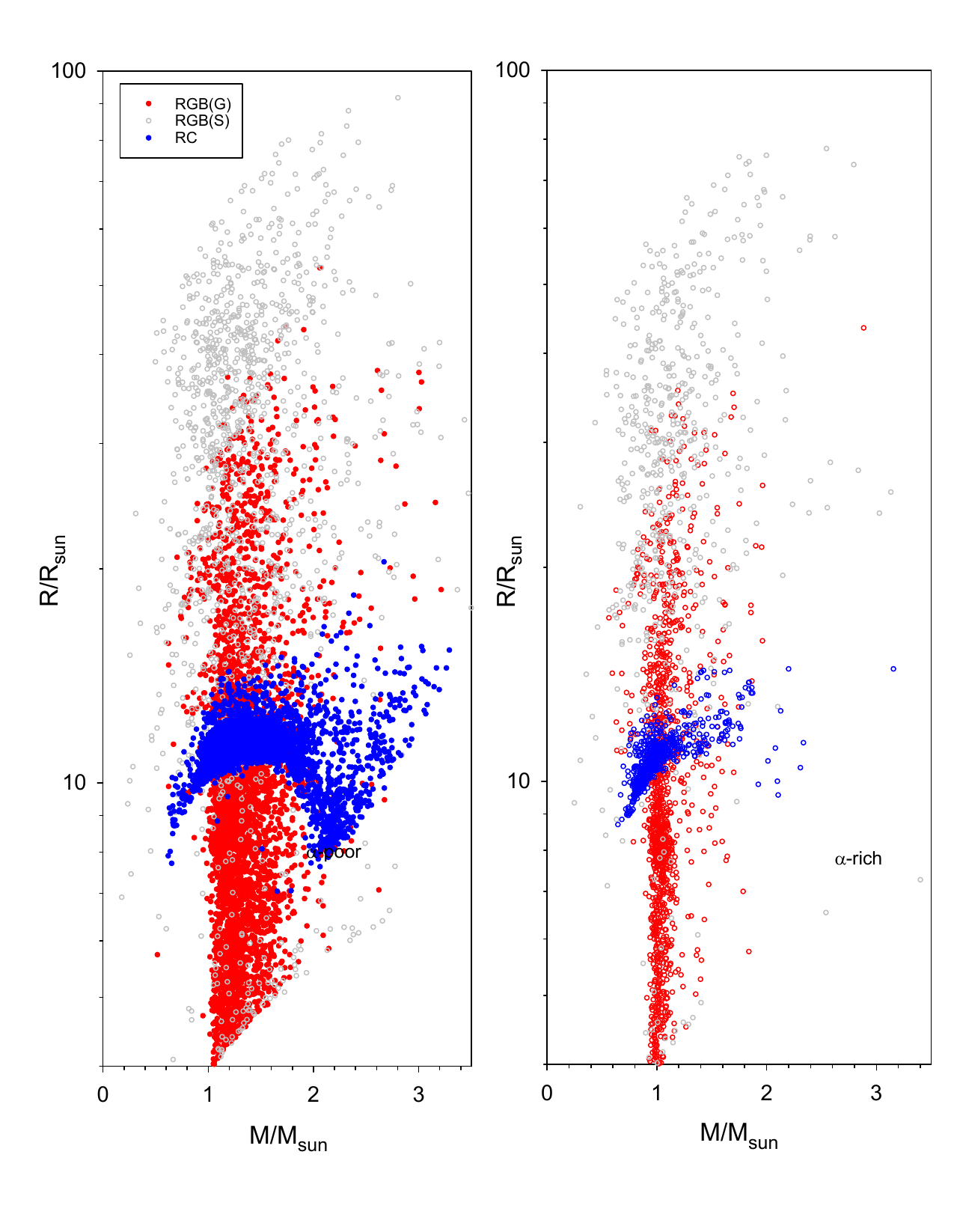}

\caption{The RGB (red, filled) and RC (blue, open) in the mass-radius plane for $\alpha$-poor stars (left) and 
$\alpha$-rich stars (right). 
Silver sample stars are shown as gray circles. The field turnoff in the RGB below the clump, mass loss in the RC, and a mixture of AGB and RGB stars above the clump are clearly seen.}

\label{fig:MRall}

\end{figure}

Figure \ref{fig:MRall} shows mass versus radius for the $\alpha$-rich (right panel) and $\alpha$-poor (left panel) 
populations. Gold RGB are in red, the RC is in blue, and Silver sample stars are in gray.  Below the level of the RC, we can see a sharp lower boundary to mass in lower RGB stars. Lower mass stars are see in and above the RC, which is clear evidence for mass loss. Above the RC we see both first ascent RGB stars and second aspect AGB stars; the latter have experienced significant mass loss, which explains the presence of luminous stars below 1 \msun. The open gray symbols are Silver sample stars, which extend the sample to lower \loggns, albeit with larger mass uncertainties.

The lower RGB, below 10 \rsun, therefore allows us to see the RGB population prior to the onset of significant mass loss. In the figures that follow, we therefore compare RC and lower RGB stars in the $\alpha$ poor and rich populations.   

\begin{figure}

\includegraphics[width=8cm]{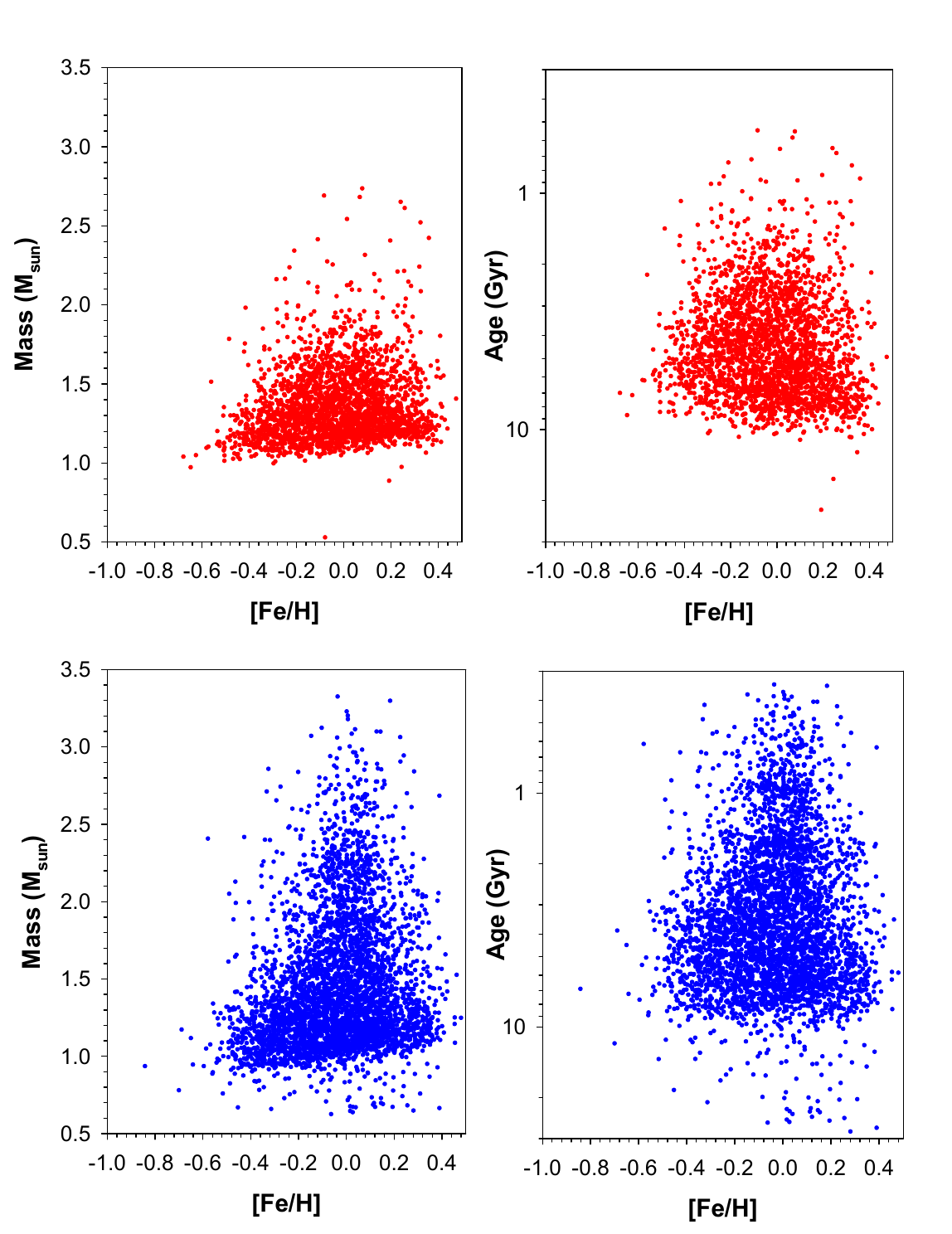}

\caption{The $\alpha$-poor RC (top) and lower RGB (bottom) in the mass-\feh plane (left panels) and the age-\feh plane (right panels). The RC spans a wider range of masses and ages, while both populations have a sharp lower age boundary corresponding to the finite age of the $\alpha$-poor population. More massive and younger stars are concentrated around solar metallicity.}

\label{fig:apoormassage}

\end{figure}

From Figure \ref{fig:MRall}, the $\alpha$-poor population
has a wide range of masses, and by extension ages. Figure \ref{fig:apoormassage}  shows the mass and age distributions of the $\alpha$-poor population. A number of features deserve comment. Higher mass RGB stars are much less common than higher mass RC stars because higher mass stars have much longer RC lifetimes relative to their RGB lifetimes; the RC therefore probes a younger population, on average, than the RGB (see \ref{fig:masstrend}). There is a well-defined upper edge in age for both the RC and RGB. The close correspondence in age for the two groups is a confirmation of our mass-loss model for the RC, as we would have obtained a significant age offset between the two populations if our assumptions about mass loss in the RC were very wrong. This is easier to quantify with the $\alpha$-rich population, which we do below. There is a small but real sub-population of very low mass stars in the RC that are not seen in the RGB. Although we report high formal ages for these stars, their low masses are the products of severe mass loss, likely from interactions with a companion \citep{Li2022}. They appear in the RC because the majority of stellar interactions on the RGB occur on the upper RGB.

There are observational selection effects disfavoring the detection of metal-poor and lower mass stars in the RC that appear as blue horizontal branch stars; such objects are too hot to excite solar-like oscillations, explaining their absence in the RC population.  See \citet{molnar2024} for a recent discussion. There is no comparable selection against very low mass or low metallicity stars on the lower RGB, so their absence is a true population feature for the \kep\ fields. Such stars are seen in APO-K2 \citep{schonhut-stasik+2023}, illustrating the important of sampling a range of Galactic stellar populations. 

Finally, a wide range of masses at fixed \feh are seen in the $\alpha$ poor RC, but high mass (and, typically, young) stars are much more likely to be seen at solar metallicity than for lower or higher metal content. This is not merely a function of there being more solar metallicity stars; there is a much larger fraction of high mass stars at solar metallicity than at lower or higher values. Our finding that the youngest stars have solar metallicity agrees with other studies in the solar circle \citep{Casagrande2011,Feuillet2018}, but with higher age precision.

This can be explained if stars born in the solar neighborhood are close to solar in metallicity, while more metal-rich and metal-poor star stars currently being born in the Galaxy are found closer and further from the Galactic center respectively. Stars migrate radially in the Galactic disk, blurring the age-metallicity distribution locally. However, there is a time lag for this process, resulting in a preference for young near-solar metallicity stars \citep{Feuillet2018, Lu2022}.







\begin{figure}

\includegraphics[width=8cm]{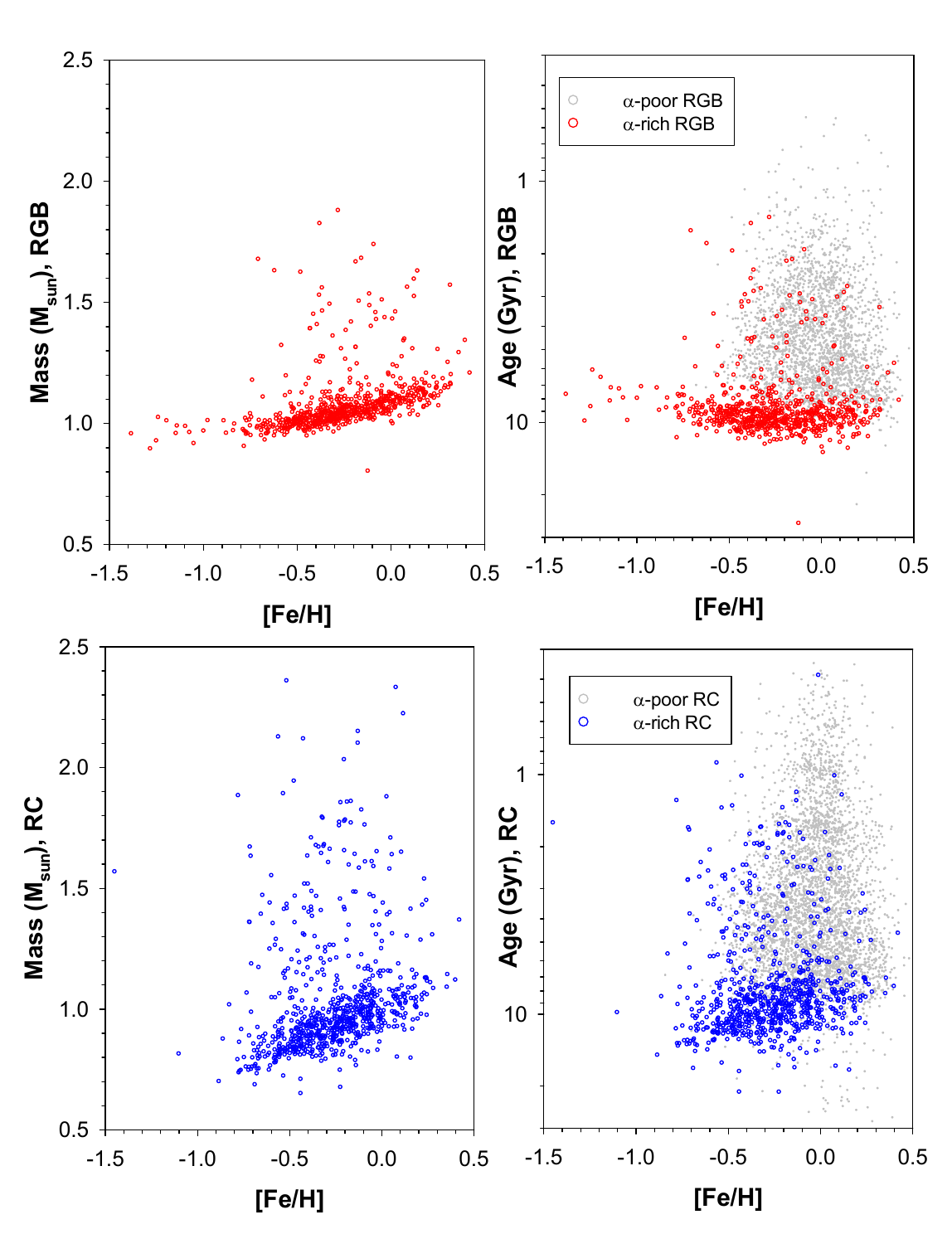}

\caption{The $\alpha$-rich populations in the mass-\feh (left) and age-\feh (right) planes for the RGB (top) and RC (bottom). The $\alpha$-poor populations in age space are shown for comparison as faint gray points in the right panels; a clear division, corresponding to the difference between the thick and thin disks, is seen. The RGB masses are also shown in the lower left panel, and the difference between them and the RC is clear evidence for mass loss.}

\label{fig:arichmassage}

\end{figure}

The $\alpha$-rich population has a distinct mass and age distribution, and it is associated with the thick disk. There is a well defined peak in the mass and age distributions. In prior studies \citep{Miglio2021} this was found to correspond to a characteristic age of 11 Gyr. Figure \ref{fig:arichmassage} shows the masses (left) and ages (right) of lower RGB (top) and RC (bottom) stars. There is a clear distinction between the ages of the $\alpha$ poor (top right, gray symbols) and the $\alpha$-rich stars. There is a sub-population of massive, and potentially young, 
$\alpha$-rich stars. As this can impact mean statistics, for the exercises that follow we use median statistics and the median absolute deviation converted to an effective sigma. Using this approach, the median age of the RGB and RC stars are $9.14 \pm 0.05$ Gyr and $9.28 \pm 0.08$ Gyr, respectively. The close correspondence between the two is a validation of our mass loss model for the RC, where we required that the median birth mass in the RC be equal to the median lower RGB mass in the $\alpha$ rich sample. After rejecting $ 5 \sigma$ age outliers, the median RGB and RC masses are, respectively, 1.03 and 0.93 \msun, for an implied median mass loss of 0.1 \msun. Reconciling the ages of the two populations would require a slight reduction, of order 0.004 \msun, in the model mass loss rates or a corresponding change in the relative RC and RGB masses, well within our uncertainties in the relative masses and radii of the two populations.


\begin{figure}

\includegraphics[width=8cm]{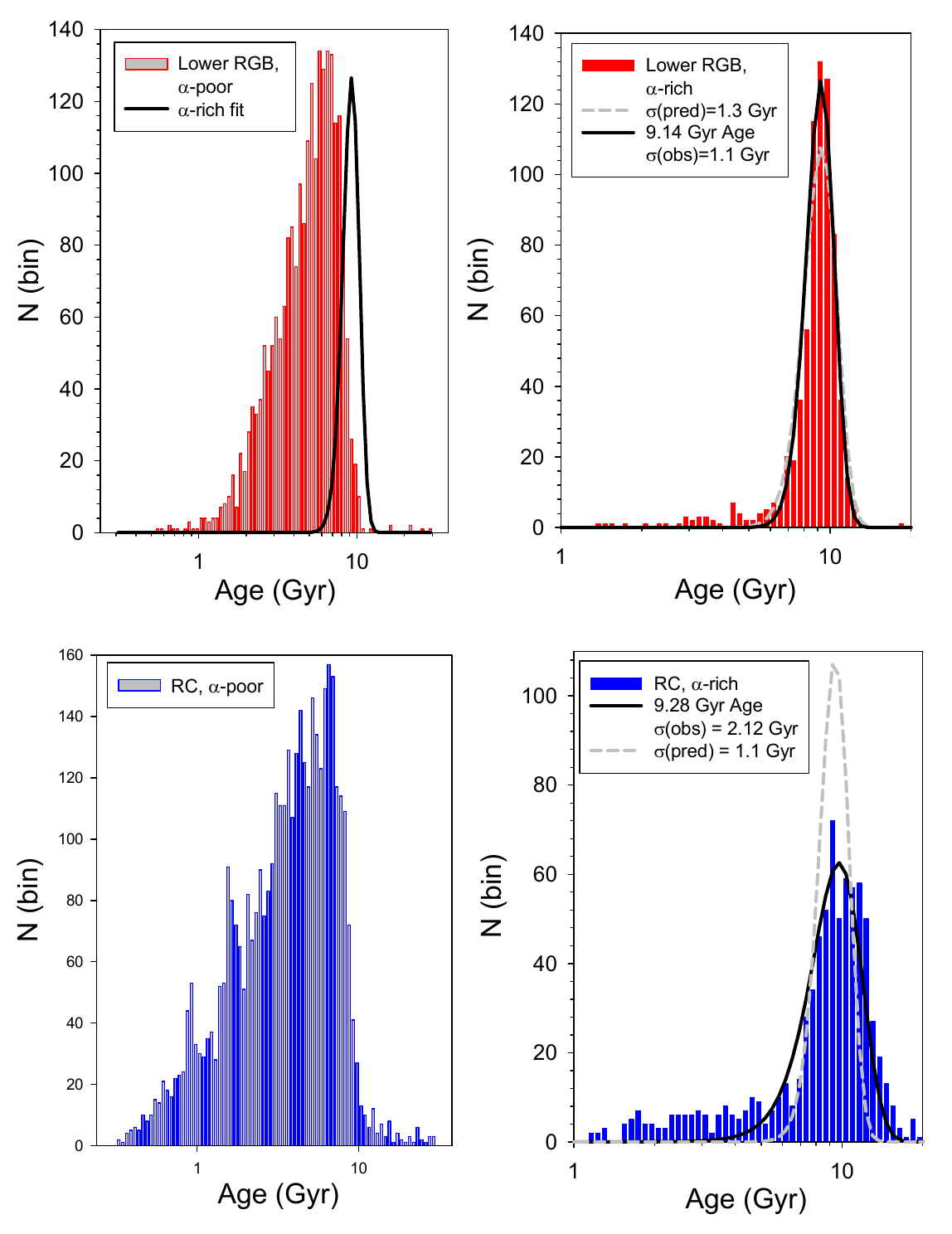}

\caption{Age distributions for different evolutionary states and chemical populations. The $\alpha$-rich populations (right) show a strong age peak, while the $\alpha$-poor ones are broader. The lower RGB (top) and RC (bottom) have similar high age cutoffs, but young stars are much more common in the RC. For the $\alpha$-rich populations the peak age and observed dispersion is overlaid for comparison (solid line); the predicted distribution is shown with a short dashed line. For the $\alpha$-rich populations the lines indicate the location of an edge in the age distribution (at 7.3 Gyr for the RGB and 6 Gyr for the RC) with the observed dispersions as derived in the right panels.}

\label{fig:agedist}

\end{figure}

After performing a $5 \sigma$ outlier rejection, we can use the width of the main age peak as a diagnostic of the uncertainties (Figure \ref{fig:agedist}).
The age dispersion in the bulk population of the lower RGB, 1.1 Gyr, is slightly below the 1.3 Gyr dispersion predicted by our error model – indicating that the random mass uncertainties are below 4\% for this sample, and that the underlying age dispersion is less than 1.1 Gyr. The RC age distribution is clearly broader, with an age $\sigma = 2.12$ Gyr. This is consistent with a dispersion in mass loss at fixed birth mass of 0.03 \msun, comparable to results seen in modeling of the horizontal branch in globular cluster stars. There is some tendency for stars below \feh $= -1$ to drift to younger ages, reflecting known problems for halo star ages from asteroseismology \citep{Epstein2014, schonhut-stasik+2023}.

\begin{figure}
    \centering
    \includegraphics[width=8cm]{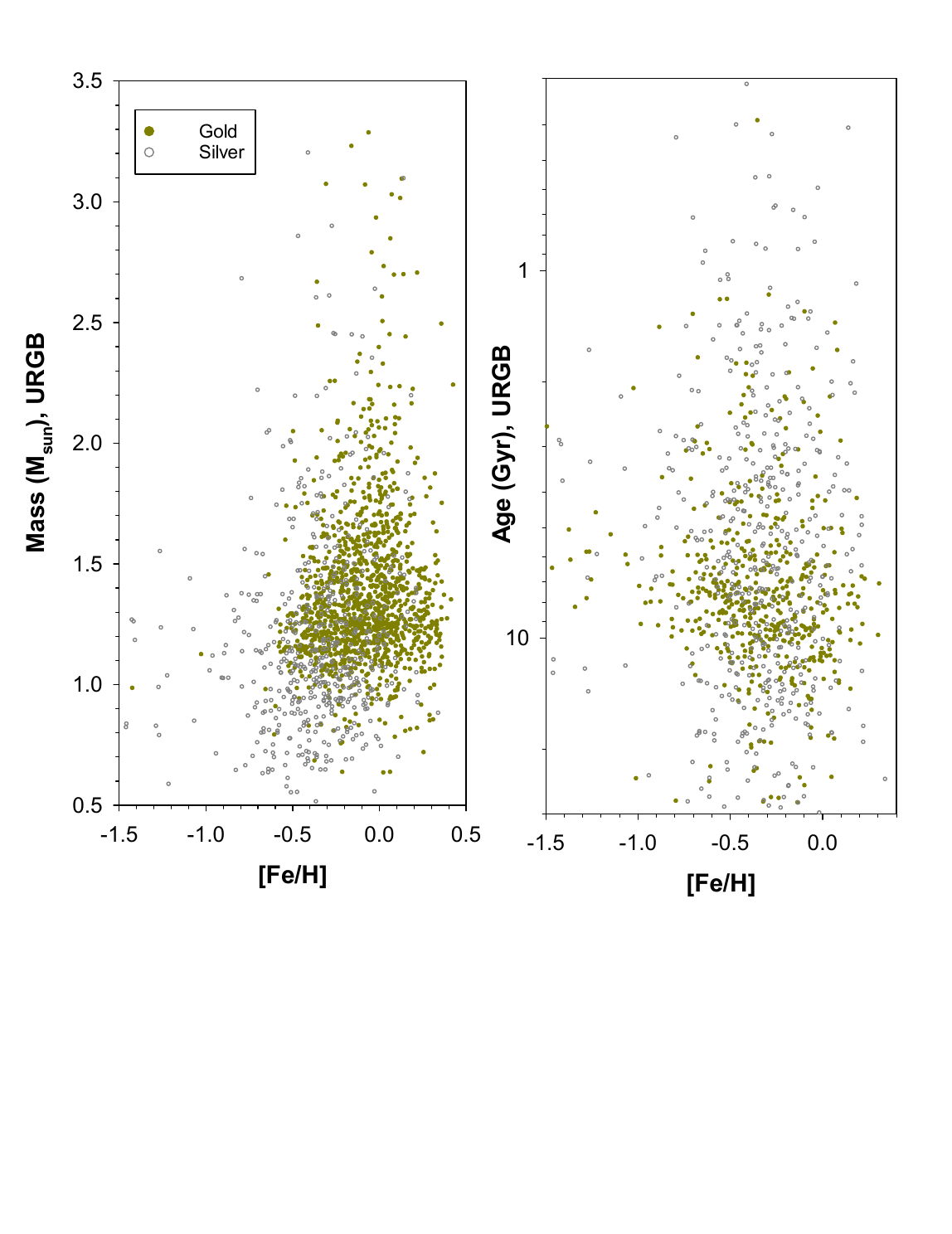}
    \caption{Upper RGB stars in the $\alpha$-rich population in the mass-\feh plane (left) and the age-\feh plane (right). Silver sample stars are dark gray symbols, and Gold sample ones are red symbols. A small number of points below 0.5 and above 3.5 \msun\ are not shown. The Silver sample includes a larger number of stars at lower metallicity, with higher scatter.}
    \label{fig:arichupper}
\end{figure}

The upper red giant branch is complicated by the observed mixture of RGB and AGB stars. The overall data quality is also lower, with a much larger fraction of Silver sample vs. Gold sample stars. However, the age distribution of these stars is an interesting astrophysical and modeling test (Figure \ref{fig:arichupper}). There is still a clear concentration of the Gold sample stars at an older age of $7.88 \pm 0.12$ Gyr; Silver sample stars have large age uncertainties and a lower mean age of $6.85 \pm 0.08$ Gyr. The primary difference between the two is that the Silver sample stars have a lower mean \loggns. If interpreted as a systematic age error, the Gold and Silver samples would be offset by 17\% and 
35\%, respectively. 

\section{Discussion and Conclusions} \label{sec:discuss}

Our goal in this paper was twofold: to develop a complete asteroseismic catalog of \kep\ giants with spectroscopy, and to critically evaluate the strengths and limitations of asteroseismic scaling relations. Out of the 15,808 stars that we studied, we report 12,448 asteroseismic masses, radii, and ages. This yield may seem surprisingly small, given that solar-like oscillations are nearly universal in cool giants. A tour through our filters, however, provides a straightforward explanation. A total of 1,356 stars are real detections, but in domains where we cannot provide valid solutions: background sources (129), stars without good spectroscopic solutions (174), oscillation frequencies that are too close to the Nyquist sampling rate (567), or stars too low for scaling relations to be valid (486). This still leaves 
2,004 stars as marginal detections or non-detections. The failure modes here are varied, but fall into a few general families. Some targets had predicted oscillation frequencies close to the Nyquist sampling rate, and may not have detectable signals with 30-minute sampling. Many targets had short time series. The \kep\ mission did not consistently observe evolved giants, particularly luminous ones, and long time series data is required to detect low oscillation frequencies. Other light curves had artifacts or highly variable background sources that confounded automated algorithms. Finally, there was an interesting minority of targets that had true astrophysical backgrounds, such as stellar activity, eclipses, or double oscillation patterns. These stars are interesting, and we have summarized their key properties in the relevant tables.


Our Gold sample of 10,036 stars is the most precise and accurate asteroseismic data set to date, making it an excellent training set for inferring ages in other surveys. The Silver sample of 2,382 stars extends the data to lower surface gravity. Our data is also a powerful test for stellar interiors models, and should be used as a reference in isochrones. The core He-burning locus, the location of the RGB and that of the RGBB as a function of mass and composition are fundamental predictions of 
stellar interiors models. Population inferences are more complex to interpret, but the sharp cutoffs in the mass distributions for thin and thick disk stars are robust, and point to stringent bounds on their formation ages. The narrow age range in the thick disk population is also a robust feature. The strong metallicity dependence of the mass distribution, similarly, will set interesting constraints on radial-mixing models.

Our approach – using multiple analysis techniques – was explicitly designed to stress test the scaling relations. Different methods had excellent internal agreement for low luminosity giants. Above $\sim 20$ \rsun, larger \textit{measurement} systematics started to arise. We saw similar effects in the mapping of \dnu\ to $\langle \rho \rangle$ from theoretical models.  The RC was more complex; in that case, there appear to be method-dependent systematics relative to the RGB at the $1 \%$ level across the board. We then did an external comparison with \gaia\ radii. The agreement at low radii was again excellent (at the $ 1 -2 \%$ fractional level in $R$). For the largest radii we saw significant trends and large (up to $50 \%$) fractional offsets.   

Putting this together, scaling relationships are remarkably precise and accurate on the lower RGB and the RC. There are some potential systematic error sources between the two phases that are important to account for, in particular for inferring mass loss during the RGB phase. Systematics between the RGB and RC could enter in at the 1, 3 and 10 \% levels in radius, mass, and age, respectively. Scaling relations remain useful for more luminous stars but require careful calibration and attentions to measurement systematics. For the most luminous stars, with \nmax\ below 1 $\rm{\mu Hz}$, we do not recommend the usage of scaling relations to infer masses and radii. This is partially driven by empirical data showing large offsets, and partially driven by theoretical work mapping out the break down in the underlying assumptions. Asteroseismology remains an interesting tool for luminous giants, but we believe that it calls for different analysis tools, such as a move to modeling of individual frequencies, and a more rigorous assessment of measurement {techniques \citep[e.g.][]{Joyce2024}.}

In terms of best practice, we have three key recommendations. First, we recommend using more than one detection method to validate results from automated pipelines. This is valuable for outlier rejection and for a good understanding of recovery rates. Secondly, theoretical stellar models are essential for interpreting the observed frequency spacings, impacting stellar parameter estimates significantly. Such models can account for non-uniform mode spacing, allowing scaling relation studies to incorporate some individual frequency properties in a compact fashion. Finally, we recommend calibration of the results against fundamental data. Even in the well-studied \kep\ fields, with excellent data, we demonstrated that significant systematics can be injected into global stellar parameters that make outcomes dependent on evolutionary state and luminosity.

Our masses and radii are tied to an absolute system, and we believe that they are precise and accurate within the quoted random and systematic uncertainties. We have provided individual measurements in case users prefer to adopt 
single pipeline measurements; in such cases we encourage users to check the outcomes against the fundamental calibration set, which we provide.

The global APOGEE \teff scale is tied to the IRFM, and for solar-type stars, there are stringent limits on deviations \citep{Zinn2019Rtest}. As discussed in the APO-K2 catalog \citep{schonhut-stasik+2023}, however, there may be \teff offsets in the metal-poor domain that could translate into mass systematics. There is some evidence for such trends in our metal-poor data. We encourage a revised look at the IRFM temperature scale in the metal-poor domain. 

Age estimates are more complex than mass or radii estimates, with clear model-dependent offsets, particularly for younger stars with convective cores on the main sequence. The RC ages also depend on the assumed mass loss in prior phases, and ages also assume a specific mapping between helium abundance and metal content. However, we see encouraging signs in our data. The maximum ages of the $\alpha$-rich and $\alpha$-poor populations are consistent between the RC and RGB, validating our mass loss model. On the lower RGB, we also see no evidence of mass or age trends with \logg. Conversely, there is some evidence for differences in mean age for $\alpha$-rich luminous stars relative to the lower RGB and RC, indicating that our radius calibration did not completely remove age systematics for such stars.

In addition to \kep, K2 is a valuable asteroseismic resource \citep{schonhut-stasik+2023}. Although the data precision is lower than that of \kep, the K2 mission sampled a more diverse set of stellar populations, particularly for more metal-poor stars. \textsl{TESS} and the upcoming \textsl{Roman} mission represent exciting opportunities for asteroseismology of stellar populations. \textsl{TESS} will provide a large sample of bright, primarily local, stars \citep{Hon21}, concentrated in the domain where scaling relations work well (the lower RGB and RC). The \textsl{Roman} Galactic Bulge Time Domain Survey will be a deep census of stars in the Galactic bulge, sampling very different stellar populations than \kep\ \citep{Huber23}.

\section{Acknowledgements}

This paper includes data collected by the \kep\ mission and obtained from the MAST data archive at the Space Telescope Science Institute (STScI). Funding for the \kep\ mission is provided by the NASA Science Mission Directorate. STScI is operated by the Association of Universities for Research in Astronomy, Inc., under NASA contract NAS 5–26555.
MHP acknowledges support from NASA grants 80NSSC24K0637 and 80NSSC18K1582.
SB acknowledges NSF grant AST-2205026. DGR acknowledges support from the Juan de la Cierva program under contract JDC2022-049054-I. PGB acknowledges support by the Spanish Ministry of Science and Innovation with the \textit{Ram{\'o}n\,y\,Cajal} fellowship number RYC-2021-033137-I and the number MRR4032204.
D.S. is supported by the Australian Research Council (DP190100666). 
MV acknowledges support from NASA grant 80NSSC18K1582 and funding from the European Research Council (ERC) under the European Union’s Horizon 2020 research and innovation programme (Grant agreement No. 101019653). 
TCB acknowledges partial support for this work from grant PHY 14-30152; Physics Frontier Center/JINA Center for the Evolution of the Elements (JINA-CEE), and OISE-1927130: The International Research Network for Nuclear Astrophysics (IReNA), awarded by the US National Science Foundation. 
The research leading to the presented results has received funding from the ERC Consolidator Grant DipolarSound (grant agreement \#101000296).
S.M.\ acknowledges support by the Spanish Ministry of Science and Innovation with the Ramon y Cajal fellowship number RYC-2015-17697, the grant no. PID2019-107061GB-C66, and through AEI under the Severo Ochoa Centres of Excellence Programme 2020--2023 (CEX2019-000920-S). S.M. and D.G.R. acknowledge support from the Spanish Ministry of Science and Innovation (MICINN) with the grant No. PID2019-107187GB-I00. 
D.G.R. acknowledges support from the Spanish Ministry of Science and Innovation (MICINN) with the Juan de la Cierva program under contract JDC2022-049054-I.

Funding for the Sloan Digital Sky Survey IV has been provided by the Alfred P. Sloan Foundation, the U.S. Department of Energy Office of Science, and the Participating Institutions. SDSS acknowledges support and resources from the Center for High-Performance Computing at the University of Utah. The SDSS web site is www.sdss4.org.

SDSS is managed by the Astrophysical Research Consortium for the Participating Institutions of the SDSS Collaboration including the Brazilian Participation Group, the Carnegie Institution for Science, Carnegie Mellon University, Center for Astrophysics | Harvard \& Smithsonian (CfA), the Chilean Participation Group, the French Participation Group, Instituto de Astrofísica de Canarias, The Johns Hopkins University, Kavli Institute for the Physics and Mathematics of the Universe (IPMU) / University of Tokyo, the Korean Participation Group, Lawrence Berkeley National Laboratory, Leibniz Institut für Astrophysik Potsdam (AIP), Max-Planck-Institut für Astronomie (MPIA Heidelberg), Max-Planck-Institut für Astrophysik (MPA Garching), Max-Planck-Institut für Extraterrestrische Physik (MPE), National Astronomical Observatories of China, New Mexico State University, New York University, University of Notre Dame, Observatório Nacional / MCTI, The Ohio State University, Pennsylvania State University, Shanghai Astronomical Observatory, United Kingdom Participation Group, Universidad Nacional Autónoma de México, University of Arizona, University of Colorado Boulder, University of Oxford, University of Portsmouth, University of Utah, University of Virginia, University of Washington, University of Wisconsin, Vanderbilt University, and Yale University.

\vfill\eject
\bibliography{Bibliography}

\appendix

\section{Asteroseismic Pipelines and Light Curve Preparation} \label{app:pipelines}

We employed a total of 10 distinct codes for inferring asteroseismic properties of the sample. Seven of these, which we designate as core methods, were designed to measure both \dnu\ and \nmax. Three were used to measure \nmax\ only. Not all methods were used on all stars; only a subset of methods were used to study the more luminous cohort of targets. In Appendix \ref{app:pipelines}, we describe how the light curves were prepared, and provide detailed descriptions of the individual methods used to infer asteroseismic properties. In Appendix \ref{app:merging} we describe how the results were combined into single values with uncertainties.

\subsection{Light Curve Preparation}

In our study, we utilize KEPSEISMIC\footnote{\url{https://archive.stsci.edu/prepds/kepseismic/}} \kep\ data from the Mikulski Archive for Space Telescopes (MAST). These light curves undergo corrections using the \kep\ Asteroseismic data analysis and calibration Software \citep[KADACS,][]{Garcia11}. The corrections involve removing outliers, addressing jumps and drifts, and stitching together data from all quarters. Additionally, to mitigate the impact of regular gaps primarily caused by instrumental operations,such as angular momentum dumps and monthly downlink Earth pointings, we employ a multi-scale discrete cosine transform to interpolate missing data \citep{2006aida.book.....S,Garcia14,Pires15}, following inpainting techniques based on a sparsity prior  \citep{Elad05}. 

Subsequently, the light curves are filtered using a 20- or 80-day high-pass filter to eliminate long drifts resulting from the \kep\ orbit. The combination of two or more high-pass filtered light curves is a common way to study low-frequency stellar signals in the \kep\ data. For example, it has been successfully applied to measure stellar rotation  
\citep[e.g.,][]{2017A&A...605A.111C,2019ApJS..244...21S,2021A&A...647A.125B,2023A&A...679L..12G}. The 20-day filter makes it difficult to extract asteroseismic signals in luminous giants with low oscillation frequencies. We therefore use the 80-day filter for stars with $\nmax < 10 \;\rm{\mu Hz}$; this threshold was chosen because signal recovery with the 20-day filter degraded drastically for lower \nmax. However, the 80-day filter is noisier, injecting scatter into measurements at high \nmax, so we adopted the 20-day filter for stars with \nmax\ above $10 \;\rm{\mu Hz}$.



\subsection{Core Asteroseismic Pipelines}

The core asteroseismic pipelines are designed to infer both \dnu\ and \nmax. 
Many methods detect a power excess above a background to infer \nmax\ (see Figure \ref{fig:fliper}), typically by smoothing, or fitting bell-shaped functions to, the discrete frequency spectrum. The discrete oscillation frequencies are then used to characterize \dnu, which is defined as the mean separation between modes of different radial order $n$ but the same spherical degree $l$. Even and odd $l$ modes cluster together in a power spectrum; in most cases $l = 0$, $1$ and $2$ can be detected, so the even ($l = 0,2 $) and odd ($l = 1$) modes can be distinguished by searching for a pattern that has close doublets alternating with single modes. The $l = 1$ (and, to a lesser degree, $l = 2$) mode frequencies can be mixed in character -- strongly influenced by core g-modes -- which can induce a large deviation from uniform spacing. These modes can also be split by rotation.

The observed mode spacing also vary with frequency; some methods focus on modes close to \nmax\ while others average results across a wider range. In practice, there can be confounding features, such as pollution of the light curves by other stellar variables in the same apertures. Six of the methods that we use here were included in APOKASC-2, although the pipelines themselves have been modified in the intervening time. We now describe the individual pipelines.

\subsubsection{COR}

The COR seismic parameters are derived from the envelope autocorrelation function (EACF), as described in \citet{Mosser2009}. The method uses the properties that the square of the autocorrelation of the time series can be calculated as the Fourier spectrum of the filtered Fourier spectrum, as initially shown by \citet{RV2006}.
It first measures the large separation \dnu\ in a fully blind manner. The reliability of the detection is given by the $H_0$ test. Then, the frequency \nmax\ is inferred from the identification of the oscillation excess power, assuming that the local stellar background around \nmax\ can be approximated by a power law in frequency. The estimate of \dnu\ can be refined, using the homologous properties of the red giant oscillation pattern, as depicted by the so-called universal red-giant oscillation pattern \citep{Mosser2011}. The COR pipeline was run on the full set of data, including luminous giants.

\subsubsection{ELS}
The ELS seismic parameters are derived using the methods described in \citet{2020Elsworth}. A key feature is the use of a layered approach, which applies many loose constraints to remove false detections. This has the advantage of a high detection rate combined with a low number of false positives. The frequency \nmax\ is determined from an MCMC fit to the power spectrum with the background represented by two Harvey-like profiles, a constant white noise component and the mode power by a Gaussian function centered on \nmax.
The effect of the integration time on the shape of the power spectrum is included.
For \dnu,  the method looks for the mode regularity by employing the power spectrum of the power spectrum in the region where the modes are most prominent. A small adjustment is applied to correct for the correlation between \dnu\ and \nmax. 

\subsubsection{SYD} \label{app:syd}
The SYD results were derived using the SYD pipeline \citep{Huber09}, largely following the approach as in previous APOKASC data releases. Here we summarise the differences. We adopted the SYD results from the catalog of 16,000 red giants by \citet{Yu18} for stars overlapping with our list 
For the rest we ran the SYD pipeline using \nmax\ values from \citet{Hon18,Hon19} as initial guesses. The results were vetted using the automated deep learning method by \citet{Reyes22}.  

\subsubsection{CAN}

The CAN pipeline was one of the five used in the APOKASC-2 paper, and was described there and in \citet{Kallinger2010}. The majority of the results here are taken from APOKASC-2. New calculations were performed for some luminous giants not present in the original APOKASC-2 effort.

\subsubsection{DIA}
This method relies on the adoption of the Bayesian inference code \textsc{Diamonds} \citep{Corsaro14}, which has been developed with the main purpose of analyzing asteroseismic data. The method is directly applied to the stellar power spectral density (PSD), which in this case has been obtained through the KADACS pipeline suite for asteroseismic-optimized datasets \citep{Garcia11,Garcia14,Pires15} starting from the raw \kep\ light curves. As presented by \cite{Corsaro17}, the PSD is first analyzed by fitting the background model introduced by \cite{Kallinger14}. The background model comprises a Gaussian function to reproduce the typical power excess of solar-like oscillations, a flat instrumental noise component, and a series of Harvey-like profiles \citep{Harvey85} accounting for the effects of stellar granulation, instrumental variations, and potential magnetic activity. The number of Harvey-like components is decided based on an assessment of the Bayesian evidence estimated by \textsc{Diamonds}, which is used to evaluate a statistical weight on the fitting model, such that the model providing the best balance between complexity (i.e., number of free parameters) and quality of the fit, is favored. The fitting of the Gaussian envelope of the background model finally provides the estimate of \nmax, along with its Bayesian credible intervals as a measure of uncertainty in the parameter. 

For obtaining an estimate of \dnu, the method evaluates a squared autocorrelation function (ACF$^2$) over the region of stellar PSD that contains the oscillation envelope. Before the ACF$^2$ is applied, the PSD is smoothed with a boxcar having width $\Delta\nu/10$, where the guess for $\Delta\nu$ is taken from a scaling relation \citep{Huber11}. The smoothing has the effect of reducing the stochasticity that is typical of this type of data, resulting in a net improvement of the ACF$^2$ signal when originating from the presence of a comb-like structure that is characteristic of the solar-like oscillations. The search range of the ACF$^2$ method is also based on the \dnu\ guess obtained from the scaling relation, and typically ranges within $\pm 30$\,\% of this value. The final estimate of \dnu\ and its 1-$\sigma$ uncertainty are obtained from the centroid of a Gaussian function fitting the ACF$^2$ around its maximum. The fitting is performed by means of a non-linear least-squares fitting method. This pipeline was run on a minority of the stars in the sample, but did include some luminous giants.

\subsubsection{GAU}
The GAU algorithm was designed to estimate \nmax, \dnu, and $A_{\rm max}$, the amplitude of the Gaussian function employed to model the oscillations' power excess. First, it reads the power spectrum of the time series, and computes the envelope of the filtered autocorrelation of the time series in the way proposed by \citet{Roxburgh_2009} and \citet{Mosser2009}. In practice, it consists of computing the power spectrum of the power spectrum multiplied by a smooth bandpass filter of 20-$\mu$Hz width that is centered on a grid of frequencies that run from  from $1\;\mu$Hz to Nyquist (283$\;\mu$Hz) every $5\;\mu$Hz. If there is a significant maximum in the envelope of the autocorrelation, its location provides both $\Delta\nu$ and an initial proxy of \nmax. Secondly, the code checks whether it picked a harmonic of $\Delta\nu$ based on the relation between $\Delta\nu$ and $\nu_{\rm max}$ from \citet{stello_2009} $\Delta\nu = \Delta\nu_\odot (\nu_{\rm max}/\nu_{{\rm max},\odot})^{0.75}$. Thirdly, it performs a background fitting of the power spectrum by including two Harvey functions for the stellar activity, a white noise level, and a Gaussian function for the oscillations. The best-fitting algorithm is a maximum likelihood estimator with Bayesian prior as initially described in \citet{Gaulme_2009}. The oscillation $\nu_{\rm max}$ and $A_{\rm max}$ are the central frequency and height of the Gaussian function. The error bars are computed from the Hessian of the likelihood estimator. The GAU pipeline was run only using the 20-day filtered data, so it was not used for stars with $\nu_{\rm max}$ below 10 $\rm{\mu Hz}$.

\subsubsection{A2Z}

The A2Z+ pipeline is a combination of the A2Z \citep{Mathur2010} and an implementation of the EACF method. 
With the A2Z method, the mean large frequency separation \dnu\ is inferred by computing the power spectrum of the power spectrum in a region of the power spectrum density where the most prominent repetition of peaks is found, above a given confidence level of 90\%. The convective background is then modeled with four different components: two Harvey functions \citep{Harvey85} to model different scales of granulation with a fixed slope of 4, a Gaussian function to model the p-mode envelope, and a white noise term. The frequency of maximum power is obtained from the fit of the Gaussian that is centered at \nmax. In addition, a complementary determination of \dnu\ and \nmax\ is included in A2Z+ following the EACF method by \citet{Mosser2009}. We then compare the results between the two methods. Stars where the values agree within 10\% are treated as firm detections, while the others are flagged and visually inspected. Finally, we implemented a refinement of \dnu\ by cross-correlating the power spectrum with a template with modes $l=0$ and 2 varying \dnu\ and the small frequency separation \citep{2016ApJ...827...50M, 2022A&A...657A..31M} to pick the \dnu\ with the highest cross-correlation value. 

\subsection{Other Methods}

It is valuable to detect an asteroseismic signal even in cases where the detailed frequency pattern cannot be characterized. The three methods in this section were not used to infer frequency spacings, but are very useful for detecting \nmax, particularly in sparse or noisy data sets. Some methods rely on machine-learning approaches, while others use other information (such as the granulation spectrum) to improve measurement precision.   

\subsubsection{HON}
This method utilizes deep learning to detect the presence of a power excess corresponding to solar-like oscillations within power spectra. As described in \citet{Hon18}, power spectra represented in logarithmic axes are converted into a binary $128\times128$ images as inputs into two convolutional neural networks. The first network classifies power spectra into those containing solar-like oscillations and those without, while the second network measures \nmax\  by identifying the region of the input image that contains the oscillations. The networks are trained with supervised learning using a labelled training set. For the classification of stars across the full \kep\ sample, the networks are trained using the \citet{Yu18} catalog as a training set, as described in \citet{Hon19}.

\subsubsection{CV}

The ``coefficient of variation'' (CV) method for detecting solar-like oscillations and reporting \nmax\ values is described in \citet{2019MNRAS.482..616B}. The power spectrum is split into frequency bins with widths that scale approximately with the expected \dnu\ if \nmax\ were centered in each bin. Within each bin, the CV metric is computed as simply the ratio of the standard deviation of the periodogram power to the mean power in the bin. These bins are narrower than the frequency scale of granulation, such that this background is essentially flat across each bin. If a bin contains no signal besides the granulation background, the CV value is expected to be near 1.0 for random noise distributed about the background as $\chi^2$ with two degrees of freedom. Candidate solar-like oscillations are identified as statistically significant excesses in CV that have widths and heights consistent with known examples of solar-like oscillating red giants. The CV method is able to effectively separate the solar-like oscillation signals from the granulation background without fitting any models to the background. The method flags light curves that show additional CV excesses that are likely caused by other types of variability or that potentially show two power excesses from solar-like oscillations. We found that the light curve processing from the KADACS pipeline caused the distribution of noise in the power spectra to differ (greater dispersion) from expectations for a $\chi^2$ distribution with two degrees of freedom, so we instead analyzed data processed by the KASOC \citep{2014MNRAS.445.2698H} filter where available (6939 stars), and we used the data from the \kep\ pipeline downloaded from MAST with some minimal processing for the rest.

\subsubsection{FLI}

\begin{figure}
    \centering
    \includegraphics[width=0.7\linewidth]{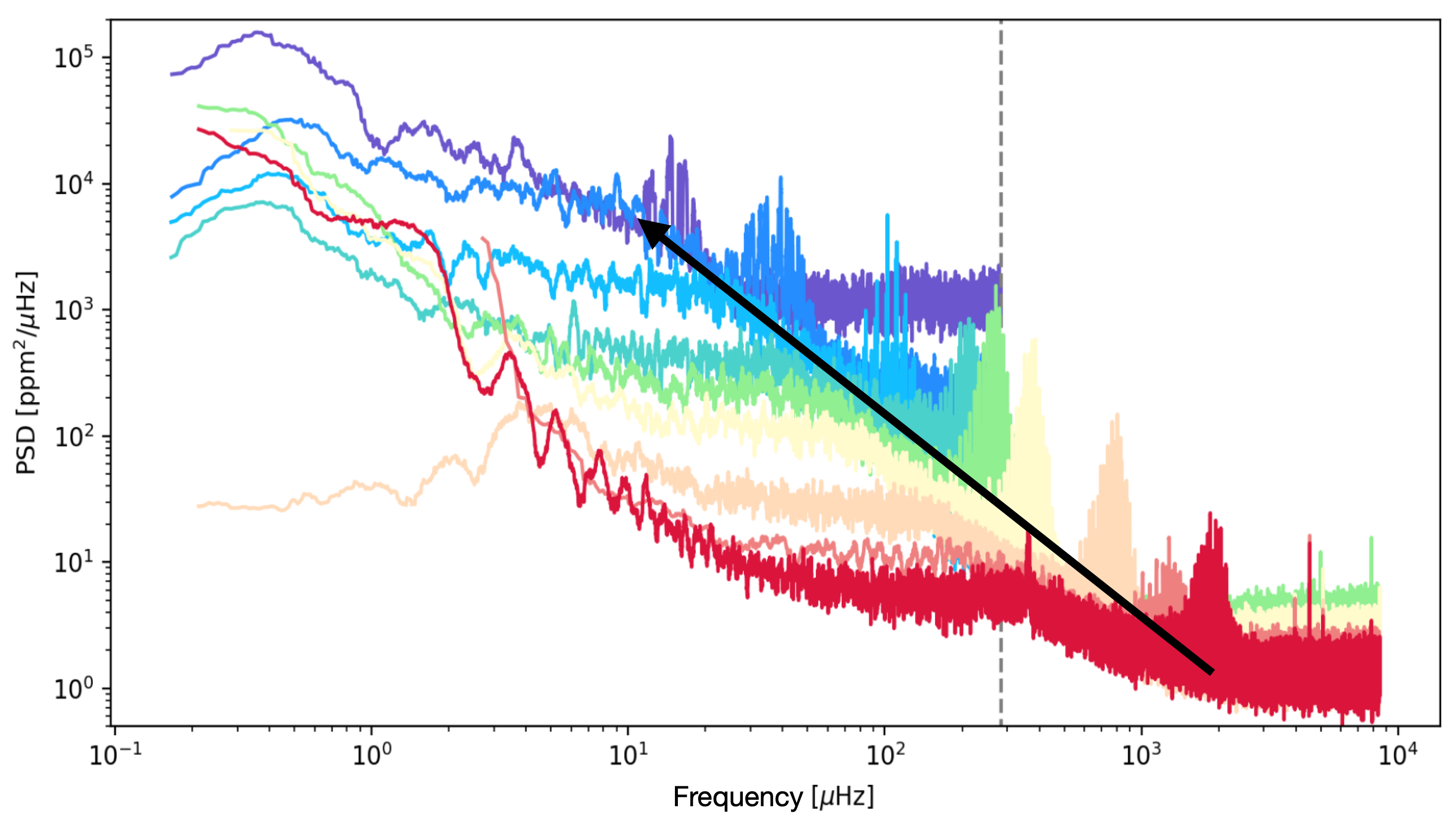}
        \caption{Power spectral density of solar-like oscillators observed by \textsl{Kepler} from the main sequence (red) to evolved red giant branch stars (purple). The vertical dashed grey line indicates the Nyquist frequency for \textsl{Kepler} long cadence observations. The arrow indicates the decrease in frequency and the increase in power density associated with stellar evolution up until the tip of the RGB. 
        }
    \label{fig:fliper}
\end{figure}

The FliPer method described in \cite{Bugnet2018} is based on the averaged power density contained in the PSD. It allows measurement of stellar surface parameters \citep{Bugnet2018} or to classify pulsators \citep{Bugnet2019}. In the case of solar-like stars, granulation occurs at characteristic frequencies and amplitude correlated with characteristic frequencies and amplitude of the stochastic oscillations it generates \citep[see Fig.~\ref{fig:fliper}, 
e.g.,][]{Kallinger2010, Mathur2010, Bastien2013}, the average power density of a solar-like star correlates with its \nmax. The FliPer method relies on a Random Forest (machine learning) algorithm, trained on thousands of \kep\ data analyzed with the A2Z+ pipeline \citep{Mathur2011} to automatically estimate \nmax{} of other \kep\ stars from the power contained in the PSD \citep[see][for more details]{Bugnet2018}. It is particularly valuable compared to classic asteroseismology for stars with \nmax{} at low frequency ($\nmax \lessapprox 10 \;\mu$Hz), close to and above the Nyquist frequency of the observations, and for stars observed with poor frequency resolution, as there is no need for mode detection to estimate the typical frequency of the oscillations.

\section{Merging Individual Results} \label{app:merging}

Our full sample includes 7,555 spectroscopic dwarfs and 15,791 spectroscopic giants. Unlike APOKASC-2, the methods used vary significantly in their precision, and we are not restricted to uniformly high quality light curves. We therefore had to filter our raw data to reject outlier measurements and to identify background sources. We began by using inferred median results and median absolute deviations for our full sample, and then identified and removed robust detections that were clearly from background sources. We followed up by applying a spectroscopic prior to reject individual false positive measurements, and then rejected measurements strongly inconsistent with the ensemble average. See Section B.2. below for the details of the outlier rejection procedure. At the end of this process, we had up to 10 \nmax\ and up to 7 \dnu\ measurements per target.

With this data in hand, we then adopted a procedure similar to that of APOKASC-2 for combining the raw data from individual pipelines into a single value for each target. We began by dividing the data into data quality categories. The highest quality cohort, the Gold sample, had a minimum of 5 independent detections of \dnu; the Silver sample had at least 2 independent \dnu\ values; and the Detected sample had a minimum of 2 independent \nmax\ measurements, but less than 2 \dnu\ values. We then placed all stars on a common zero-point by applying small offsets to data from each pipeline, using the Gold RGB sample as a reference data set.  We follow by using the agreement between individual pipelines and the ensemble average to give each pipeline a weight; this was done separately for the Gold RGB, Gold RC, and Silver samples. We could then construct weighted averages for our central values; for the ``Detected'' sub-sample, we report only \nmax. Details of our procedure are described below.

\subsection{Identifying Background Sources}

We started with data from 13 pipelines. Three of the results corresponded to alternative measurements using the same underlying method as another entry. As these data are strongly correlated, we adopted only one measurement per technique. 
This left us with 10 independent measurements for \nmax, and 7 for \dnu. This is because the other 3 methods (Bell, Hon and FliPer) either did not provide \dnu, or did so purely in a statistical manner. 

Each light curve is extracted from a number of pixels centered on the target. These \kep\ ``postage stamps'' can be relatively large, and it is possible for asteroseismic signals to be associated with a different target than the associated KIC ID. In some cases the sources are separable spatially, but this need not be true. The most straightforward discriminant is whether the measured \nmax\ is in the rough domain expected given the spectroscopic data. We therefore used APOGEE DR16 and DR17 spectroscopic surface gravities and effective temperatures to predict a ``spectroscopic'' \nmax. The median uncertainty in the spectroscopic measurements is 0.065 dex (Table \ref{table:uncertainties}); to be conservative we broadened the prior to 3 $\sigma$ in both directions, for a minimum total permitted range of 0.39 dex. As discussed in \citet{Jonsson2020}, there were corrections applied to the derived spectroscopic values that depended on evolutionary state. To ensure that our results were not biased by the evolutionary states assigned, we further broadened the priors by the difference between the spectroscopic gravities that would have been inferred for red clump and red giant branch states in the domain where both families are observed. Finally, we broadened the prior to account for \logg differences between DR16 and DR17 for stars with data in both.   

We then computed the \nmax\ median and median absolute deviation for all targets. Stars with measurements inconsistent with the spectroscopic prior at the 5 sigma level are shown in Figure \ref{fig:specout}. A total of 129 asteroseismic detections were classified as background sources with this technique. This included 63 targets classified as giants and 66 classified as dwarfs or subgiants.  Their properties are discussed in Section \ref{sec:outliers}.

\subsection{Median Outlier Rejection: Spectroscopic and Ensemble Priors.}

\begin{figure}
\includegraphics[width=8cm]{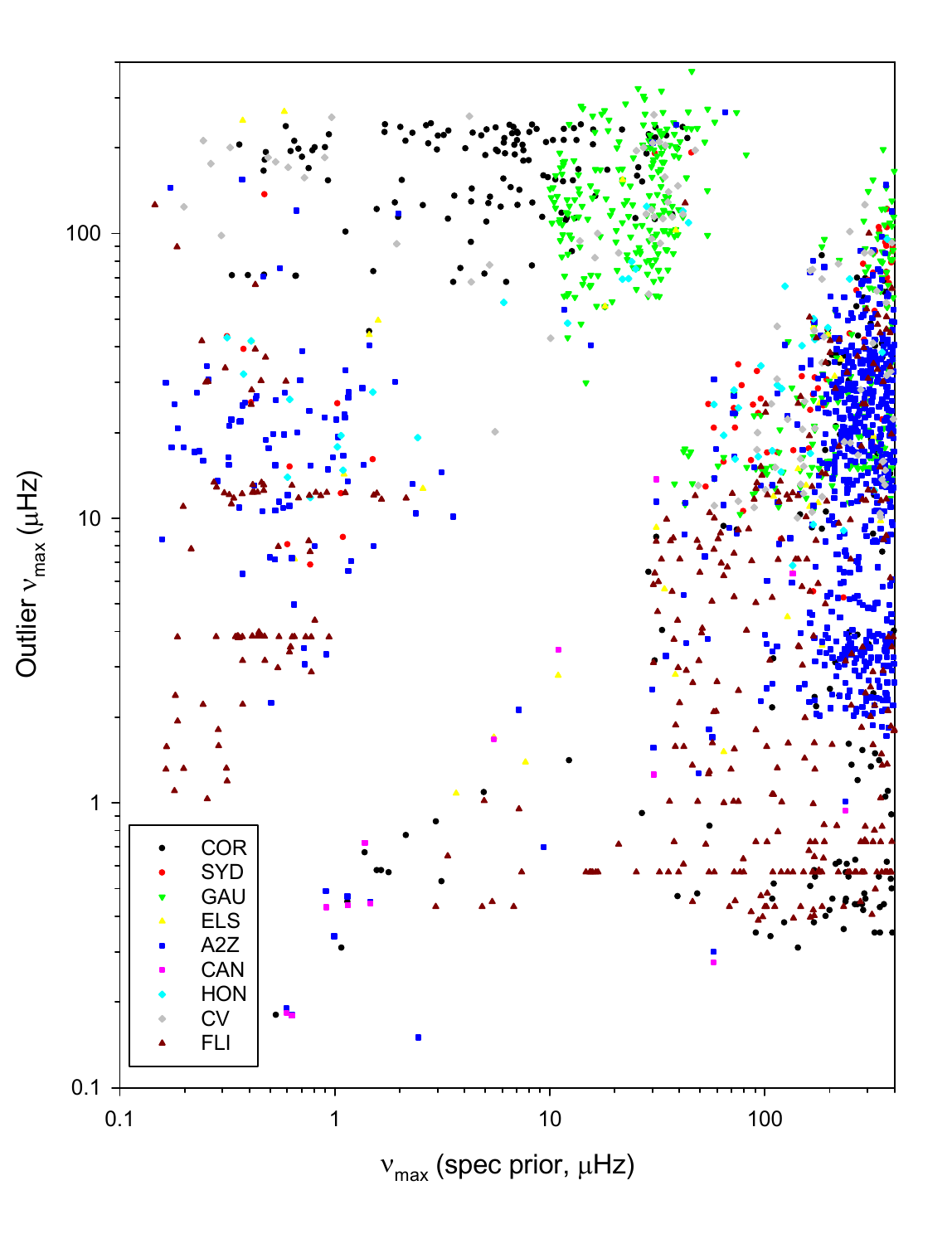}
\includegraphics[width=8cm]{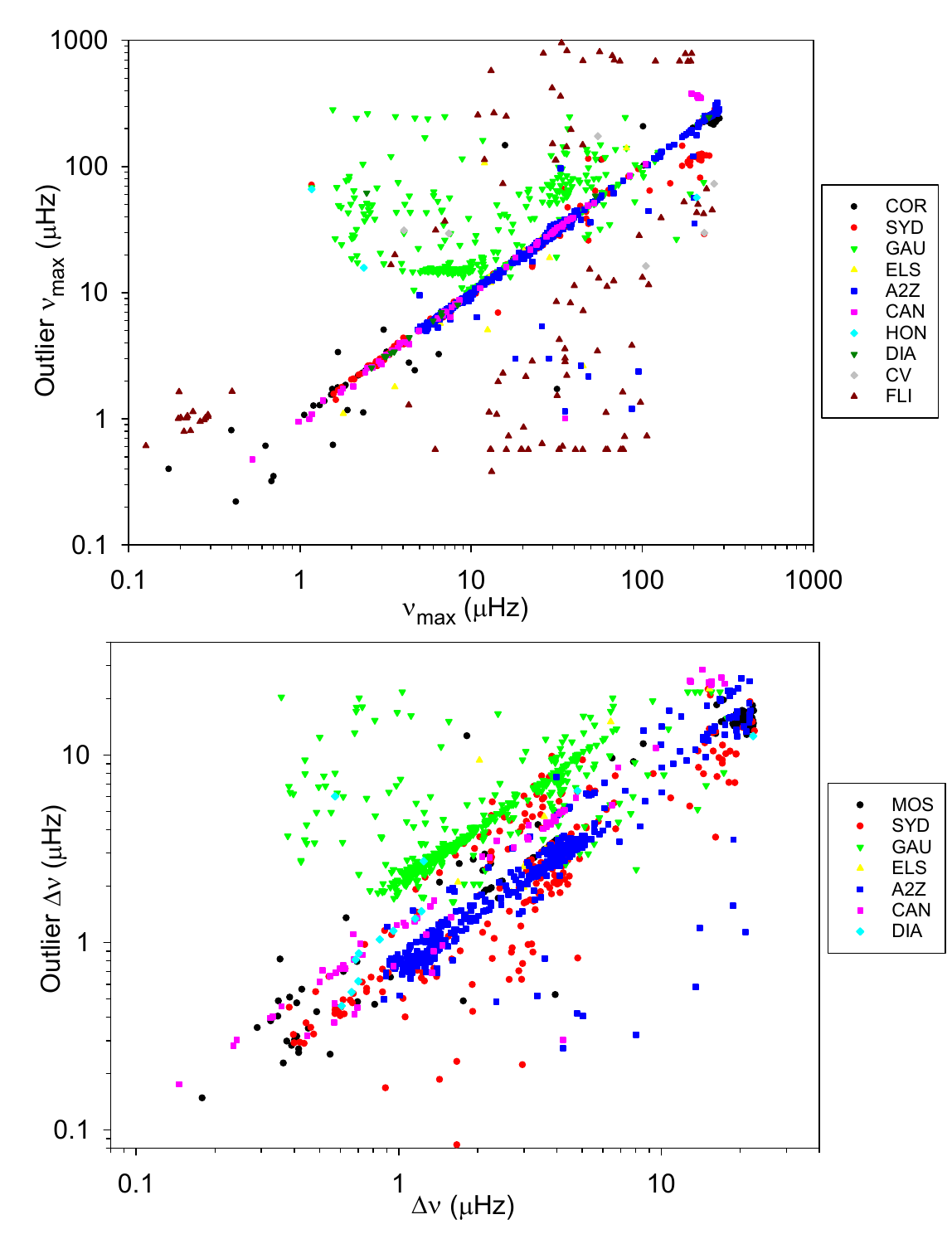}
\centering
\label{fig:specout}
\caption{Measurements flagged as outliers in APOKASC-3. The left panel shows data with \nmax\ inconsistent with the spectroscopic prior. The right panels show measurements inconsistent with other pipelines in \nmax\ (top right) and \dnu\ (bottom right). Measurements consistent with the prior are not shown for visual clarity. The population of \nmax\ measurements close to the unity line were rejected because their corresponding \dnu\ measurements disagreed. The large majority of rejected measurements are strongly inconsistent with the prior.}
\end{figure}

Having checked for global consistency with spectroscopy, we now turn to the validation of individual measurements. We are using automated techniques to collapse an observed frequency pattern into two global figures of merit. In real data, however, there are confounding factors that can give rise to spurious measurements. For example, a background classical pulsator might inject a single spurious frequency signal into a normal red giant pattern; even faint sources can have a detectable variability amplitude. Other stars have unusual oscillation amplitudes or pattern that can confound detection algorithms. Outlier rejection is an efficient tool for identifying both individual failure modes and targets with light curves that are difficult to interpret. We therefore removed individual \nmax\ measurements inconsistent with our spectroscopic prior, as defined above.  The results are illustrated in Figure \ref{fig:specout}. If we excluded a \nmax\ detection, we also excluded any corresponding \dnu\ measurement.   

Our final pass involved checking whether the measurements that we included were consistent with the full ensemble of data.  Our averaging technique will be biased in the presence of large method to method differences, which would manifest as a translation of systematic errors into random ones. To avoid this, in some cases the trends that we identified caused us to restrict the \nmax\ domain for techniques that differed significantly from the mean. The GAU results were not run with the longer 80-day filter, so we did not use them for stars with \nmax\ below 10 $\rm{\mu Hz}$. The A2Z pipeline had significant systematic differences in \dnu\ relative to other methods for stars with \nmax\ less than 5 $\rm{\mu Hz}$, due to the way it computes \dnu\ on a broad frequency range instead of computing a local value centered on \nmax. We therefore did not include them in this domain. 

With these data removed, there is both a well-behaved core and an excess of outliers. We therefore performed a final outlier rejection test for individual pipeline results relative to the median. We performed this test only for targets with 3 or more detections, and we removed measurements discrepant from the median at more than five $\sigma$. This test was employed for both \nmax\ and \dnu, with both measurements excluded if either failed the outlier test.  Figure \ref{fig:specout} compares individual measurements excluded by this method with the ensemble average.  (Some of the data appear close to the median in \nmax\ or \dnu; such stars were failures in the other value.)


Table \ref{table:Pipedet} summarizes our measurement and detection statistics for all methods. For each pipeline, the total number of raw \nmax\ detections per method is in row 1; the number of \nmax\ values excluded as spectroscopic outliers is in row 2; the number of \nmax\ values excluded as ensemble outliers is in row 3, and the corresponding number of \dnu\ outliers excluded is in row 4. The total number of filtered \nmax\ and \dnu\ measurements per pipeline are in rows 5 and 6.

\begin{table*}
    \centering
    \begin{tabular}{c c c c c c c c c c c}
         Category & COR & ELS & GAU & SYD & A2Z & DIA & CAN & CV & HON & FLI  \\
        \hline
 Detected (\nmax) & 12860 & 12204 & 11122 & 13147 & 13676 & 2069 & 8294 & 10483 & 12590 & 14113 \\
 Reject (\loggns) & 988 & 56 & 3583 & 94 & 3203 & 0 & 19 & 303 & 58 & 1289\\
 Reject (ens) & 15 & 6 & 279 & 19 & 14 & 1 & 12 & 6 & 3 & 113\\
        \hline
 Detected (\dnu) & 11867 & 11222 & 10547 & 12127 & 10995 & 1899 & 7164 & 0 & 0 & 0\\
 Reject (ens) & 120 & 19 & 517 & 220 & 540 & 13 & 77 & N/A & N/A & N/A\\
    \hline
    \end{tabular}
    \caption{Detection statistics by pipeline. Columns correspond to different pipelines. The first row identifies the pipelines. Rows 2, 3, and 4 are the number of \nmax\ detections, measurements inconsistent with the spectrosopic prior, and measurements inconsistent wit the ensemble median respectively. Rows 5 and 6 present \dnu\ detections and measurements rejected as inconsistent with the ensemble median respectively.}
    \label{table:Pipedet}
\end{table*}
    
Table \ref{table:alldata} summarizes the net outcome of our filtering process. We includes the raw data, the spectroscopic prior range used, star-by-star detection statistics, and quality codes for individual measurements.  

\begin{table*}
    
    \centering
    \begin{tabular}{c c}
         Label & Contents  \\
          
 \hline
 KIC ID & Number in the \kep\ Input Catalog \\
\nmax\ Spec Min, Max & Lower and Upper Bounds, Spectroscopic Prior  \\
NDET & Total number of \nmax\ detections prior to quality cuts \\
NNMAX & Number of valid \nmax\ measurements after outlier rejection \\
NDNU & Number of valid \dnu\ measurements after outlier rejection \\

NCOR to NFLI & \nmax\ data quality flag for pipelines COR, ELS, GAU, SYD, A2Z, DIA, CAN, CV, HON \& FLI, respectively \\
\nmax\ COR to FLI & \nmax\ values for pipelines COR, ELS, GAU, SYD, A2Z, DIA, CAN, CV, HON \& FLI, respectively \\
DCOR to DCAN & \dnu\ data quality flag for pipelines COR, ELS, GAU, SYD, A2Z, DIA, \& CAN, respectively \\
\dnu\ COR to CAN & \dnu\ values for pipelines COR, ELS, GAU, SYD, A2Z, DIA \& CAN, respectively \\
    \hline
    
    \end{tabular}

    \caption{Raw Asteroseismic Measurements and Filtering Results. We include the raw measurements from all 10 input pipelines for \nmax\ and all 7 for \dnu. We also give the spectroscopic prior and the permitted range for the spectroscopic filter, the number of measurements with any \nmax\ detections, the number of filtered \nmax\ detections, and the number of filtered \dnu\ detections. The codes NCOR to NFLI reflect the category of the data for each star.  A 0 entry denotes no data, 1 is a valid filtered measurement, 2 is one that failed the spectroscopic prior, and 6 is one that failed the ensemble prior test. A 9 entry denotes data excluded from usage because it fell outside the range of validity for that pipeline. The DCOR to DCAN codes give data quality for \dnu\ measurements, using the same notation as that used for \nmax.}
    \label{table:alldata}
\end{table*}

\subsection{Data Quality: Gold, Silver, Detection, No Detection.}

There is a strong correlation between the number of consistent measurements from different techniques and the measurement scatter.  Two of our 7 core methods were not used on the full data set, so we define our Gold sample as stars with five or more \dnu\ detections.
We required at least 2 measurements to treat a detection as valid. If there were 2 or more \dnu\ data, the target was included in the Silver sample; stars with 0 or 1 \dnu\ values, but at least 2 \nmax\ ones, were treated as Detections. Other stars were classified as Non-Detections. Our detection statistics are discussed in Section \ref{sec:recovery}.  

\subsection{Zero-Points and Pipeline Weights}

Each asteroseismic pipeline has both a solar zero-point and a measurement uncertainty based on the quality of the power spectrum. However, in the APOKASC-2 paper we measured the ensemble average for all methods and could infer both the relative measurement zero-points and the scatter of each method around the mean. We found that the formal uncertainties were not correlated with how well an individual pipeline predicted the ensemble average, and the relative stellar zero-points were not the same as the relative solar zero-points. Given these results, we therefore adopted an empirical approach. Table \ref{table:zeropt} presents the scale factors and weights. 

For stars with high quality measurements, we compare individual pipeline values to the ensemble mean, and define relative scale factors for each method. This ensures that the mix of detection methods does not bias the zero-point of the average.  For \dnu, the fractional zero-point corrections are at most $\pm 0.004$; for \nmax, they are at most $\pm 0.008$. The solar \nmax\ reference value in APOKASC-2, 3076 $\rm{\mu Hz}$, was inferred by requiring agreement between fundamental masses in open cluster stars and asteroseismic values. In the current paper, we use fundamental radii to anchor the asteroseismic radii, and by extension masses, onto a fundamental system. This new correction factor is inferred relative to the APOKASC-2 solar reference value, and described in Section \ref{sec:nmax}. For \dnu, we use the APOKASC-2 solar value of 135.1416 $\rm{\mu Hz}$ for the initial base zero-point.

For pipeline weights, we divided our sample into three groups. Gold sample stars had 5 or more \dnu\ and \nmax\ detections, and we further split the Gold sample into separate RC and RGB cohorts. Silver sample stars had a minimum of 2 \dnu\ detections and a maximum of 4. Most of these stars are on the upper RGB. Table \ref{table:zeropt} lists the standard deviation of each method in each group. We note that the scatter here is somewhat inflated by a modest population of large outliers; with 5 $\sigma$ outlier rejection, the formal errors are roughly two-thirds as large, with similar relative performance in different pipelines. However, to be conservative in our error budget, we used the larger uncertainties.

\begin{table*}
    \centering
    \begin{tabular}{c c c c c c c c c c c}
         Category & COR & ELS & GAU & SYD & A2Z & DIA & CAN & CV & HON & FLI  \\
        \hline
 Scale (\nmax) & 0.9961 & 0.9995 & 1.0014 & 1.0003 & 1.0005 & 1.0018 & 1.0031 & 0.9911 & 1.0011 & 0.9729 \\
 N(\nmax) & 5124 & 5130 & 4710 & 5144 & 5143 & 975 & 3643 & 4565 & 5125 & 5136 \\
 $\sigma$(Gold, RGB) & 0.0115 & 0.0140 & 0.0282 & 0.0158 & 0.0224 & 0.0245 & 0.0116 & 0.0511 & 0.0568 & 0.1352 \\
 $\sigma$(Gold, RC) & 0.0110 & 0.0137 & 0.0146 & 0.0167 & 0.0274 & 0.0110 & 0.0111 & 0.0618 & 0.0533 & 0.1429 \\
 $\sigma$(Silver) & 0.0859 & 0.0525 & 0.1744 & 0.0736 & 0.0816 & 0.0465 & 0.0549 & 0.5360 & 0.2039 & 1.3767 \\
 \hline
  Scale (\dnu) & 1.0045 & 0.9983 & 0.9989 & 0.9985 & 1.0000 & 0.9968 & 1.0000 &   n/a &   n/a & n/a \\
 N (\dnu) & 5113 & 5077 & 4622 & 5133 & 4987 & 973 & 3611 & 0 & 0 & 0\\
$\sigma$(Gold, RGB) & 0.0158 & 0.0099 & 0.0239 & 0.0153 & 0.0222 & 0.0329 & 0.0135 & n/a & n/a & n/a \\
 $\sigma$(Gold, RC) & 0.0148 & 0.0116 & 0.0176 & 0.0144 & 0.0227 & 0.0136 & 0.0185 & n/a & n/a & n/a \\
 $\sigma$(Silver) & 0.0932 & 0.0567 & 0.1740 & 0.1056 & 0.1221 & 0.0521 & 0.0903 & n/a & n/a & n/a \\
     \hline
    \end{tabular}
    \caption{Relative zero-point scales and sample sizes by pipeline.}
    \label{table:zeropt}
\end{table*}
    
\begin{figure}
\includegraphics[width=0.5 \textwidth]{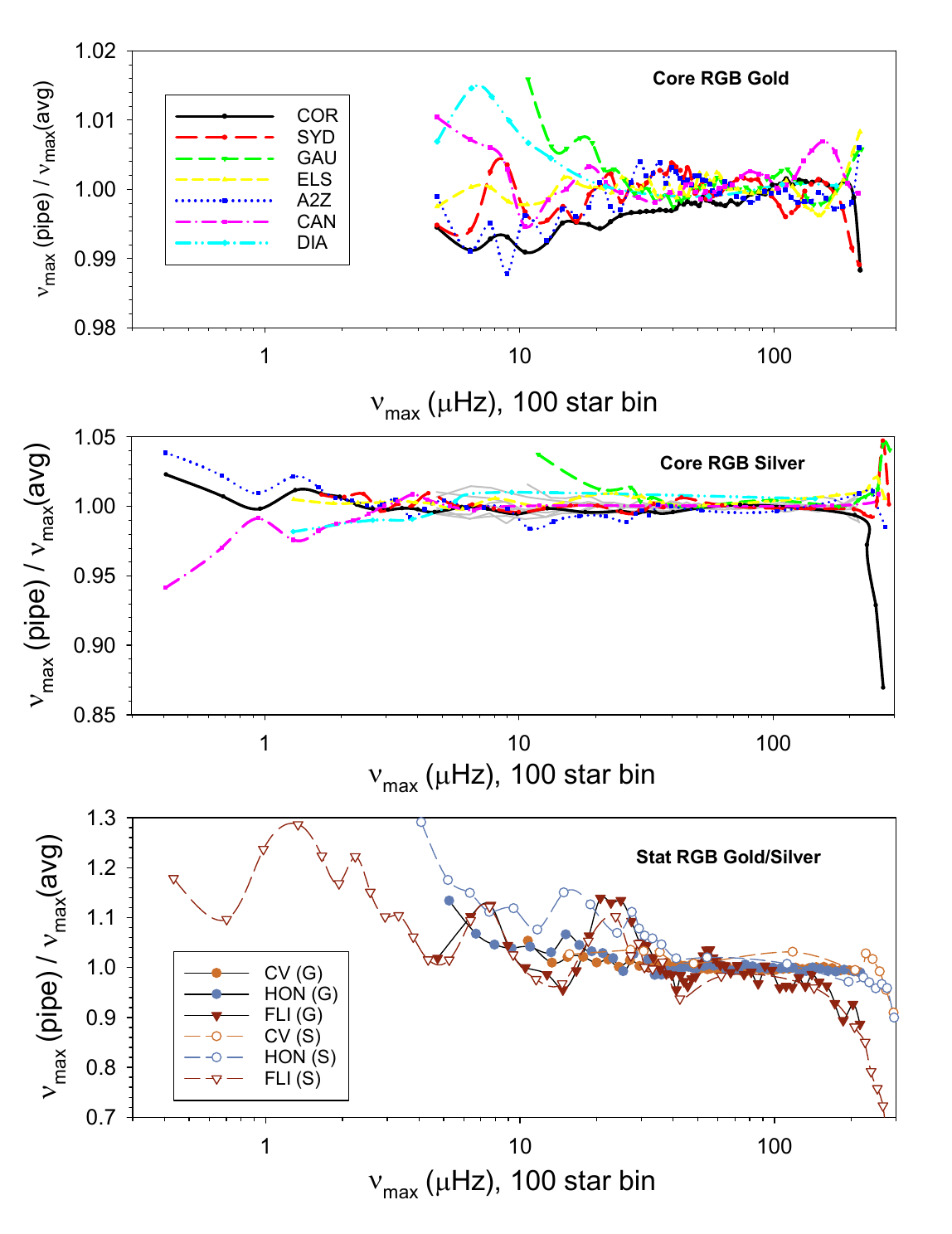}
\includegraphics[width=0.5 \textwidth]{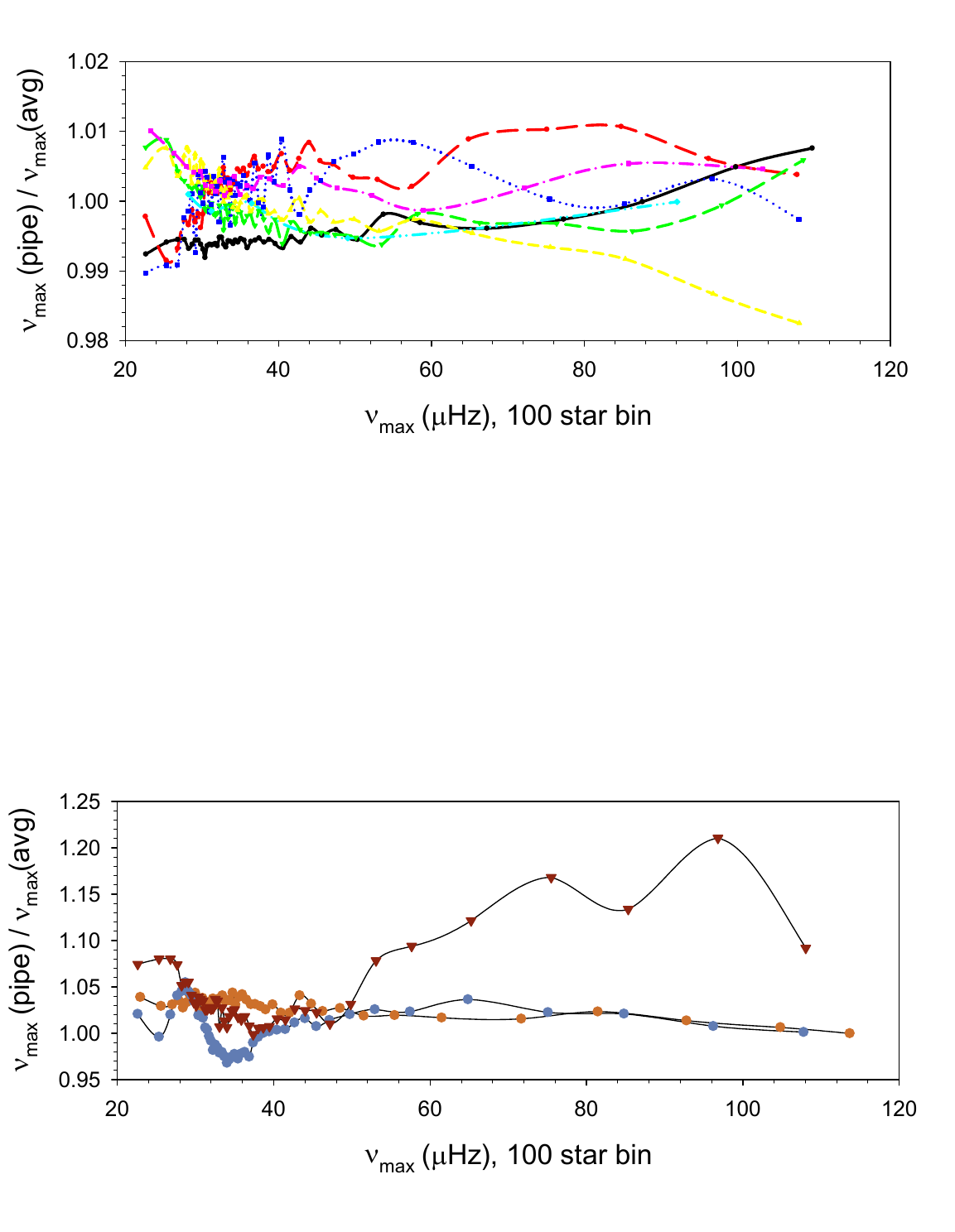}
\caption{Pipeline systematic offsets in \nmax\ for RGB (left) and RC (right). We distinguish our 7 core methods and show results for the Gold sample (top) and Silver sample (middle). Our \nmax-only methods are shown in the bottom panels.  Different symbols and line styles demote different pipelines (see the legend). For \nmax\ in the RGB Gold sample, fractional differences are at the $\pm 0.004$ level, rising to $\pm 0.012$ at low and high \nmax; a comparable $\pm 0.009$ range is seen for the RC. The Silver sample and the \nmax-only methods have larger scatter, especially in the low \nmax\ domain.}
\label{fig:nmaxtrends}
\end{figure}

\begin{figure}
\includegraphics[width=0.5 \textwidth]{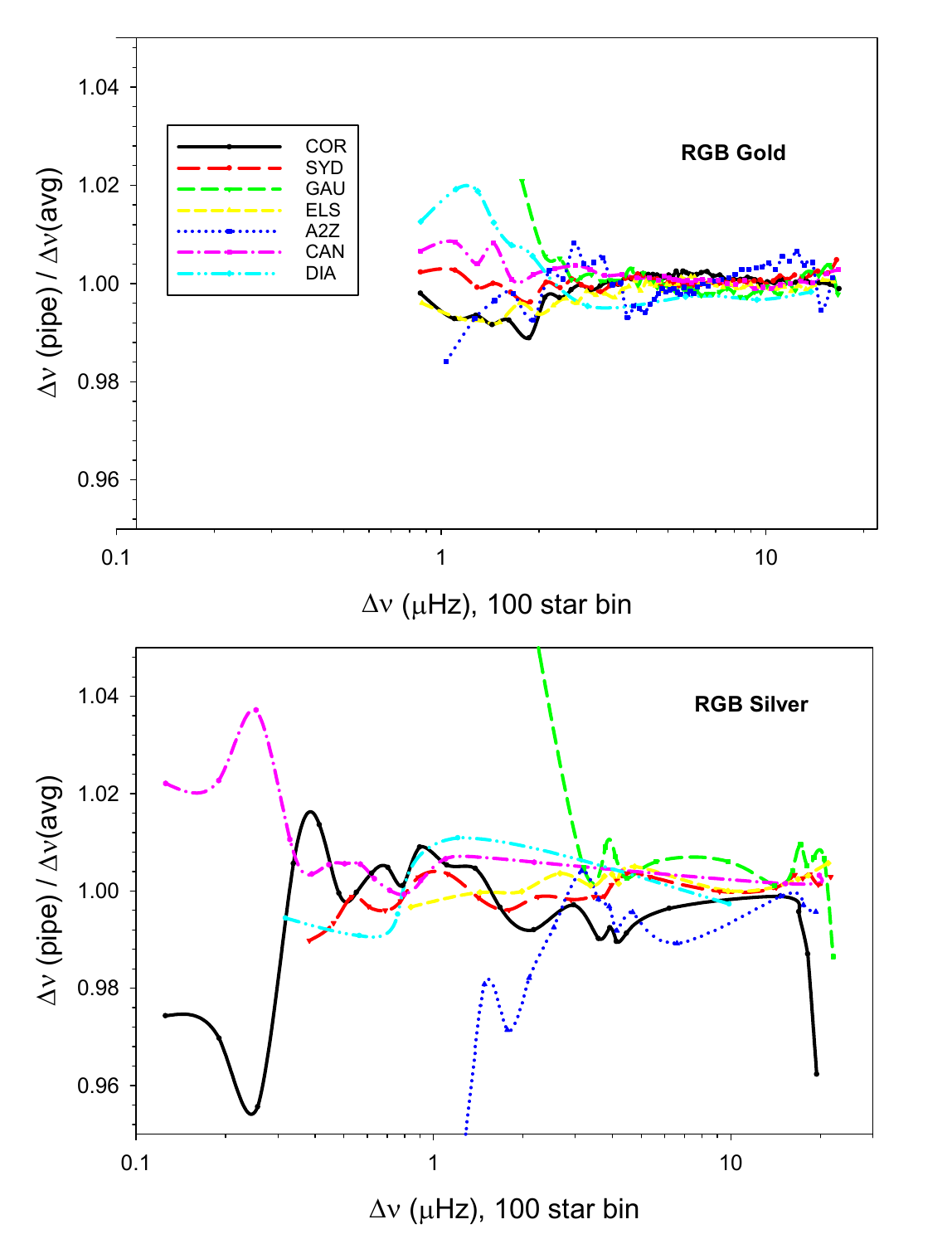}
\includegraphics[width=0.5 \textwidth]{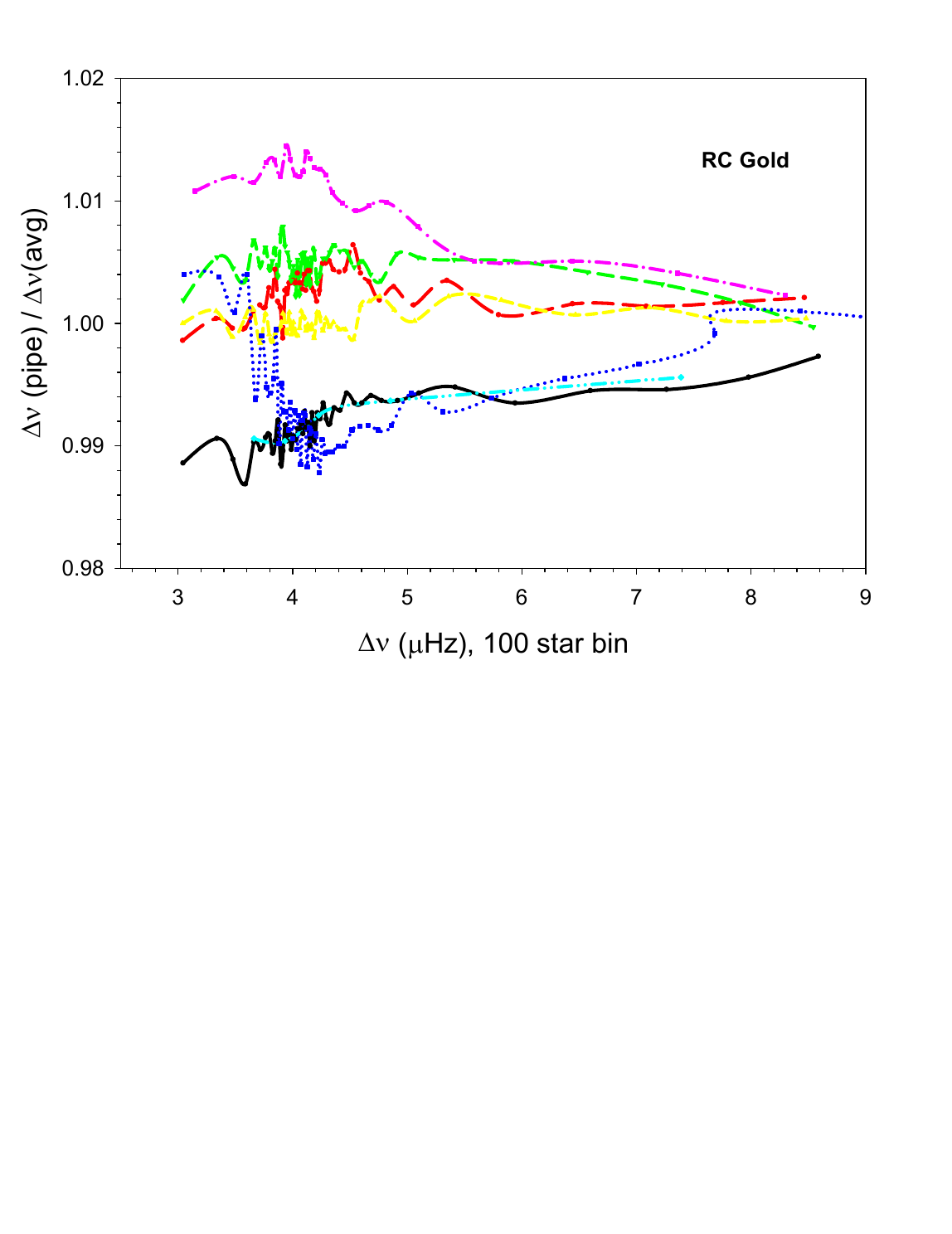}
\caption{Pipeline systematic offsets in \dnu\ for RGB (left) and RC (right). We show results for the Gold sample (top) and Silver sample (bottom). Different symbols and line styles demote different pipelines (see the legend). For \dnu\ in the Gold RGB sample, fractional differences are at the $\pm 0.004$ level, rising to $\pm 0.02$ at low \dnu; a $\pm 0.01$ range is seen in the primary RC, with a  smaller $\pm 0.005$ range in the secondary RC. Note that the Silver RC sample is small, and not shown here.}
\label{fig:dnutrends}
\end{figure}

To study systematic uncertainties in measurements, we rank-ordered data in mean \nmax\ and \dnu. Figure \ref{fig:nmaxtrends} shows the ratio of measurements from each technique in 100-star bins to the mean, as functions of \nmax\ for the RC and RGB stars. The corresponding data for \dnu\ is shown in Figure \ref{fig:dnutrends}.

Table \ref{table:useddata} presents our individual and averaged measurements.  The zero-point adjusted (see Table \ref{table:zeropt}) individual pipeline entries that were accepted as valid (State 1 from Table \ref{table:alldata}) are given here.
We then present several statistical characterizations of the data. We present both straight averages of measurements and ones corrected to the same mean zero point. The weighted averages used the weights given in Table \ref{table:zeropt}, and the uncertainties are the formal standard error of the mean. We also give median values and the median absolute deviation converted to an effective dispersion by multiplying the MAD by 1.4826, appropriate for a normal distribution. For the work that follows, we will use the weighted mean averages in most cases, switching to a median for cases with only 2 detections, which is equivalent to a simple average. We discuss tests of our error model in Section \ref{sec:uncertainties}.

\begin{table*}
    
    \centering
    \begin{tabular}{c c}
         Label & Contents  \\
          
 \hline
 KIC & Number in the \kep\ Input Catalog \\
NDet, NFDet & Number of raw and filtered \nmax\ detections \\
NDNDet & Number of filtered \dnu\ detections\\
NmaxCor, SigNmaxCor & Zero-point adjusted average \nmax\ and fractional $\sigma$\\
NmaxWtCor, SigNmaxWtCor & Weighted mean \nmax\ and fractional standard error of the mean\\
NmaxMed, SigNmaxMed & Median \nmax\ and Median absolute deviation converted to fractional $\sigma$ \\
NMAXCOR to NMAXFLI & \nmax\ values for pipelines COR, ELS, GAU, SYD, A2Z, DIA, CAN, CV, HON \& FLI, respectively \\
DNuCor, SigDNuCor & Zero-point adjusted average \dnu\ and fractional $\sigma$\\
DNuWtCor, SigDNuWtCor & Weighted mean \dnu\ and fractional standard error of the mean\\
DNuMed, SigDNuMed & Median \dnu\ and Median absolute deviation converted to fractional $\sigma$\\
DNUCOR to DNUCAN & \dnu\ values for pipelines COR, ELS, GAU, SYD, A2Z, DIA \& CAN, respectively \\
    \hline
    
    \end{tabular}

    \caption{Filtered asteroseismic measurements and averages.}
    \label{table:useddata}
\end{table*}

\end{document}